 \documentclass[twocolumn, twocolappendix]{aastex63}
\usepackage{amsmath,amstext}
\usepackage[T1]{fontenc}
\usepackage{apjfonts} 
\usepackage[figure,figure*]{hypcap}
\usepackage{commath, mathrsfs}
\usepackage{color, upgreek}

\usepackage[version=4]{mhchem}
 
\newcommand{\keV}{\, {\rm keV}}
\newcommand{\eV}{\,{\rm eV}}
\renewcommand{\sec}{{\,\rm s}}

\newcommand{\kms}{\,{\rm km\,s^{-1}}}

\newcommand{\AU}{ \, {\rm au}}
\newcommand{\au}{\AU}
\newcommand{\Msun}{{\rm \, M_{\odot}}}

\newcommand{\yr}{\, {\rm yr} }

\newcommand{\Myr}{\, {\rm Myr} }

\newcommand{\dd}{{\rm d}}

\newcommand{\cm}{{\, \rm cm}}

\newcommand{\K}{{\rm K}}
\newcommand{\erg}{{\rm \, erg}}

\ifx\arcsec\undefined
\newcommand{\arcsec}{\, {\rm arcsec}}
\fi

\newcommand{\rad}{{\,\rm rad}}

\renewcommand{\arraystretch}{1.4}

\newcommand{\braket}[1]{\left(#1\right)}

\newcommand{\e}[1]{\times 10^{#1}}

\newcommand{\beqas}{\begin{eqnarray*}}
\newcommand{\eeqas}{\end{eqnarray*}}

\ifx\del\undefined
\newcommand{\del}[2]{\frac{\dd #1}{\dd #2}}
\fi

\ifx\appref\undefined
\newcommand{\appref}[1]{Appendix~\ref{#1}}
\fi
\newcommand{\fref}[1]{Figure~\ref{#1}}
\newcommand{\tref}[1]{Table~\ref{#1}}

\ifx\secref\undefined
\newcommand{\secref}[1]{Section~\ref{#1}}
\fi
\newcommand{\eqnref}[1]{Eq.(\ref{#1})}

\newcommand{\mum}{\, {\rm \mu m}}

\newcommand{\cs}{c_{\rm s}}

\ifx\dif\undefined
\newcommand{\dif}[2]{\frac{{\rm d}#1}{{\rm d}#2}}
\fi

\makeatletter
\newcommand*{\rom}[1]{\expandafter\@slowromancap\romannumeral #1@}
\makeatother
\ifx\ion\undefined
\newcommand\ion[2]{#1$\;${\small\rmfamily\rom{#2}}\relax}
\fi

\newcommand{\nh}{n_{\text{\rm H}}}

\newcommand{\Kelvin}{{\rm \, K}}

\ifx\fh\undefined
\newcommand{\fh}{f_{\rm h}}
\fi

\ifx\rmxaa\undefined
\newcommand{\rmxaa}{}
\fi

   \newcommand{\mach}{\mathcal{M}}

	\renewcommand{\mum}{{\rm \,\upmu m}}
	
	\newcommand{\km}{{\,\rm km}}
	\renewcommand{\Msun}{{\,M_\odot}}
	\renewcommand{\rad}{{\,\rm rad}}
	
\shorttitle{Classification of Photoevaporative Wind}
\shortauthors{Nakatani}

\begin{document}

\title{
Broadening the Canonical Picture of EUV-Driven Photoevaporation of Accretion Disks 
}
\author[0000-0002-1803-0203]{Riouhei Nakatani}
\affiliation{NASA Jet Propulsion Laboratory, California Institute of Technology, 4800 Oak Grove Dr, Pasadena, CA 91109, USA}
\affiliation{RIKEN Cluster for Pioneering Research, 2-1 Hirosawa, Wako, Saitama 351-0198, Japan}
\email{ryohei.nakatani@jpl.nasa.gov}
\author[0000-0001-8292-1943]{Neal J. Turner}
\affiliation{NASA Jet Propulsion Laboratory, California Institute of Technology, 4800 Oak Grove Dr, Pasadena, CA 91109, USA}
\author[0000-0003-3882-3945]{Shinsuke Takasao}
\affiliation{Department of Earth and Space Science, Graduate School of Science, Osaka University, Toyonaka, Osaka 560-0043, Japan}

\begin{abstract}
Photoevaporation driven by hydrogen-ionizing radiation, also known as extreme-ultraviolet (EUV), profoundly shapes the lives of diverse astrophysical objects.
Focusing here mainly on the dispersal of protoplanetary disks, we construct an analytical model accounting for the finite timescales of photoheating and photoionization.
The model offers improved estimates for the ionization, temperature, and velocity structures versus distance from the central source, for a given EUV emission rate and spectral hardness. 
Compared to the classical picture of fully-ionized and isothermal winds with temperatures $\approx 10^4\Kelvin$ and speeds $\approx 10\kms$, our model unveils broader hydrodynamical and thermochemical states of photoevaporative winds. 
In contrast to the classical picture, T~Tauri stars with EUV luminosities around $10^{30}\erg\sec^{-1}$ have non-isothermal ionized winds at lower temperatures than the classical value if the spectrum is soft, with an average deposited energy per photoionization less than about 3.7~eV. 
Conversely, if the spectrum is hard, the winds tend to be atomic and isothermal at most radii in the disk.  
For lower EUV intensities, even with soft spectra, atomic winds can emerge beyond $\sim 10\au$ through advection. 
We demonstrate that the analytical model's predictions are in general agreement with detailed radiation-hydrodynamics calculations. 
The model furthermore illustrates how the energy efficiency of photoevaporation varies with the intensity and spectral hardness of the EUV illumination, as well as addressing discrepancies in the literature around the effectiveness of X-ray photoevaporation.
These findings highlight the importance of considering the finite timescales of photoheating and photoionization, both in modeling and in interpreting observational data.

\end{abstract}


\section{Introduction}

\newcommand{\begelman}{BMS83}


Hydrogen-ionizing radiation, also called extreme-ultraviolet (EUV;$13.6\lesssim h\nu \lesssim 100 \eV$), is widely recognized for its crucial role in various astrophysical contexts, including planetary atmospheres \citep[e.g.,][]{2009_MurrayClay, 2016_Owen}, protoplanetary disks \citep[e.g,.][]{1994_Hollenbach}, molecular clouds \citep[e.g.,][]{1954_Kahn, 1989_Bertoldi, 1990_Bertoldi, 2019_Nakatani}, and galactic minihalos \citep[e.g.,][]{2004_Shapiro, 2005_Iliev, 2020_Nakatani}.   
EUV radiation induces not only photoionization of hydrogen-dominated gas but also associated heating, driving photoevaporative winds that can disperse structures within a finite time. 
Essentially, this means that ionizing radiation has exerted a significant influence on the evolution and formation processes of planets, stars, and galaxies since the dawn of the cosmic reionization epoch. 

Classically, it is commonly assumed that EUV-heated gas is fully ionized and exhibits temperatures of $\approx 10^4\Kelvin$, regulated by a balance between heating and cooling processes, resulting in supersonic photoevaporative winds. 
However, this scenario is valid only when the timescales of heating and ionization are much shorter than the dynamical timescales, and thus it could vary widely depending on factors such as the magnitude of gravity exerted by the object involved, its configuration, and EUV radiation environments, e.g., EUV flux and spectrum. 
Consequently, a diverse range of modes is possible for EUV-driven photoevaporation. 
Given that EUV photoevaporation occurs in diverse environments, there is a pressing need for a systematic and comprehensive understanding of this process. However, such an understanding is not yet fully developed.

In the context of protoplanetary disk dispersal, which constitutes the primary focus of this study, EUV-driven photoevaporation is considered as one of the mechanisms responsible for the final-stage dispersal of the gas disks \citep{2014_Alexander, 2022_Pascucci},
consequently determining the time limit available for (gas giant) planet formation.
Therefore, a comprehensive understanding of EUV photoevaporation is of great importance for advancing the planet formation theory. 

The current canonical picture of EUV-driven photoevaporation in protoplanetary disks can be outlined as follows: 
EUV energy is injected into the \ion{H}{1} layer of the disk just ahead of the ionization front, where the EUV optical depth $\tau_\mathrm{EUV}$ is approximately unity (\fref{fig:schematic}), 
Subsequently, the gas undergoes immediate photoionization to form ionized thermal winds emanating from the disk surface. 
These winds typically attain around $\approx 10^4\Kelvin$, a value determined by the thermal balance between photoionization heating and various radiative cooling within the \ion{H}{2} region, including metal forbidden lines, Ly$\alpha$ emissions, and radiative recombination.  
The resulting sound speed corresponds to $\approx 10\kms$, providing a typical speed of the transonic flows. 

This conceptual framework was initially proposed in the context of disk photoevaporation around massive stars \citep{1993_Hollenbach, 1994_Hollenbach} and has been applied to the EUV photoevaporation of protoplanetary disks in general, including around young low-mass stars \citep[e.g.,][]{1993_Shu_b}. 
However, the validity of this generalization is not straightforward, given that the typical EUV emission rate $\Phi_\mathrm{EUV}$ of young low-mass stars \citep[$\lesssim10^{42}\sec^{-1}$; e.g.,][]{2005_AlexanderClarkePringle, 2014_Pascucci}
is likely several orders of magnitude less than that of massive stars \citep[$\Phi_\mathrm{EUV} \approx  10^{48}$--$10^{49}\sec^{-1}$; e.g.,][]{1987_Maeder, 1994_Hollenbach}. 
Qualitatively, the lower $\Phi_\mathrm{EUV}$ implies a longer time required for heating and ionizing the gas. 
Consequently, the canonical picture, which implicitly assumes an infinitely short photoheating and photoionization timescale, is expected to be applicable only in cases where $\Phi_\mathrm{EUV}$ is sufficiently high (see also discussions in Appendix~A of \citet{1994_Hollenbach}).

Moreover, the EUV spectrum of massive stars notably includes a significant contribution from photospheric emission with an effective temperature of $ \approx 30000$--$50000\Kelvin$. 
This spectral attribute is softer in comparison to the EUV spectrum of young low-mass stars, 
where metal line emissions from $\sim 10^5\Kelvin$ plasma are predominant \citep{2021_ShodaTakasao}. 
Since the absorption cross-section of the gas for EUV is generally lower for higher photon energies, the difference in spectrum hardness can profoundly influence even the qualitative chemo-dynamical aspects of EUV-driven photoevaporation. 

These considerations regarding application limits raise pertinent questions: Under what parameter space (EUV emission rate, spectral hardness, and distance from the source) does the canonical picture hold true? And what chemical and hydrodynamical states might an EUV-driven wind exhibit in the other parameter spaces?
How might these insights be relevant to photoevaporation induced by alternative photoheating processes, such as far-ultraviolet (FUV) grain photoelectric heating and X-ray photoionization heating?
To tackle these questions, we discuss how the picture of EUV photoevaporation evolves contingent upon the stellar mass $M_*$, EUV emission rate $\Phi_\mathrm{EUV}$, and EUV spectrum while incorporating the finite timescales of photoheating and photoionization.
We develop an analytical model based on phenomenology that adequately treats these timescales as well as the dynamical timescale of winds and recombination timescale. 
Our phenomenological model draws inspiration from the analytical model for the Compton-heated winds of active galactic nuclei presented in \citet{1983_Begelman} (hereafter \begelman{}).

This paper is organized as follows.
We begin by outlining the basic setup and introducing fundamental quantities for our model in \secref{sec:analytic_theory}.
Following this, in \secref{sec:classification}, we provide a qualitative overview of the key findings from our model, reserving detailed quantitative derivations for \secref{sec:cs}. 
These sections delve into discussing the typical temperature, wind speed, and ionized state EUV-driven winds would have across various regions of the parameter space. 
In \secref{sec:comparison_HD_simulations}, we compare model predictions with hydrodynamics simulations incorporating self-consistent radiative transfer and nonequilibrium thermochemistry to assess the predictability. 
To demonstrate the practical utility of the model in understanding the physical properties of photoevaporating disks, we explore the radial extension of the \ion{H}{2} region in \secref{sec:limiting_radius} and analyze the energy efficiency of EUV heating in driving winds in \secref{sec:work_efficiency}. 
In \secref{sec:discussion}, we extend the application of our model to photoevaporation driven by far-ultraviolet (FUV) and X-ray radiation.
Based on this application, we address a possible cause for a long-standing issue in the field --- the divergent conclusions regarding the effectiveness of X-ray photoevaporation. 
Additionally, caveats and model limitations are presented in this section. 
Finally, we provide a summary of our major findings and outline potential avenues for future prospects in \secref{sec:summary}. 

\section{Fundamental Quantities}   \label{sec:analytic_theory}

The purpose of this study is to construct an updated picture of EUV-driven thermal winds. 
To that end, this section focuses on introducing the fundamental setup of the model and the key quantities for the energetics and thermochemical structures of the system, such as the gravitational timescale and the rates of photoheating and photoionization. 
A list of the symbols used in our model is available in Appendix~\ref{sec:list_of_variables}.

\begin{figure}
    \centering
    \includegraphics[clip, width = \linewidth]{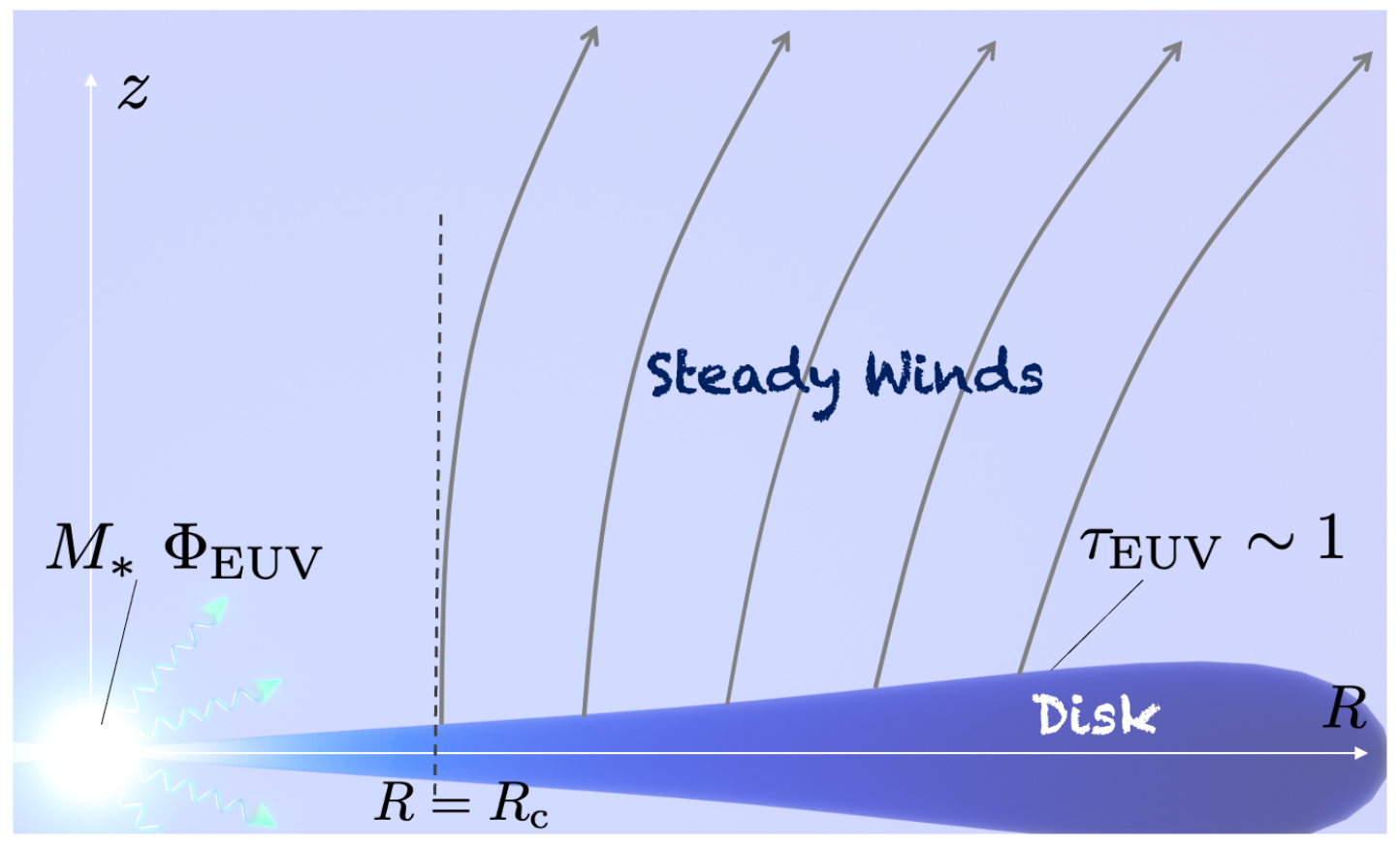}
    \caption{Schematic picture of a disk under EUV-driven photoevaporation. 
    The radiation source with a mass of $M_*$ and an EUV emission rate of $\Phi_\mathrm{EUV}$ is illuminating the disk. 
    The EUV photon hits the disk surface, where the optical depth to EUV is approximately unity, and photoionizes atomic hydrogen there, depositing excessive energy remaining after photoionization to the gas. 
    As a result, photoevaporative winds arise from the disk surface. 
    Vigorous winds are possible only beyond the critical radius $R_\mathrm{c}$ owing to weak gravity at a distance. 
    In a classical picture, the winds are ionized and have an equilibrium temperature of $\approx 10^4\Kelvin$. 
    The corresponding isothermal sound speed is $c_\mathrm{s} \approx 10\kms$, which sets a typical flow speed of the winds due to the thermally-driven nature of photoevaporation. 
    }
    \label{fig:schematic}
\end{figure}

The basic setup of our model is schematically depicted in \fref{fig:schematic}: 
a geometrically thin disk predominantly composed of hydrogen is irradiated by the central EUV source. 
Hydrogen exists in atomic form in the upper layers of the disk due to the photodissociation of molecular hydrogen, which is the dominant species in the deeper layers. 
The energy of EUV photons is deposited to the gas at the atomic layer by heating associated with photoionization, \ce{H + EUV -> H+ + e-},  
and the gas can escape from the gravitational binding of the central object if it has reached a sufficiently high temperature. 
This escaping gas gives rise to steady thermal winds emanating from the so-called base, corresponding to the optically thick surface of the disk where $\tau_\mathrm{EUV}\sim 1$.

\newcommand{\avesigma}{\bar{\sigma}_0}
\newcommand{\aveenergy}{\bar{E}_0}
\newcommand{\ceq}{c_\mathrm{eq}}
\newcommand{\phieuv}{\Phi_\mathrm{EUV}}
\newcommand{\vg}{v_\mathrm{g}}

\subsection{Critical Radius}
\label{sec:critical_radius}
The winds can have a temperature of $\approx 10^4\Kelvin$ (thermal equilibrium temperature) at most because cooling due to forbidden lines and Ly$\alpha$ emission quickly becomes effective as the temperature approaches the equilibrium value. 
The corresponding sound speed is $\ceq  \approx 10\kms$, and this sets the order of a maximum achievable speed for EUV-driven winds.

We can define a threshold radius below which EUV-heated gas cannot escape from the gravitational binding even when it achieves the maximum flow speed $\ceq $. 
\citet{1994_Hollenbach} introduce such a threshold radius (termed ``gravitational radius'') as  
\[
    R_{\rm g} \equiv \frac{GM_*}{\ceq ^2}. 
\]
through dimensional analysis comparing the thermal energy and gravitational potential energy.  
The gravitational radius exactly corresponds to the Compton radius $R_{\rm IC}$ in \begelman{}. 
\begelman{} predicted that the actual threshold radius separating the wind-excited and wind-inhibited regimes might be present at the cylindrical distance from the source of $R \approx 0.1$--$1R_{\rm IC}$ instead of $R = R_\mathrm{IC}$.  
This is directly examined by hydrodynamics simulations in \citet{1996_Woods}, and they found the actual threshold radius to be $\approx 0.2R_{\rm IC}$.

\citet{2003_Liffman} suggests that the threshold radius of $0.2R_{\rm IC}$ corresponds to the radius at which the sum of photoheated gas's enthalpy and mechanical energy equals zero. 
Although the physical origin of the threshold radius is still worth exploring, in this study, we follow \citet{2003_Liffman} to define the ``critical radius'' as
\begin{equation}
    R_{\rm c} 
    \equiv \frac{R_{\rm g}}{2c_p}, 
    \label{eq:critical_radius}
\end{equation}
where 
$c_p$ is a normalized specific heat at constant pressure
\[
    c_p \equiv \frac{\gamma}{\gamma - 1 }
\]
with $\gamma$ being the specific heat ratio of the gas. 
\eqnref{eq:critical_radius} is a modified version of $R_\mathrm{g}$, including correcting factors from enthalpy and centrifugal force due to gas's angular momentum.

We define the gravitational velocity $\vg $ as 
\begin{equation}
    \vg  
    \equiv \sqrt{\frac{GM_*}{2c_p R}}
    = \ceq    x ^{-1/2}, 
    \label{eq:vg_ceq}
\end{equation}
so that $\ceq  = \vg $ at $R = R_\mathrm{c}$. 
Here, $ x $ is a normalized distance 
\[
     x  \equiv \frac{R}{R_\mathrm{c}}. 
\]
The corresponding gravitational timescale is 
\begin{equation}
    t_{\rm g} \equiv 
    \frac{R}{\vg }
    = 
    \sqrt{\frac{2c_p R^3}{GM_*}}, 
    \label{eq:gravitational_timescale}
\end{equation}
The normalized radius $x$ corresponds to the normalized radius ``$\xi$'' in \begelman{} except for the factor of $1/2c_p$ and the equilibrium temperature. 
We adopt denoting ``$x$'' for the normalized distance to highlight these differences and to avoid confusion with the ionization parameter, which is often denoted by $\xi$. 

In this paper, we focus on the region where vigorous winds possibly form, 
and thus we only consider the region with $ x  \geq 1$, which is a necessary condition for EUV-heated gas to escape freely without being inhibited by gravity. 
We leave modeling the inner $x < 1$ region for future studies, since its nearly hydrostatic structure requires a different approach from the one presented in this study, as discussed in \secref{sec:discussion:model_limitation}.

\subsection{EUV Heating Rate}
\label{sec:euv_heating_rate}
We calculate the frequency-dependent (specific) photon number flux from the point radiation source with radial ray tracing, 
\[
    F_\nu (r, N_{\rm HI})= \frac{\Phi_\nu }{4\pi r^2} e^{-\sigma_\nu N_{\rm HI}},
\]
where $\nu$ is EUV frequency, 
$\Phi_\nu$ is the specific EUV emission rate of the radiation source, 
$r$ is the spherical distance from the source, 
$\sigma_\nu $ is the absorption cross-section of atomic hydrogen for EUV, 
and $N_{\rm HI}$ is the radial column density of atomic hydrogen along an EUV ray defined as 
\[
    N_\mathrm{HI} \equiv \int \dd r \, y_\mathrm{HI} \nh 
\]
with $y_\mathrm{HI}$ and $\nh$ being atomic hydrogen abundance and hydrogen nucleus number density, respectively. The atomic hydrogen number density is computed by $n_\mathrm{HI} = y_\mathrm{HI}\nh$. 
Energetic EUV photons can also be absorbed by \ion{He}{1} and \ion{He}{2} in general, but we ignore their contributions. 
This simplification has a minimal impact on the results of our order-of-magnitude phenomenological discussions in this study. 
We will delve into the effect of this simplification in more detail in \secref{sec:discussion:model_limitation}.

The expression of $F_\nu$ implicitly assumes the direct EUV component dominating over the diffuse component. 
This assumption is based on the findings of \cite{2013_Tanaka}, where the authors performed full 2D radiative transfer calculations and observed that the direct EUV dominates over the diffuse EUV.
This is in contrast to the classical findings of \citet{1994_Hollenbach}, where they employed the 1+1D radiative transfer and concluded that the inner ionized atmosphere completely cuts off the direct EUV within the gravitational radius. 
This discrepancy likely results from the difference in the treatment of the diffuse component transfer.
Similarly to \citet{2013_Tanaka}, the direct EUV has been observed to reach the outer radii in more recent self-consistent radiation hydrodynamics simulations \citep{2018_Nakatani, 2018_Nakatanib, 2021_Komaki, 2022_NakataniTakasao}, even when the critical radius is well resolved. 
A nearly hydrostatic, ionized atmosphere is formed at inner radii, agreeing with the prediction of the previous hydrostatic models, but it does not terminate the radial EUV rays within the gravitational radius. 
These updated results justify our assumption on $F_\nu$ above. 

As for photoheating, the excess energy of EUV photons over the ionization potential energy ($h\nu_1 \approx 13.6\eV$) is responsible for EUV heating, and thus the specific heating rate is 
\begin{equation}
\begin{split}
    \Gamma_{\rm EUV} \equiv 
    &    \frac{y_\mathrm{HI}}{m} \int_{\nu_1}^{\nu_\mathrm{max}} \sigma_\nu (h\nu - h\nu_1) \frac{\Phi_\nu }{4\pi r^2} 
    e^{-\sigma_\nu N_{\rm HI}}\, \dd \nu \\
    &   =   \frac{y_\mathrm{HI} }{m}  F_0
    \delta \langle \sigma \rangle
            \langle \Delta E \rangle_i  
\end{split}
\label{eq:gamma_EUV}
\end{equation}
where $m$ is the total gas mass per hydrogen nucleus, which can include the mass of other minor elemental species, like Helium, and is related to the gas density $\rho$ through $m \equiv \rho/\nh$; 
$\nu_\mathrm{max}$ is the upper limit of the EUV frequency, typically set to $\sim 100\eV$; 
and $F_0 $ is the unattenuated total EUV number flux at the spherical distance $r$, 
\[
    F_0 \equiv  \int_\mathrm{\nu_1}^\mathrm{\nu_\mathrm{max}} \dd \nu \, {F_\nu (r, 0)}  = \frac{\phieuv }{4\pi r^2}. 
\]
Here, $\phieuv $ is the EUV emission rate of the radiation source. 
The dimensionless quantity $\delta$ is the ratio of the attenuated flux to the unattenuated flux $F_0$ at $N_{\rm HI}$, 
\[
\delta (N_{\rm  HI})\equiv 
\dfrac{\int \dd \nu \,  F_\nu (r, N_\mathrm{HI}) }{F_0}
=
\dfrac{\int \dd \nu \, \Phi_\nu \exp \braket{- \sigma_\nu N_{\rm HI}}}{\phieuv }. 
\]
This parameter describes the fraction of the EUV photons remaining at $N_{\rm HI}$ without being attenuated. 
The average cross-section of the remained EUV, $\langle \sigma \rangle $, is defined by 
\[
\langle \sigma \rangle (N_{\rm  HI}) \equiv 
\frac{\int \dd \nu \, \sigma_\nu F_\nu (r, N_\mathrm{HI})}{\int \dd \nu \, F_\nu (r, N_\mathrm{HI}) }, 
\]
and $\langle \Delta E \rangle_i$ is an average deposited energy per ionization at $N_{\rm HI}$
\[
\langle \Delta E \rangle_i (N_{\rm  HI}) \equiv 
\frac{\int \dd \nu \, \sigma_\nu (h\nu - h\nu_1) F_\nu (r, N_\mathrm{HI})}{\int \dd \nu \, \sigma_\nu F_\nu (r, N_\mathrm{HI})}. 
\]
All of $\delta$, $\langle \sigma \rangle$, and $\langle \Delta E \rangle_i$ are dependent only on $N_{\rm HI}$ and a priori calculable, once $\Phi_\nu$ is given. 
The former two are monotonically decreasing functions because the remaining EUV photons decrease as they go through the gas, and the average energy of the remaining photons increases to have a smaller cross-section on average. On the other hand, $\langle \Delta E \rangle_i$ is a monotonically increasing function since the average energy of remaining EUVs increases as they go through the gas. 
The product $ \delta \langle \sigma\rangle \langle \Delta E\rangle_i$ is a monotonically decreasing function with respect to $N_{\rm HI}$, 
which simply indicates that the less attenuated, the more energy is available for heating. 

For convenience, we rewrite the product $ \delta \langle \sigma\rangle \langle \Delta E\rangle_i$ in the dimensionless form of
\[
 \begin{split}
     \chi_\mathrm{e} & \equiv \delta 
 \braket{\frac{\langle\sigma\rangle}{\avesigma }}
 \braket{\frac{\langle \Delta E \rangle_i }{\aveenergy }}
 \\ & 
 =\int \dd \nu \frac{\Phi_\nu}{\phieuv} \frac{\sigma_\nu}{\avesigma} \frac{h\nu - h\nu_1}{\aveenergy} e^{-\sigma_\nu N_\mathrm{HI}}. 
 \end{split}
\]
where
\[
\begin{gathered}
 \avesigma  \equiv \left.\langle \sigma \rangle\right|_{N_{\rm HI} = 0}
 = \phieuv^{-1} \int \dd \nu \, \sigma_\nu \Phi_\nu 
 \\
 \begin{aligned}
     \aveenergy  & 
     \equiv \left. \langle \Delta E \rangle_i\right|_{N_{\rm HI} = 0}
     \\
    & = \braket{\int \dd \nu \, \sigma_\nu \Phi_\nu  }^{-1} \int \dd \nu \, \sigma_\nu \Phi_\nu \braket{h\nu - h\nu_1}
 \end{aligned}
 \\
\end{gathered} 
\]
With these notations, \eqnref{eq:gamma_EUV} can be expressed in a simpler form 
\begin{equation}
    \Gamma_\mathrm{EUV}
    = \frac{y_\mathrm{HI}}{m}
    \frac{\phieuv }{4\pi r^2}\avesigma  \aveenergy  \chi_\mathrm{e}. 
    \label{eq:gamma_EUV:2}
\end{equation}
The attenuation factor $\chi_\mathrm{e}$ depends only on $N_\mathrm{H}$ and is a monotonically decreasing function ranging in $0 < \chi_\mathrm{e} \leq 1$; the upper and lower limits are achieved when $N_\mathrm{HI} = 0$ and $N_\mathrm{HI}\rightarrow \infty$, respectively.

In this paper, we perform our analysis under the idealization that a large portion of the wind region is optically thin ($\chi_\mathrm{e} \sim 1$), following the approach of \begelman{}.
While this idealization has its limitations and somewhat restricts the applicability of our model, it allows us to streamline our analysis. 
Importantly, despite this limitation, constructing a model based on this simplification is a crucial initial step towards developing a more comprehensive model available for a broader range of spectra.
Furthermore, the insights obtained from our simplified model lay the groundwork for addressing more general cases and prove valuable in systematically interpreting the outcomes of such broader contexts. 
The idealization would serve as a reasonable zeroth-order approximation in cases where spectra have minimal dispersion in photon energy, so that most photons are intensively absorbed at a certain column density. For instance, this includes relatively soft spectra, where most photons have energies close to the Lyman limit, or harder spectra with delta-function-like shapes. 
(See \secref{sec:discussion:model_limitation} for the potential impacts of this adopted approximation.)

\subsection{Photoionization and Recombination Timescales}
\label{sec:sub:photoionization_recombination_timescales}
The EUV-driven winds undergo photoionization, \ce{H + EUV -> H+ + e-}, and radiative recombination, \ce{H+ + e- -> H + photon}, after launched. 
The photoionization rate coefficient is the number of photons absorbed per unit time per hydrogen atom
\begin{equation}
\begin{split}
    k_\mathrm{ioni} 
    & \equiv \int \dd \nu \,  \sigma_\nu  \frac{\Phi_\nu}{4\pi r^2} 
    e^{- \sigma_\nu N_{\rm HI}}
    = \frac{\Phi_{\rm EUV}}{4\pi r^2} \avesigma  \chi_\mathrm{i}
    ,
\end{split}
\label{eq:kioni}
\end{equation}
where $\chi_\mathrm{i} \equiv \delta  \langle\sigma\rangle/\bar{\sigma}_0 $ is another $N_\mathrm{HI}$-dependent, a priori calculable EUV attenuation factor, and $0 < \chi_\mathrm{i}\leq 1$. 
Again, similarly to $\chi_\mathrm{e}$, we use the assumption of $\chi_\mathrm{i} \sim 1$ in this study, as mentioned below \eqnref{eq:gamma_EUV:2}.

The photoionization timescale is the inverse of the photochemical reaction rate coefficient
\begin{equation}
    t_{\rm ioni}
\equiv k_\mathrm{ioni}^{-1}
= 
\braket{\frac{\Phi_{\rm EUV}}{4\pi r^2} \avesigma }^{-1} .
\label{eq:tioni}
\end{equation}
This quantity sets the average timescale on which a single hydrogen atom takes to be photoionized.

Next, we introduce the recombination timescale 
\[
    t_\mathrm{rec} \equiv \braket{\nh \alpha} ^{-1}
\]
with $\alpha$ being the case~B recombination coefficient. 
We give $\alpha$ in the form of
\begin{equation}
\alpha 
    =   \alpha_\mathrm{eq} \braket{\frac{\cs }{\ceq }}^{-2 \beta}, 
\label{eq:recombination_coeff}
\end{equation}
where $\alpha_\mathrm{eq}$ is the case~B recombination coefficient at $T\approx 10^4\Kelvin$ and, typically, $\alpha_\mathrm{eq} \sim 2\e{-13}\cm^3\sec^{-1}$; $\beta$ is the power-law exponent for temperature and typically $0.5 \lesssim \beta \lesssim 0.75 $ \citep[cf. the UMIST database,][]{2012_UMIST}.

In contrast to $t_\mathrm{ioni}$, the recombination timescale is dependent on the gas density (and temperature) of the winds. 
\citet{1994_Hollenbach} and \citet{2013_Tanaka} characterize $\nh$ at the base of EUV-driven winds by Str\"omgren condition, i.e., ionization-recombination equilibrium along line of sights from the radiation source, as 
 \begin{equation}
 \begin{split}
     n_\mathrm{base} & = C \sqrt{\frac{3\phieuv }{4\pi r^3 \alpha }}    \\
     & \approx 7.6\e{5} \cm^{-3}
     \braket{\frac{\phieuv}{10^{40}\sec^{-1}}}^{1/2}
     \braket{\frac{r}{1\au}}^{-3/2}
     \\
     &\quad
     \times 
     \braket{\frac{\alpha}{2\e{-13}\cm^3\sec^{-1}}}^{-1/2}
     \braket{\frac{C}{0.4}}
 \end{split}
    \label{eq:base_density}
 \end{equation}
Here, $C$ is an artificial dimensionless factor to match the numerical results, originating from the geometry of the ionized region \citep[typically, $0.1\lesssim C\lesssim 0.4$;][]{1994_Hollenbach, 2004_Font, 2013_Tanaka}. 

Substituting this density into $t_\mathrm{rec}$,
the recombination timescale is expressed as
\begin{equation}
    t_\mathrm{rec} 
    =  \braket{\frac{3C^2\phieuv  \alpha_\mathrm{eq} }{4\pi r^3}}^{-1/2} \braket{\frac{\cs }{\ceq }}^\beta. 
    \label{eq:trec_base_density}
\end{equation}
Strictly, this equation specifically provides the recombination timescale near the launch point in the case where recombination dominates the opacity for EUV in the wind region. 
Typically, as the evaporating gas flows away from the disk, its density decreases due to expansion, leading to a longer recombination timescale and consequently a higher ionization degree.
Moreover, the base density could vary depending on the EUV spectral hardness in practice. 
Eqs.\eqref{eq:base_density} and \eqref{eq:trec_base_density} do not account for such variations and are probably applicable only under conditions where recombination dominates the opacity over absorption by advected atomic hydrogen. 
This condition typically holds for softer spectra. 
Regardless of these uncertainties, in this study, we uniformly use \eqnref{eq:trec_base_density} as a rough estimate for the recombination timescale to the height of $\sim R$ and for any spectra. 
We discuss the potential impacts of this simplification in \secref{sec:discussion:model_limitation}.

The ratio of the ionization timescale to the recombination timescale is a key metric for characterizing the ionization degree of the gas. 
By using Eqs.\eqref{eq:tioni} and \eqref{eq:trec_base_density}, this can be expressed as 
\begin{equation}
    \begin{split}
    \frac{t_\mathrm{ioni}}{t_\mathrm{rec}}
    =  q^{1/2} 
    \braket{\frac{t_\mathrm{ioni} }{R/\ceq}}^{1/2} 
    \braket{\frac{\cs }{\ceq }}^{-\beta}
    \quad \braket{ = \frac{3C^2 \chi_\mathrm{i}^{-1}}{ n_\mathrm{base}\avesigma R}}
    . 
\end{split}
\label{eq:tioni_trec_ratio}
\end{equation}
with $q$ being a dimensionless constant
\begin{equation}
    q \equiv \frac{3C^2\alpha_\mathrm{eq}}{\ceq \avesigma}. 
    \label{eq:q}
\end{equation}
This constant is uniquely set once an EUV spectrum is given; a harder spectrum results in a smaller average cross-section $\avesigma$, leading to a larger $q$. 
We will explain this spectral hardness parameter is one of the parameters used for the classification of EUV photoevaporation in \secref{sec:sub:euv_classes}. 

To interpret the physical meaning of the $q$ parameter in a more direct manner, it is convenient to evaluate the ratio of $t_\mathrm{ioni}$ to $t_\mathrm{rec}$ with $\cs = \ceq$: 
\[
    \frac{t_\mathrm{ioni}}{t_\mathrm{rec, eq}}
    = 
    q^{1/2}
    \braket{\frac{t_\mathrm{ioni} }{R/\ceq}}^{1/2},
\]
where $t_\mathrm{rec,eq} \equiv t_\mathrm{rec}(\cs = \ceq)$.
The above equation gives 
\[
    q =  \braket{\frac{R/\ceq}{t_\mathrm{rec,eq}}}  \braket{\frac{t_\mathrm{ioni} }{t_\mathrm{rec,eq}}}
    =  \braket{\alpha_\mathrm{eq} n_\mathrm{base} \frac{R}{\ceq} }\braket{k_\mathrm{ioni} t_\mathrm{rec, eq}}^{-1}.
\]
Hence, $q$ is the number of recombination while the gas traveling to the height of $R$, with respect to the number of photoionization during one recombination event. 
Thus, $q $ assesses the efficiency of reproducing atomic hydrogen through recombination against photoionization. 

An increase in spectral hardness leads to a decrease in $\avesigma$, thereby relatively weakening photoionization compared to recombination by extending the photoionization timescale.
This effectively enhances the relative importance of atomic hydrogen reproduction, meaning a large $q$.



\section{Classification of photoevaporating disks}
\label{sec:classification}

In this section, we summarize a qualitative overview of the major findings derived from our phenomenological model. 
Our model is designed to characterize the typical temperature, namely the isothermal sound speed $\cs $, typical flow speed, and the ionization state of winds launched at a distance $R$ under the influence of a radiation source emitting EUV with a rate of $\phieuv $. 
Since photoevaporative winds are thermally driven, the sound speed can also dictate the typical flow velocity. 
We identify three distinct classes in EUV photoevaporation, each contingent upon the hardness of the EUV spectrum. 

Here, we provide the general pictures for each class, outlining the key parameters employed in their classification.
We begin by discussing the possible ionization states of winds, which form the foundational basis of our model (\secref{sec:sub:wind_types}).
Subsequently, we delve into the heating timescale, which stands as the most important quantity to define the hydrodynamical states of winds (\secref{sec:sub:t_h}). 
Expanding on this foundation, the typical temperature and velocity of winds are estimated (\secref{sec:typical_flow_velocities}), followed by the introduction of the EUV photoevaporation classes (\secref{sec:sub:euv_classes}). 
Finally, we illustrate the distinctive characteristics of the three classes (one each in \secref{sec:sub:H-class}--\ref{sec:sub:Sr-class}). 
For detailed derivations of the quantities, parameter spaces, and classes reviewed in this section, the readers are referred to \secref{sec:cs}.  

As mentioned in Sections~\ref{sec:euv_heating_rate} and \ref{sec:sub:photoionization_recombination_timescales} and similarly to the approach of \begelman{}, this study focuses specifically on the cases where a significant portion of the wind region is optically thin (i.e., $\chi_\mathrm{e}\sim  1$ and $\chi_\mathrm{i} \sim 1$ in Eqs.\eqref{eq:gamma_EUV:2} and \eqref{eq:kioni}). 
While this simplifying approach does place limitations on the applicability, the constructed model provides a foundation and valuable insights for addressing more general cases.
We discuss the potential impacts of this simplification in \secref{sec:discussion:model_limitation}.
Throughout this section, we adopt a specific value of $\beta = 0.75$ for the power-law index of the recombination rate coefficient (\eqnref{eq:recombination_coeff}) for simplicity.

\subsection{Wind Types: Ionized VS Atomic}
\label{sec:sub:wind_types}
\begin{figure*}
    \centering
    \includegraphics[clip,width=\linewidth]{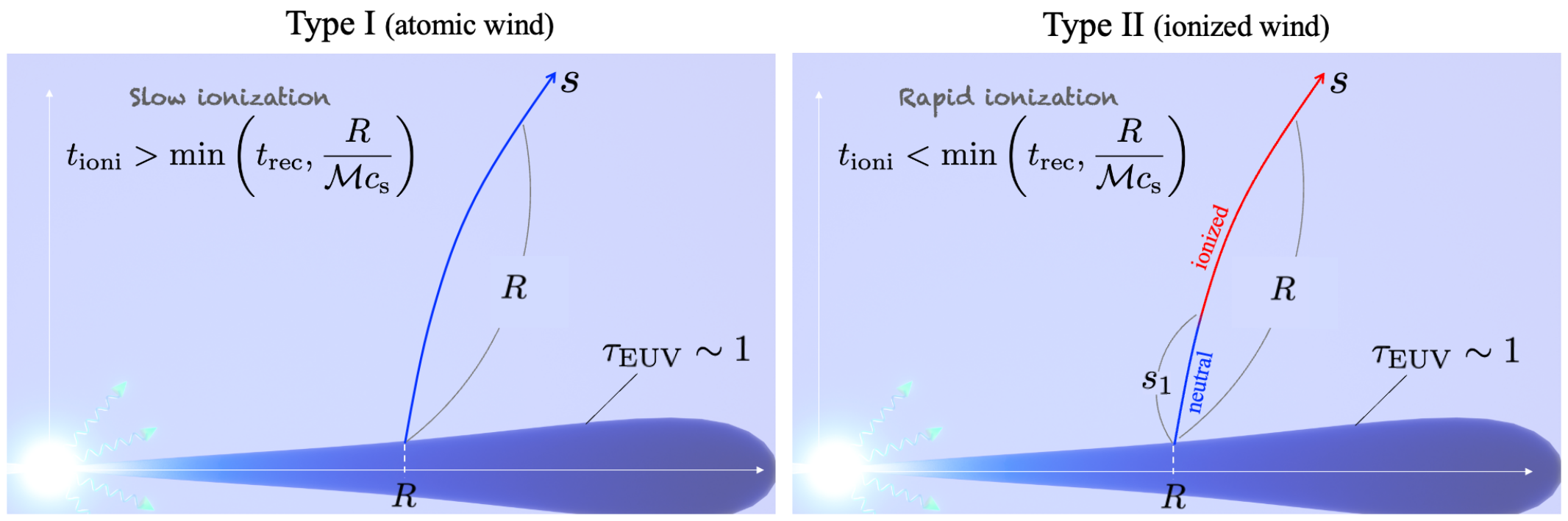}
    \caption{Two possible configurations of EUV-driven winds. 
    (left) The wind is essentially atomic while traveling to the height of $\sim R$. 
    This would occur when $\phieuv $ is small so that the ionization timescale is longer than either the wind crossing timescale or recombination timescale within the wind. 
    We term this type of atomic wind as Type~I. 
    The variable $s$ here is coordinates along streamlines. 
    (right) In contrast to Type~I, the photoionization timescale is shorter than the recombination timescale and wind crossing timescale. 
    In this case, the energy-injected gas can be ionized at some point $s_1$ between the launching point and the height $\sim R$. 
    Hydrogen atoms within the launched gas can survive for a photoionization timescale $t_\mathrm{ioni}$ on average, and thus the transition height can be given as $s_1\sim \mach \cs  t_\mathrm{ioni}$. 
    We refer to this ionized flow as Type~II. 
    Note that the neutral region ($s \lesssim s_1$) can be geometrically very thin, but here, we draw it vertically extended for visualization purposes. 
    }
    \label{fig:streamline_types}
\end{figure*}
To discuss the ionization state of winds, we introduce the coordinate $s$ along a streamline of steady winds, with $s = 0$ denoting the launch point (\fref{fig:streamline_types}). 
Initial energy injection takes place in the \ion{H}{1} layer near the launch point. 
If photoionization proceeds sufficiently rapidly, the launched gas can undergo complete photoionization, transitioning into the ionized wind before reaching a height of $s \sim R$ above the disk. 
In essence, for an ionized wind to form, the photoionization timescale must be shorter than both the recombination timescale and wind crossing timescale.

This consideration leads us to categorize EUV-driven thermal winds into two types: 

{\it (I) Winds characterized by atomic composition}, i.e., chemically neutral, within $s\leq R$ (left panel in \fref{fig:streamline_types}). 
This can arise when either the photoionization timescale is longer than the recombination timescale or the crossing timescale of the launched gas
\begin{equation}
    t_\mathrm{ioni} > \mathrm{min} \braket{t_\mathrm{rec}, \frac{R}{\mach \cs }}, 
    \label{eq:typeI_condition}
\end{equation}
where $\mach$ is a representative Mach number in the wind within $s \leq R$. 
Qualitatively, this wind type is expected when $\phieuv $ is relatively low yet capable of gradually depositing energy into the gas to drive winds or when the spectrum is relatively hard, so $y_\mathrm{HI}$ in the ionization-recombination equilibrium is nearly unity. 

{\it (II) Winds transitioning into an ionized state at a certain height}, $s = s_1 (< R)$, and being atomic below that (right panel in \fref{fig:streamline_types}). 
This scenario aligns with the canonical picture of EUV photoevaporation. 
It occurs when $\phieuv $ is sufficiently high, causing the ionization timescale ($t_\mathrm{ioni}$) to be shorter than both the recombination and crossing timescales. 
\begin{equation}
    t_\mathrm{ioni} < \mathrm{min} \braket{t_\mathrm{rec}, \frac{R}{\mach \cs }}. 
    \label{eq:typeII_condition}
\end{equation}
Henceforth, we refer to the first and second types as Type~I and Type~II, respectively. 
While the picture of Type~II is more conventional for EUV photoevaporation, 
the numbering convention is chosen to facilitate immediate association with the chemical states of the winds. Specifically, Type~I corresponds to \ion{H}{1} winds, and Type~II corresponds to \ion{H}{2} winds.

\subsection{Characteristic Sound Speed and Heating Timescale}   \label{sec:sub:t_h}

The magnitude of photoheating determines the typical temperature and speed of winds. 
Photoheating can be gauged by the attainable sound speed as a result of heating.
In general, the wind's sound speed does not necessarily reach the equilibrium value $\ceq$ when the heating rate is small due to a finite photoheating rate.  

Following \begelman{}, we evaluate the attainable sound speed by the characteristic sound speed $c_\mathrm{ch}$ defined as 
\begin{equation}
    c_p c_\mathrm{ch}^2 =  \bar{\Gamma}_\mathrm{EUV} \frac{R}{c_\mathrm{ch}}
    \label{eq:c_ch_def}
\end{equation}
where $\bar{\Gamma}_\mathrm{EUV}$ is a time-average specific heating rate
\[
    \bar{\Gamma}_\mathrm{EUV} \equiv \braket{\frac{R}{c_\mathrm{ch}}}^{-1} \int \Gamma_\mathrm{EUV} \frac{\dd s}{c_\mathrm{ch}}. 
\]
The integral is taken from the base to the height $s = R$, treating $c_\mathrm{ch}$ as a constant while taking into account the variability of $y_\mathrm{HI}$ in $\Gamma_\mathrm{EUV}$ (cf. \eqnref{eq:gamma_EUV}); thus, the form of $c_\mathrm{ch}$ differs by the wind types as we will see below. 
The RHS of Eq.\eqref{eq:c_ch_def} indicates the total specific energy deposited until the gas reaches the height of $R$.  
\footnote{
Defining $c_\mathrm{ch}$ using the volume-integrated photoheating rate divided by the area of a flow bundle on the right-hand side of \eqnref{eq:c_ch_def} can give a more accurate estimation of wind's sound speed. 
Nevertheless, using the path integral instead, as in \eqnref{eq:c_ch_def}, suffices for the order-of-magnitude estimate in the present study. 
}

The characteristic sound speed corresponds to the isothermal sound speed that the launched gas would attain if all the specific energy integrated during its ascent to the height $\sim R$ were converted into the specific enthalpy. 
It is important to note that the characteristic sound speed serves as a metric of the heating magnitude and does not necessarily reflect the actual sound speed of the gas $\cs $, as cooling and gravity play roles in determining the hydrodynamics of winds. 
We will elaborate on this aspect later in \secref{sec:typical_flow_velocities}. 

In \begelman{}, $c_p $ does not appear on the left-hand side (LHS) of \eqnref{eq:c_ch_def}.
This coefficient would not be uniquely determined, and one could potentially use another constant as long as it is $\mathcal{O}(c_p)$; this introduces some arbitrariness into the model. 
Nevertheless, 
such an arbitrariness does not significantly alter the results and conclusions presented in this paper.

Correspondingly to \eqnref{eq:c_ch_def}, the heating timescale is defined as 
\[
    t_\mathrm{h} \equiv 
    \frac{c_p c_\mathrm{ch}^2}{\bar{\Gamma}_\mathrm{EUV}}
    = \frac{R}{c_\mathrm{ch}}. 
\]
If the heating timescale is longer than the crossing timescale of an isothermal wind, $R/\ceq$, the wind does not have long enough time to be heated to the equilibrium temperature by reaching $\sim R$, within which the bulk of energy deposition occurs. 
Moreover, if $t_\mathrm{h}$ is further longer than the gravitational timescale, the wind is significantly inhibited by gravity.

Since the specific heating rate $\Gamma_\mathrm{EUV}$ (\eqnref{eq:gamma_EUV}) depends on the \ion{H}{1} abundance $y_\mathrm{HI}$, the expression for $c_\mathrm{ch}$ varies depending on whether the wind encompasses an ionized region or not. 
For atomic winds (Type~I; left panel in \fref{fig:streamline_types}), the hydrogen abundance can be approximated to $y_\mathrm{HI}\approx 1$ in $\Gamma_\mathrm{EUV}$ of \eqnref{eq:c_ch_def}, and $c_\mathrm{ch}$ reduces to 
\begin{equation}
    c_\mathrm{ch} =
    c_\mathrm{ch}^\mathrm{I} 
    \equiv  \braket{\frac{\Gamma_\mathrm{EUV}R}{c_p}}^{1/3}
    = \ceq  \braket{\frac{\varphi}{x}}^{1/3},
\label{eq:cchI_def}
\end{equation}
where $\varphi$ is the dimensionless EUV emission rate, which is $\phieuv $ normalized by the critical EUV emission rate, 
\begin{equation}
    \Phi_\mathrm{c } 
    \equiv 
    \frac{4\pi m \ceq GM_*}{2\avesigma  \aveenergy }, 
    \label{eq:critical_luminosity}
\end{equation}
Again, $x$ is the distance from the source in units of the critical radius $R_\mathrm{c}$ (\eqnref{eq:critical_radius}):
\[
    x \equiv \frac{R}{R_\mathrm{c}}. 
\]
In this study, we focus on the region with $x > 1$, where strong winds can be anticipated. 
We defer investigation of the inner region with $x < 1$ for future studies (see \secref{sec:discussion:model_limitation}).

Note that the critical EUV emission rate is defined in such a way that it depends solely on spatially invariant constants determined by the properties of the radiation source and the disk gas.
It is physically interpreted as the EUV emission rate when the power required to evaporate the gas against the effective gravity at the equilibrium sound speed, $m \ceq  GM_*/2 R^2$, equilibrates with the heating rate yielded by the local flux, $\aveenergy  \avesigma  \Phi_\mathrm{c}/4\pi R^2$ (see also \secref{sec:work_efficiency}). 


It is noteworthy that the specific internal energy $\sim \braket{c_\mathrm{ch}^\mathrm{I}}^2$ is not proportional to $\Gamma_\mathrm{EUV}$ but exhibits a slightly slower increase, scaling as $\Gamma_\mathrm{EUV}^{2/3}$. 
This originates from the drawback of increasing the gas temperature: while higher $\phieuv$ can increase the internal energy, the resulting short crossing time ($R/c_\mathrm{ch}$) leads to a shorter integration time of $\Gamma_\mathrm{EUV}$ and, thereby, a reduced total deposited energy during the gas traveling to the height of $\sim R$, where the bulk of photoheating and photoionization takes place.
This also implies that the fraction of the deposited energy going into the kinetic energy of the gas (and being drained by cooling) increases with $\phieuv$ (\secref{sec:work_efficiency}).

For Type~II winds (right panel of \fref{fig:streamline_types}),
the total energy gain is the sum of the initially injected energy until the launched atomic hydrogen gets ionized, plus the additional energy resulting from the reionizations of recombined hydrogen within the \ion{H}{2} region.
The initial deposited energy is estimated as the product of $t_\mathrm{ioni}$ and $\Gamma_\mathrm{EUV}$ with $y_\mathrm{HI}\approx 1$, which is equivalent to $\aveenergy$. 
To quantify this, we can introduce a spectral hardness parameter by normalizing the initially injected energy with the enthalpy of the gas at the equilibrium temperature:
\begin{equation}
\begin{split}
    \varepsilon  & \equiv \frac{\aveenergy }{m c_p \ceq ^2}
    \\ & 
    \approx 1 
    \braket{\frac{\aveenergy}{3.7 \eV}}
    \braket{\frac{m}{1.4m_\mathrm{H}}}^{-1}
    \braket{\frac{c_p}{5/2}}^{-1}
    \braket{\frac{\ceq}{10\kms}}^{-2}
    . 
\end{split}
    \label{eq:tildes}
\end{equation}
When $\varepsilon  > 1$, it signifies a hard spectrum, indicating that the gas could potentially be heated to a temperature higher than the equilibrium temperature by a single photoionization event.  

The additionally supplied energy through reionizations is obtained by calculating the integration (\eqnref{eq:c_ch_def}) from $s = s_1$ to $s = R$ with approximating $y_\mathrm{HI}$ in $\Gamma_\mathrm{EUV}$ to the ionization-recombination equilibrium abundance $y_\mathrm{HI}\sim t_\mathrm{ioni}/t_\mathrm{rec}$ (\eqnref{eq:tioni_trec_ratio}).
Hence, the expression for $c_\mathrm{ch}$ for Type~II winds are derived by solving this implicit equation with respect to $c_\mathrm{ch}^\mathrm{II}$: 
\[
  \braket{\frac{c_\mathrm{ch}^\mathrm{II}}{\ceq }}^2 
    = \varepsilon   
    + \braket{\frac{\varphi \varepsilon   q}{ x }}^{1/2}
    \braket{\frac{c_\mathrm{ch}^\mathrm{II}}{\ceq }}^{-1.75}
    \braket{1 - \varphi^{-1}\varepsilon   x \frac{c_\mathrm{ch}^\mathrm{II}}{\ceq }}
    ,
\]
The first and second terms of the RHS correspond to the initially injected energy and additional energy through reionizations, respectively. 
We refer the readers to \secref{sec:typeII} (\eqnref{eq:chII_original_def:dimensionless}) for the derivation of this equation. 

The root of the above implicit equation can be described by simple expressions in some special cases. 
When $t_\mathrm{rec}$ is sufficiently longer than $R/\cs $ so that energy addition due to reionizations can be ignored, $c_\mathrm{ch}^\mathrm{II}$ is approximated to
\[
    c_\mathrm{ch}^\mathrm{II} \approx c_1 \equiv \varepsilon ^{1/2} \ceq ,
\]
by dropping the second term on the RHS of the above equation. 
On the other hand, when recombination is sufficiently rapid, $t_\mathrm{rec} \ll R/\cs $, the reionization term dominates, and $c_\mathrm{ch}^\mathrm{II} $ is approximated to 
\begin{equation}
    c_\mathrm{ch}^\mathrm{II} \approx c_2  \equiv \ceq  \braket{\frac{\varphi_\mathrm{II}}{x}}^{1/7.5}, 
    \label{eq:cchII_def}
\end{equation}
where $\varphi_\mathrm{II}$ is the EUV emission rate normalized by the secondary critical EUV emission rate for recombination-dominated Type~II winds, 
\[
    \Phi_\mathrm{c,rec}^\mathrm{II} \equiv \frac{\Phi_\mathrm{c}}{\varepsilon  q}. 
\]

Again, both $c_\mathrm{ch}^\mathrm{I}$ and $c_\mathrm{ch}^\mathrm{II}$ represent different forms of $c_\mathrm{ch}$ and thus serve as indicators of the heating timescale, not necessarily providing the actual sound speed of the wind, $\cs $, which is presented in the following section.

\subsection{Typical Flow Velocities}
\label{sec:typical_flow_velocities}
When photoheating operates on a timescale shorter than gravitational deceleration, such that $t_\mathrm{h} < t_\mathrm{g}$, winds can be rapidly heated to have $\cs  $ exceeding the escape velocity $\vg $ without feeling the gravity, implying that the wind does not experience a large pressure drop. 
In such instances, the typical flow velocity of the freely photoevaporating winds can approximate the isothermal sound speed, i.e., $\mach \sim 1$. 

In a case of strong photoheating where $c_\mathrm{ch} > \ceq $, the gas undergoes immediate heating to a high temperature where cooling is activated. 
This regulates the gas temperature to maintain $\cs \approx \ceq $. 
For $x > 1$, this renders vigorous isothermal winds at the equilibrium temperature with a typical velocity of $\ceq $. 
The temperature equilibration occurs within a quite short distance, $s \approx \ceq  t_\mathrm{h} = R (\ceq/c_\mathrm{ch})$.

In cases where $\vg  < c_\mathrm{ch} < \ceq $, gravitational deceleration is negligible, but heating is not rapid enough to heat the gas to the equilibrium temperature within $s \lesssim R$. 
Here, neither cooling nor gravity significantly influences the wind, resulting in forming free winds with $\cs  \approx c_\mathrm{ch}$ and $\mach \sim 1$.

Conversely, when the heating timescale is longer than the gravitational timescale, or equivalently $c_\mathrm{ch} < \vg $, gravity can inhibit winds significantly. 
This takes place typically at small distances with either low EUV emission rates or hard spectra.
Regardless of the emphasized gravitational effect, photoheated gas will not be completely hydrostatic, as the gas does not have an alternative option other than forming winds to process the continuously deposited energy due to its too low temperature to activate cooling. 
In these escaping winds, a steady state is achieved such that the sound speed at a height of $\sim R$ is maintained at $\sim \vg $.
The typical flow speed of such gravity-inhibited winds can be estimated using the self-consistent condition (cf. \eqnref{eq:c_ch_def}):
\[
    c_p \vg ^2 \approx \int \Gamma_\mathrm{EUV} \frac{\dd s}{\mach \vg }. 
\]
This reflects that when the heating timescale is long, the flow speed is self-regulated so that the gas can take a long time to reach $\sim R$ to obtain $\sim c_p \vg ^2$ worth of energy with a low heating rate.
Essentially, the above definition results in an estimated Mach number of 
\begin{equation}
    \mach_\mathrm{g} \equiv \mach \approx \braket{\frac{t_\mathrm{g}}{t_\mathrm{h}}}^3 ,
    \label{eq:mach_g}
\end{equation}
indicating that winds are subsonic at $\sim R$. 
The location of the sonic point is considerably farther from the base, $\sim (t_\mathrm{g}/t_\mathrm{h})^{-6} R $ \citep{1983_Begelman}.  

The specific form of the Mach number (\eqnref{eq:mach_g}) for gravity-inhibited Type~I winds is derived as 
\[
    \mach   \sim \mach_\mathrm{g}^\mathrm{I}
    \equiv  \varphi x^{1/2},
\]
and that for gravity-inhibited Type~II winds is 
\[
    \mach   \sim    
    \mach_\mathrm{g}^\mathrm{II}
    \equiv
        \braket{\frac{\varphi q x^\beta}{\varepsilon  }}^{1/2}
        \left[
        x^{-1} \varepsilon ^{-1} - 1 
        + \braket{\frac{\varepsilon  q  x ^{\beta + 1} }{\varphi}}^{1/2} 
        \right]^{-1}. 
\]
In either wind type, the Mach number increases with the EUV emission rate and reaches unity at $t_\mathrm{h} = t_\mathrm{g}$.

\begin{table*}[]
    \centering
    \begin{tabular}{l l C C C}
    Regime & Description & \text{Condition} & \text{Typical Sound Speed: } \cs & \text{Typical Mach Number: } \mach \\ \hline
    A & Isothermal wind&  c_\mathrm{ch} > \ceq  & \sim \ceq  & \sim 1\\
    B & Free wind &  \vg  < c_\mathrm{ch} < \ceq  & \sim c_\mathrm{ch} & \sim 1 \\
    C & Gravity-inhibited wind & c_\mathrm{ch} < \vg  & \sim \vg  & \sim (t_\mathrm{g}/t_\mathrm{h})^3  
    \end{tabular}
    \caption{
    Summary of wind regimes. 
    }
    \label{tab:regime_summary}
\end{table*}
The hydrodynamical characteristics of the isothermal wind, free wind, and gravity-inhibited wind regimes can be summarized as \tref{tab:regime_summary}. 
In \begelman{}, the corresponding parameter spaces are labeled as Regions~A, B, and C, respectively.
We follow this notation, but in our model, each regime bifurcates depending on the wind's chemical state, as we will see in the following sections (see also \tref{tab:sound_spees}). 
The three regimes cover different areas on the $x$--$\varphi$ phase diagram, depending on the EUV photoevaporation class. 

\subsection{EUV Photoevaporation Classes}
\label{sec:sub:euv_classes}
Varying spectral hardness results in differing relative magnitudes of photoheating and photoionization. 
For instance, a soft spectrum yields a relatively low amount of energy deposited into the gas per photoionization. 
In such cases, while winds are readily photoionized, efficiently heating them proves challenging.
Hence, differences in spectral hardness suggest distinct combinations of the thermal and ionization states for winds at given $\phieuv$ and $R$.

\begin{figure}
    \centering
    \includegraphics[clip, width = \linewidth]{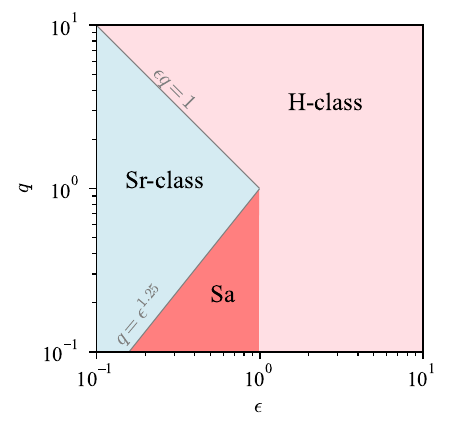}
    \caption{Illustration of the spectrum-hardness diagram for the EUV photoevaporation classes (cf. \tref{tab:sq_plane}). When the spectrum parameter $\varepsilon $ (\eqnref{eq:tildes}) and ratio of recombination to photoionization $q$ (\eqnref{eq:q}) for a given EUV spectrum is positioned in $\varepsilon  > 1$ or $\varepsilon  q > 1$, it is categorized into the hard-spectrum class (H-class). 
    The other cases are classified into the soft-spectrum classes. 
    For $q < \varepsilon ^{1.25}$, recombination is slow, and advection is more effective in maintaining atomic hydrogen within the launched gas. We call this class Sa-class. 
    The opposite class $q > \varepsilon ^{1.25}$, where recombination is rapid, is termed as Sr-class.
    Here, $\beta = 0.75$ is used.
    }
    \label{fig:sqplane}
\end{figure}
By evaluating Eqs.\eqref{eq:typeI_condition} and \eqref{eq:typeII_condition}, 
we find the distinct classes of EUV photoevaporation that are uniquely determined by the two spectrum-hardness parameters, $\varepsilon$ and $q$ (Eqs.\eqref{eq:tildes} and \eqref{eq:q}), 
\[
\begin{gathered}
    \varepsilon  \equiv \frac{\aveenergy }{m c_p \ceq ^2}
    \approx 1 
    \braket{\frac{\aveenergy}{3.7 \eV}}
    \braket{\frac{m}{1.4m_\mathrm{H}}}^{-1}
    \braket{\frac{c_p}{5/2}}^{-1}
    \braket{\frac{\ceq}{10\kms}}^{-2}
    \\
    \begin{split}
     q  \equiv \frac{3C^2 \alpha_\mathrm{eq} }{\ceq \avesigma } 
     & \approx 0.1 \braket{\frac{C}{0.4}}^2 
     \braket{\frac{\alpha_\mathrm{eq}}{2\e{-13}\cm^3\sec^{-1}}} 
     \\ & \quad \times 
     \braket{\frac{\ceq }{10\kms}}^{-1}
     \braket{\frac{\avesigma }{10^{-18}\cm^2}}^{-1}
 \end{split}
    . 
\end{gathered}
\]
\begin{table}[htbp]
    \centering
    \begin{tabular}{l  C  r}
    Class name & \text{Parameter space} & Relevant source
    \\ \hline
    H-class &  \left\{(\varepsilon , q)~|~\varepsilon  > 1 ~ \cup ~ \varepsilon  q > 1 \right\}
    & T Tauri star
    \\
    Sa-class&  \left\{(\varepsilon , q)~|\varepsilon   < 1, ~ q < \varepsilon ^{1.25} \right\}
    & MS IM star
    \\
    Sr-class&    \left\{(\varepsilon , q)~| \varepsilon  q < 1, ~ q > \varepsilon ^{1.25} \right\}
    & None(?)
    \end{tabular}
    \caption{EUV photoevaporation classes and corresponding parameter spaces. $\beta = 0.75$ is used here. (See also \fref{fig:sqplane}.)
    The right column shows potentially relevant radiation sources to each class, where ``MS IM star'' denotes main-sequence intermediate-mass star. 
    }
    \label{tab:sq_plane}
\end{table}
The class of EUV photoevaporation is identified by the location of the given spectrum's $\varepsilon $ and $q$ on the $\varepsilon $--$q$ phase diagram (\fref{fig:sqplane}).
The corresponding parameter spaces are summarized in \tref{tab:sq_plane}.
The hard spectrum class is referred to as ``H-class,'' while the soft spectrum classes are denoted as ``Sa-class'' or ``Sr-class.'' 
The lower-case letters of the soft spectrum classes stand for ``advection'' and ``recombination,'' respectively. 
We will delve into the characteristics of each class in Sections~\ref{sec:sub:H-class}--\ref{sec:sub:Sr-class}. 

For young low-mass stars, in particular, EUV spectra might be classified into the H-class since $\varepsilon $ easily exceeds unity if there is a contribution of energetic EUV originating from the stellar magnetic activity. 
However, the possibility remains that the spectra of low-mass stars are classified into the soft-spectrum classes if those low-mass stars have dominating accretion-originated soft EUV. 
A blackbody with the effective temperatures of $\lesssim 5\e{4}\Kelvin$ falls into the soft spectrum class.

A necessary condition to be in the soft-spectrum classes is that the average deposited energy $\aveenergy$ is less than $m c_p \ceq^2 \approx 3.7\eV$. 
This indicates most EUV photons have energies close to the Lyman limit. 
Such extremely soft spectra would be possible for stars without convective zones, like main-sequence stars with intermediate masses, where EUV emission is mostly covered by moderately hot photospheres.

\subsection{H-Class}
\label{sec:sub:H-class}
\begin{figure}
    \centering
    \includegraphics[clip, width  =\linewidth]{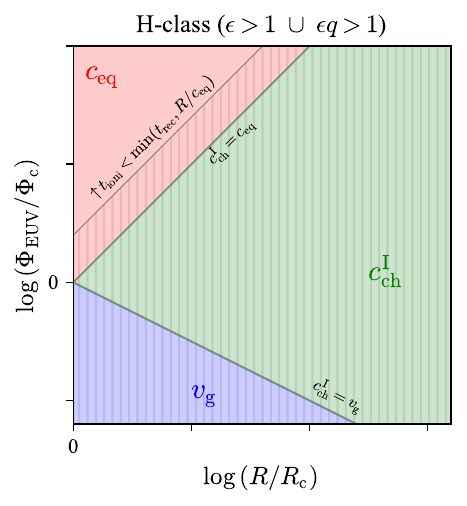}
    \caption{
    Phase diagram for the H-class. 
    The horizontal axis is normalized radius $x = R/R_\mathrm{c}$, and the vertical axis is normalized EUV emission rate $\varphi = \phieuv /\Phi_\mathrm{c}$. 
    The red- and blue-shadings are the isothermal-wind (Regime~A) and gravity-inhibited (Regime~C) regimes, respectively. 
    The parameter space of steadily-heated, free winds (Regime~B, or equivalently Regime~B-I) is colored in green. 
    Different colors indicate the typical sound speed $\cs $
    (see also \tref{tab:regime_summary} for the chemical and hydrodynamical characteristics of each regime.)
    The gray vertical stripes represent the Type~I regime, where photoevaporative winds are atomic. 
    The remaining area is Type~II. 
    Regime~A has both Type~I and II regimes, while the others do not. 
    This is a unique feature for the H-class. 
    }
    \label{fig:H-class}
\end{figure}
This scenario corresponds to cases where the EUV spectrum is relatively hard (see \tref{tab:sq_plane}). 
In such instances, EUV can deposit a relatively large amount of energy into the gas per photoionization, enabling efficient heating of the gas. 
In other words, this class has a relatively short heating timescale for a photoionization timescale. 
Stars with active surface magnetic activities are anticipated to exhibit EUV spectra falling under this class. 

The fast-ionization condition (\eqnref{eq:typeII_condition}) is rewritten to 
\begin{equation}
    \varphi > \mathrm{max}\braket{1, q} \times \varepsilon  x.
    \label{eq:typeII_condition_H_class}
\end{equation}
In this parameter space, the wind becomes ionized (Type~II). 
The remaining parameter space corresponds to Type~I winds, depicted by the gray vertical lines in \fref{fig:H-class}.

Photoheating is notably intense when \(c_\mathrm{ch}^\mathrm{I} > \ceq \), or equivalently \(\varphi > x\), instantly raising the gas temperature to the equilibrium value. 
Consequently, the wind becomes isothermal with $\cs  \approx \ceq $ in this parameter space. 
The corresponding area is visually highlighted by the red-shading in \fref{fig:H-class} and is labeled as Regime~A, following Region~A of \begelman{} (\tref{tab:sound_spees}). 
Regime~A occupies the upper left corner of the $x$--$\varphi$ plane, reflecting the tendency for the heating timescale to be generally shorter at smaller $R$ and with higher $\phieuv $, i.e., where EUV flux is intense. 
\begin{table*}[htbp]
    \centering
    \begin{tabular}{l l c c C C c}
    Regime
    &
    Subregime
    & 
    Dynamical state 
    &
    Chemical state 
    & 
    \cs 
    &
    \mach
    &
    Color 
    \\
    \hline \hline
    A
    &
    \begin{tabular}{l}
        A-I\\
        A-II
    \end{tabular}
    &
    Isothermal Wind 
    &
    \begin{tabular}{c}
        Atomic (Type~I)\\
        Ionized (Type~II)
    \end{tabular}
    & 
    \sim \ceq 
    & 
    \sim 1
    &
    \begin{tabular}{c}
        red  (w/ gray vertical stripes)\\
        red  (w/o gray vertical stripes)  
    \end{tabular}
    
    \\  \hline
    B
    &
    \begin{tabular}{l}
        B-I\\
        B-II
    \end{tabular}
    &
    \begin{tabular}{c}
        Steadily Heated, Free Wind \\
        Free Wind 
    \end{tabular}
    &
    \begin{tabular}{c}
        Atomic (Type~I)\\
        Ionized (Type~II)
    \end{tabular}
    &
    \begin{tabular}{C}
        \sim c_\mathrm{ch}^\mathrm{I} \\
        \sim c_\mathrm{ch}^\mathrm{II} 
    \end{tabular}
    &
    \sim 1
    &
    \begin{tabular}{c}
    green\\
    orange
    \end{tabular}
    \\  \hline
    C
    &
    \begin{tabular}{l}
         C-I\\
         C-II
    \end{tabular}
    &
        Gravity-inhibited wind
    & 
    \begin{tabular}{c}
         Atomic (Type~I)\\
         Ionized (Type~II)
    \end{tabular}
    & 
    \sim \vg 
    &
    \begin{tabular}{C}
        \sim \mach_\mathrm{g}^\mathrm{I}\\
         \sim \mach_\mathrm{g}^\mathrm{II}
    \end{tabular}
    &
    \begin{tabular}{c}
        blue (w/ gray vertical stripes) \\
        blue (w/o gray vertical stripes)
    \end{tabular}
    \\
    \hline
    \end{tabular}
    \caption{
    Qualitative characteristics of the chemical and hydrodynamical states for winds in each regime. 
    Each of the fundamental three regimes (Regimes~A, B, and C; see \tref{tab:regime_summary}) has two subregimes according to the wind types. 
    The listed ``Color'' in the last column is used in Figures~\ref{fig:H-class}, \ref{fig:Sa-class}, \ref{fig:Sr-class}, and \ref{fig:classmaps} to represent the corresponding parameter space on the phase diagrams, differentiating the typical sound speeds of winds $\cs$. 
    Note that Regime~A-I is only possible in the H-class, while Regimes~B-II and C-II are present only in the Sa- and Sr-classes. 
    }
    \label{tab:sound_spees}
\end{table*}
Ionized winds (Type~II) can exist only in Regime~A in this class. 
It is also possible that the winds manifest as an isothermal atomic flow for a lower $\phieuv $. 
With the short heating timescale in this regime, the effect of gravity is minimal, and winds can freely escape at a typical flow speed of $\sim \ceq $.

If $\phieuv $ is small, or the distance $R$ is large, the heating timescale gets long, and in the ultimate case, $t_\mathrm{h}$ becomes longer than the gravitational timescale.
This condition is equivalent with $c_\mathrm{ch}^\mathrm{I} < \vg $, and in this case, the wind is significantly inhibited by gravity. 
The energy deposited by photoheating charges to the gas 
keeps the isothermal sound speed at $\cs  \approx \vg $. 
This regime is represented by $\varphi < x^{-1/2}$, or equivalently $c_\mathrm{ch}^\mathrm{I} < \vg $, shown by the blue-shading in \fref{fig:H-class} (Regime~C). 
The fast-ionization condition is not met there, meaning that gravity-inhibited winds are atomic in this class. 
Correspondingly, the typical Mach number of gravity-inhibited winds in the H-class is $\sim \mach_\mathrm{g}^\mathrm{I}$. 

In the other regime, $x^{-1/2} < \varphi < x$, neither gravity nor cooling significantly affects the system.
This regime corresponds to Region~B of \begelman{}, but we label it as Regime~B-I in this study to highlight the fact that the winds are atomic, and the sound speed is given by the Type~I characteristic sound speed $c_\mathrm{ch}^\mathrm{I}$.
Similarly to Regime~A, the heating timescale is sufficiently short compared to the gravitational timescale, and thus winds can freely escape at $\mach\sim 1$ without being influenced by gravity. 
However, the heating timescale is not short enough so that gas temperature can reach the equilibrium temperature while traveling to the height of $\sim R$, 
and hence $\cs  \approx c_\mathrm{ch}^\mathrm{I}$. 


\subsection{Sa-class}
\label{sec:sub:Sa-class}

\begin{figure}
    \centering
    \includegraphics[clip, width = \linewidth]{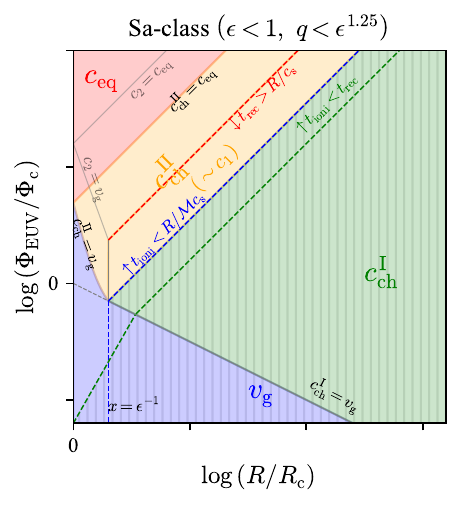}
    \caption{
    Phase diagram for the Sa-class, which is overall similar to \fref{fig:H-class}. 
    However, Regime~B is separated into two subcategories: Regime~B-I (green) and Regime~B-II (orange).
    Both are free wind regimes, but the former is for atomic winds (Type~I), and the latter is for ionized winds (Type~II).
    The blue-dashed line indicates where the ionization timescale is equal to the crossing timescale $t_\mathrm{ioni} = R/\mach \cs $; the ionization timescale is shorter above this line. 
    The green-dashed line is where the ionization and recombination timescales are equal; 
    again, the ionization timescale is shorter above this line. 
    The boundary where the recombination and wind crossing timescale is equal is represented by the red-dashed line, below which the recombination timescale is longer. 
    This class is visually distinguished from the other classes by the parameter space having the presence of the orange area attaching to the blue-dashed line and the red-dashed line positioning in the Type~II regime.
    }
    \label{fig:Sa-class}
\end{figure}
This class is the case when the spectrum is soft, and recombination occurs relatively slowly in the wind. 
The deposited energy per photoionization is small, and one single photoionization is not sufficient to supply the energy worth of the gas's internal energy at the equilibrium temperature, as opposed to the H-class. 
A relatively high $\phieuv $ is required to heat the gas at a certain level than is needed in the H-class.
Due to the slow recombination, winds in this class do not necessarily undergo secondary photoionization following the initial photoionization at the launching point until the escaping gas travels to the height of $\sim R$.
Stars lacking surface magnetic activities, such as main-sequence intermediate-mass stars, would have EUV spectra classified into this class.

The fast-ionization condition is rewritten to 
\begin{equation}
    \left\{ 
    \begin{array}{ll}
         \varphi > \varepsilon  q x^{1 + \beta}
         & \quad \mathrm{for}~ 1\leq  x  < \varepsilon  ^{-1} \\
         \varphi >  \varepsilon  ^{3/2}  x & \quad \mathrm{for}~  x  > \varepsilon  ^{-1} \\
    \end{array}
    \right.
    \label{eq:Sa-class:TypeIIcondition}
\end{equation}
which is equivalent to $t_\mathrm{ioni} < R/\cs $ for $x < \varepsilon ^{-1}$ (green dashed line) and $t_\mathrm{ioni} < R/\cs $ for $x > \varepsilon ^{-1}$ (blue dashed line). 
In this parameter space, the wind becomes ionized (Type~II), while the remaining parameter space (the area covered by gray vertical lines in \fref{fig:Sa-class}) represents the atomic-wind regime (Type~I). 
For $x>\varepsilon ^{-1}$, atoms in the winds are maintained primarily through advection from the upstream region rather than reproduced by recombination, and vice versa for $x < \varepsilon ^{-1}$.
The blue dashed line, where $t_\mathrm{ioni} = R/\cs $, is vertical at $x = \varepsilon ^{-1} $ in \fref{fig:Sa-class}. 
This implies that steady winds cannot develop without additional energy deposition through reionizations in $x < \varepsilon $, since the initially injected energy is never sufficient for heating the gas to yield an escaping speed.

Contrary to the H-class, isothermal winds of this class (Regime~A; the red-shade in \fref{fig:Sa-class}) are always in an ionized state.
The corresponding parameter space is where the heating timescale is sufficiently short, i.e., $c_\mathrm{ch}^\mathrm{II} > \ceq $, or equivalently 
\[
    \varphi  > f_\mathrm{eq} ( \varepsilon , q) x,
\]
where 
\[
    f_\mathrm{eq} (\varepsilon , q) 
    \equiv \frac{4 \varepsilon  q }
    {\braket{\sqrt{(\varepsilon ^{-1}-1)^2 + 4q} - (\varepsilon ^{-1}-1)}^2}.
\]
The factor $f_\mathrm{eq}$ is larger than unity for soft spectrum classes ($\varepsilon  < 1, ~\varepsilon  q< 1$), meaning that on the $x$--$\varphi$ plane, the $c_\mathrm{ch}^\mathrm{II} = \ceq $ line is located above $c_\mathrm{ch}^\mathrm{I} = \ceq $ line, which separated Regimes~A and B in the H-class. 
This results from more efficient heating in an atomic gas than in an ionized gas since the absorber per gas mass is higher. 
The reduced heating rate necessitates a higher $\phieuv $ to heat the gas to the equilibrium value ($\cs  = \ceq $), compared to what would be required if the gas were atomic ($\varphi > x$, or equivalently $c_\mathrm{ch}^\mathrm{I} > \ceq $). 

In the limit of $(\varepsilon^{-1}-1)^2 \gg 4 q$, which is often the case for Sa-class spectra (\tref{tab:sound_spees}), $f_\mathrm{eq}$ is approximated to 
\[
    f_\mathrm{eq} \approx \frac{(1 - \varepsilon)^2}{\varepsilon q} \quad    \braket{\text{for } (\varepsilon^{-1}-1)^2 \gg 4 q}. 
\]

A distinct characteristic of the soft-spectrum classes is that ionized winds can be inhibited by gravity. 
This occurs when the heating timescale is shorter than the gravitational timescale. 
For Type~II winds, this condition is expressed by $c_\mathrm{ch}^\mathrm{II}< \vg $, which is rewritten to 
\[
\begin{split}
    \varphi&  < \varphi_\mathrm{g} (x; \varepsilon , q, \beta)
    \\ 
    & \equiv 
    \braket{
    \frac{2\sqrt{q\varepsilon }x^{(\beta+1)/2}}
    {\sqrt{(x^{-1}\varepsilon ^{-1}-1)^2 + 4 q x^{(2\beta + 1)/2}} - (x^{-1}\varepsilon ^{-1}-1)}
    }^2
    , 
    \end{split}
\]
Hence, the parameter space for the gravity-inhibited, Type~II winds is expressed by 
\[
    \left\{
    (x, \varphi)~|~
    x < \varepsilon ^{-1},
    ~
    \varphi > \varepsilon  q x^{\beta + 1},
    ~
    \varphi < \varphi_\mathrm{g}
    \right\}.
\]
In this parameter space, typical sound speed and flow speed are $\sim \vg $ and $\sim \mach_\mathrm{g}^\mathrm{II}$, respectively. 
This space has a shape that looks like extruding the blue-shaded area upward on the $x$--$\varphi$ plane within $x < \varepsilon ^{-1}$, compared to Regime~C of the H-class. 
Again, this distinction highlights the lower heating efficiency in an ionized wind than in an atomic wind.
Therefore, in order to deposit sufficient energy for the gas to form steadily escaping winds, shortening the recombination timescale by increasing flow density, i.e., increasing $\phieuv  $, is needed. 
This facilitates the (re)production of atomic hydrogen, which is subsequently reionized in the wind, resulting in secondary energy deposition.

Another unique characteristic of the soft-spectrum classes is having the parameter space of free winds that are ionized (orange-shaded area in \fref{fig:Sa-class}).  
In this regime, the heating timescale is sufficiently shorter than the gravitational timescale while not being short enough to heat the gas to reach the equilibrium temperature. 
The photoionization timescale is also sufficiently short to make the photoheated gas ionized in this regime. 
As a result, the typical sound speed and Mach number of the ionized winds are given by $\sim c_\mathrm{ch}^\mathrm{II}$ and $\sim 1$, meaning lower-temperature ionized winds in contrast to the canonical picture.

The soft-spectrum classes are distinguished by the rapidity of recombination within free-ionized winds. 
It differentiates the magnitude of the additionally deposited energy through reionizations with respect to the initially injected energy. 
The Sa-class has relatively slow recombination and thus always has a parameter space where the contribution of additional energy deposition is minimal. 
This is especially the case in the vicinity of the Type~I and Type~II boundary in Regime~B (blue-dashed line in \fref{fig:Sa-class}). 
The sound speed and flow speed can be well approximated to $\sim c_1$ there.
In the zeroth order, this approximation is valid as long as the recombination timescale is longer than the wind crossing timescale. 
This parameter space corresponds to the area bounded by the blue- and red-dashed lines in \fref{fig:Sa-class}.



\subsection{Sr-class}
\label{sec:sub:Sr-class}
\begin{figure}
    \centering
    \includegraphics[clip, width = \linewidth]{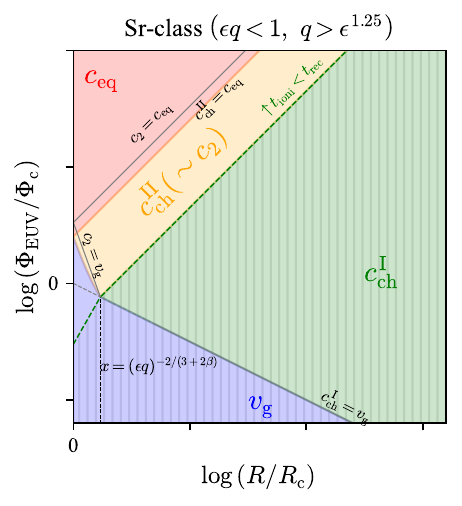}
    \caption{
    Same as \fref{fig:Sa-class}, but we omit the blue-dashed line ($t_\mathrm{ioni} = R/\cs $) and red-dashed line ($t_\mathrm{rec} = R/\cs $), which position below the green-dashed line (cf. the bottom left panel in \fref{fig:classmaps}). 
    Having the orange area attached to the green-dashed line is a visual uniqueness of the parameter space for this class. 
    }
    \label{fig:Sr-class}
\end{figure}
This is the other soft-spectrum class in addition to the Sa-class. 
Overall characteristics are the same as those of the Sa-class, but this class has a shorter recombination timescale. 
In this class, the boundary between the Type~I and Type~II regimes is set by $t_\mathrm{rec} = t_\mathrm{ioni}$ (green dashed line in \fref{fig:Sr-class}) instead of $t_\mathrm{ioni} = R/\cs $.
Thus, the slow-recombination regime (the area bounded by the blue and red dashed lines in \fref{fig:Sa-class}) is absent. 
This absence means that ionized winds are invariably associated with (multiple) reionizations of recombined hydrogen to heat the gas. 
However, this class may not manifest with realistic EUV spectra, at least in the context of protoplanetary disk photoevaporation.  

This class is distinguished by the $x$--$\varphi$ plane having the $t_\mathrm{ioni} = t_\mathrm{rec}$ line (the green dashed line), positioning above both the $t_\mathrm{ioni} = R/\cs $ (blue dashed) and $t_\mathrm{rec} = R/\cs $ (red dashed) lines (cf. \fref{fig:classmaps} in \secref{sec:typeI:regime}). 
Note that we omit the blue and red dashed lines in \fref{fig:Sa-class} for clarity.


\section{Derivations of Parameters and Classes}
\label{sec:cs}

In this section, we provide the detailed derivations of the parameters and EUV photoevaporation classes reviewed in \secref{sec:classification}. 
We start with the estimation of the sound speed and flow speed for Type~I winds (\secref{sec:c_ch}), introducing photoionization-relevant parameters (\secref{sec:sub:photoionization_parameters}), and identifying the parameter space where these winds are anticipated (\secref{sec:typeI:regime}). 
Following this, we extend our analysis to Type~II winds in Sections~\ref{sec:typeII} and \ref{sec:typeII:regime}.


\subsection{Characteristic Sound Speed: Type I}  \label{sec:c_ch}

Following \begelman{}, the characteristic sound speed (\eqnref{eq:c_ch_def}) is defined as 
\[
    c_p c_\mathrm{ch}^2 = \int \Gamma_\mathrm{EUV} \frac{\dd s}{c_\mathrm{ch}}.
\]
For Type~I winds, this can be approximated to 
\[
    c_p c_\mathrm{ch}^2 \approx \frac{R}{c_\mathrm{ch}} \Gamma_\mathrm{EUV}, 
\]
assuming that $y_\mathrm{HI}$ in $\Gamma_\mathrm{EUV}$ (\eqnref{eq:gamma_EUV}) is nearly unity for the Type~I winds uniformly.
The above equation can be solved with respect to $c_\mathrm{ch}$, 
\[
    c_{\rm ch} = \braket{\frac{\Gamma_\mathrm{EUV} R }
                {c_p}}^{1/3} 
                \quad \braket{y_\mathrm{HI} \approx 1}, 
\]
retrieving \eqnref{eq:cchI_def}. 
Then, the heating timescale is given by 
\[
    t_\mathrm{h } \equiv \frac{c_p c_\mathrm{ch}^2}{\Gamma_\mathrm{EUV}}
     = \frac{R}{c_\mathrm{ch}}, 
\]
and it characterizes the strength of the photoheating. 
We stress again that the characteristic sound speed $c_\mathrm{ch}$ is a virtual sound speed that a gas parcel would achieve if the photoheating energy is purely deposited to the internal energy of freely flowing gas. 
In other words, $c_\mathrm{ch}$ does not necessarily give the actual isothermal sound speed of the gas $\cs $, which is also influenced by the effects of cooling and gravity, as mentioned in \secref{sec:classification}.

We assume that $r$ in $\Gamma_\mathrm{EUV}$ can be approximated to $r \approx R$ within $s \leq R$ for our order-of-magnitude estimate, 
and then \eqnref{eq:c_ch_def} is rewritten to 
\begin{equation}
\begin{split}
c_{\rm ch}
& = \ceq  \braket{\frac{\phieuv y_\mathrm{HI}\chi_\mathrm{e}}{\Phi_\mathrm{c}  x }}^{1/3} 
  \quad \braket{y_\mathrm{HI} \approx 1}
\end{split}
\label{eq:cch_VS_ceq},
\end{equation} 
where
\[
\begin{split}
    \Phi_\mathrm{c}& \equiv \frac{ 4\pi m  \ceq GM_*} {2  \avesigma  \aveenergy }
    \\
    &   \approx
            1.7\e{38}\sec^{-1}
            \braket{\frac{M_*}{1\Msun}}
            \braket{\frac{m}{1.4m_{\rm H}}}  
            \braket{\frac{\ceq }{10\kms}}  
    \\      &   \times            
            \braket{\frac{\avesigma  \aveenergy }{7\e{-18}\cm^2 \eV}}^{-1}
\end{split}
\]
is the critical EUV emission rate (\eqnref{eq:critical_luminosity}). 
The critical EUV emission rate is spatially independent but depends only on the stellar property, namely the spectral shape of EUV and the stellar mass, and the disk gas property, namely gas mass per hydrogen nucleus and the equilibrium sound speed. 

Using the EUV emission rate normalized by the critical EUV emission rate
\[
    \varphi \equiv \frac{\phieuv }{\Phi_\mathrm{c}} ,
\]
we denote the characteristic sound speed of \eqnref{eq:cch_VS_ceq} as 
\[
    c_\mathrm{ch}^\mathrm{I} 
    \equiv \ceq  
    \braket{\frac{\varphi}{ x }}^{1/3} 
\]
to emphasize this being $c_\mathrm{ch}$ for Type~I winds (cf. \eqnref{eq:cchI_def}). 
In this paper, we use $c_\mathrm{ch}^\mathrm{I}$ as the characteristic sound speed of the Type~I wind instead of the more strict expression \eqnref{eq:cch_VS_ceq} for simplicity. 
{
This means that we analyze under the idealization that a large portion of the wind region is optically thin ($\chi_\mathrm{e} \sim 1$), following \citet{1983_Begelman}. 
While this idealization restricts our model's applicability, it simplifies our analysis. 
This idealization is a reasonable approximation for spectra with minimal photon energy dispersion, where most photons are intensively absorbed at a certain column density. 
Examples include relatively soft spectra dominated by photons near the Lyman limit or harder spectra with delta-function-like shapes.
Regardless of these limitations, the adopted approach is an essential first step towards developing a more comprehensive model available for a wider range of spectra.
Besides, the insights gained from our simplified model provide a foundation for tackling more general cases. 
In \secref{sec:discussion}, We delve deeper into the potential impacts of this adopted approximation.
}

The Type~I characteristic sound speed $c_\mathrm{ch}^\mathrm{I}$ increases with $\phieuv $ and eventually exceeds $\ceq $ when $\phieuv  > \Phi_\mathrm{c} x $. 
This means the photoheating is so strong that the gas temperature could potentially exceed the equilibrium value $\ceq $. 
However, as explained in \secref{sec:typical_flow_velocities}, it does not happen for real; instead, cooling gets activated to force the actual sound speed $\cs $ to stay at $\ceq $. 
In other words, coolants act as a resistance force for gas temperature so as not to exceed the terminal value $\ceq $. 

On the contrary, when heating is weak, the characteristic sound speed can get lower than $\vg $, or equivalently $t_\mathrm{h} > t_\mathrm{g}$. 
This appears as if the gas would be bounded so that it forms a hydrostatic atmosphere; however, this is not the case. 
Even if the gas is indeed in hydrostatic equilibrium at first, 
the energy will be continuously charged to the gas to increase the temperature gradually, as the deposited energy cannot be processed by radiative cooling at this low temperature. 
At some point, $\cs $ will eventually reach the escaping velocity $\vg $, forming slow winds.  
The system would achieve a steady state such that the sound speed is maintained of the order of $\vg  (= \sqrt{GM/2c_p R})$ with a small flow velocity of
\[
    v \sim \braket{\frac{t_\mathrm{g}}{t_\mathrm{h}}}^3 \vg  = \braket{\frac{c_\mathrm{ch}^\mathrm{I}}{\vg }}^3 \vg  
    = \varphi x^{1/2} \vg ,
\]
which indicates
\[
    \mach_\mathrm{g}^\mathrm{I}  = \varphi x^{1/2} \quad\braket{\text{for Type~I}}
\]
(cf. \eqnref{eq:mach_g}).
This gravity-inhibited-wind regime occurs when $c_\mathrm{ch}^\mathrm{I} < \vg $, i.e., when the ratio of  
\begin{equation}
    \frac{c_\mathrm{ch}^\mathrm{I}}{\vg }
    = \braket{\varphi  x ^{1/2}}^{1/3}
    \label{eq:cchI_vg}
\end{equation}
gets lower than unity. 

\subsection{Photoionization-Related Parameters}
\label{sec:sub:photoionization_parameters}
As explained in \secref{sec:sub:wind_types}, 
the picture of the Type~I wind is valid only when the photoionization timescale is longer than the recombination or crossing timescale of the wind (left panel of \fref{fig:streamline_types}). 
This condition is expressed by the slow photoionization condition, \eqnref{eq:typeI_condition}, 
\[
    t_\mathrm{ioni} > \mathrm{min}\braket{t_\mathrm{rec}, R/\mach \cs }. 
\]
The Type~I characteristic sound speed $c_\mathrm{ch}^\mathrm{I}$ (\eqnref{eq:cchI_def}) is available only in this parameter space, for which we will identify the expressions in the following sections. 
To that end, we introduce photoionization-relevant quantities in this section.

After the photo-energy is injected into a new \ion{H}{1} layer close to the $\tau_\mathrm{EUV}\sim 1$ surface, a launched hydrogen atom takes $t_\mathrm{ioni}$ to be photoionized in general. 
The flowing hydrogen atom can survive to a distance of  
\[
\begin{split}
    \Delta s_\mathrm{suv} & \equiv \mach \cs  t_\mathrm{ioni} \\
    & \approx 1.9 \au \braket{\frac{v}{10\km \sec^{-1}}}
    \braket{\frac{\phieuv }{10^{40} \sec^{-1}}}\\
    & \quad \times 
    \braket{\frac{r}{10\au}}^{-2}
    \braket{\frac{\avesigma }{10^{-18} \cm^2}} 
    \braket{\frac{\mach}{1}}
\end{split}
\] 
at the shortest. 
We note that $\Delta s_\mathrm{suv}$ is a {\it potential} thickness of the photoheated \ion{H}{1} layer and 
does not necessarily represent the actual thickness of it. 
When recombination is strong, i,e., $t_\mathrm{rec} < R/\mach \cs $,  
the recombination timescale determines the thickness of the photoheated \ion{H}{1} layer,
as implied by the slow photoionization condition, \eqnref{eq:typeI_condition}.

The condition, $\Delta s_\mathrm{suv} > R$, is a sufficient condition for winds to be Type~I. 
This is rewritten to 
\begin{equation}
    \frac{R}{\mach \cs t_\mathrm{ioni}} 
    = 
    \mach^{-1}
    \braket{\frac{\cs }{\ceq }}^{-1}
    \frac{\phieuv }{\Phi_\mathrm{c,i}  x }\chi_\mathrm{i}  < 1, 
    \label{eq:tioni_tcross}
\end{equation}
and $\Phi_\mathrm{c,i}$ is the photoionization critical EUV emission rate
\[
  \begin{split}
    \Phi_\mathrm{c,i} & \equiv \frac{4 \pi GM_* }{ 2c_p \ceq  \avesigma }  
    = \Phi_\mathrm{c} \frac{\aveenergy }{c_p m \ceq ^2}\\
    & \approx 3.3\e{38} \sec^{-1} 
    \braket{\frac{M_*}{1\Msun}}
    \braket{\frac{c_p}{5/2}}^{-1}
    \\& \times
    \braket{\frac{\ceq }{10\kms}}^{-1}
    \braket{\frac{\avesigma }{1\e{-18}\cm^2}}^{-1}.
  \end{split} 
\]  
Again, the critical EUV emission rate of photoionization depends only on the stellar parameters and disk gas properties. 
Similarly to $\chi_\mathrm{e}$, $\chi_\mathrm{i}$ is expected to be of the order of unity in the photoheated layer. 
Therefore, we ignore $\chi_\mathrm{i}$ in the following sections to remove complexity from our order-of-magnitude estimation (again, see \secref{sec:discussion:model_limitation} for the potential impact of this simplification).  

Here, we retrieve one of the spectral hardness parameters, $\varepsilon $ (\eqnref{eq:tildes}), 
\[
    \begin{split}
    \varepsilon   & 
    \equiv \frac{ \Phi_\mathrm{c,i} }{ \Phi_\mathrm{c} }
     = \frac{\aveenergy }{c_p m \ceq ^2}
    \\ 
    & 1
    \approx \braket{\frac{\aveenergy }{3.7 \eV}}
    \braket{\frac{c_p}{5/2}}^{-1}
    \braket{\frac{m}{1.4 m_\mathrm{H}}}^{-1}
    \braket{\frac{\ceq }{10\kms}}^{-2}.
    \end{split}
\]
This parameter characterizes the spectral hardness for the advection of atomic hydrogen.
The inverse $\varepsilon  ^{-1}$ is physically interpreted as the number of photoionization necessary to heat the gas to the equilibrium temperature, or equivalently to have an energy of $c_p m \ceq ^2$. 
If $\varepsilon  >1$, the spectrum is heating-oriented; 
the gas can get the heat exceeding $c_p m \ceq ^2$ with only single photoionization of atomic hydrogen. 
Conversely, for $\varepsilon   < 1$, the spectrum is ionization-oriented;
multiple photoionization is needed to heat the gas to $c_p m \ceq ^2$. 
A relatively large amount of energy is consumed through photoionization rather than heating in this case. 
We stress that $\varepsilon  $ is determined by the spectral shape of EUV and is independent of the absolute value of the incident flux or $\phieuv $.

Using the dimensionless quantities, 
the recombination timescale \eqnref{eq:trec_base_density} is rewritten to 
\begin{equation}
    t_\mathrm{rec} 
    = \frac{R_\mathrm{c}}{\ceq }
    \braket{
    \frac{q\varepsilon  ^{-1} \varphi}{ x ^3}
    }^{-1/2}
    \braket{
    \frac{\cs }{\ceq }
    }^\beta. 
    \label{eq:trec_reduced}
\end{equation}
Here, we find the other spectral hardness parameter, $q$ (\eqnref{eq:q}),
 \[
 \begin{split}
     q & \equiv \frac{3C^2 \alpha_\mathrm{eq} }{\ceq \avesigma } \\
     & \approx 0.096 \braket{\frac{C}{0.4}}^2 
     \braket{\frac{\alpha_\mathrm{eq}}{2\e{-13}}} 
     \\ & \quad \times 
     \braket{\frac{\ceq }{10\kms}}^{-1}
     \braket{\frac{\avesigma }{10^{-18}\cm^2}}^{-1}
 \end{split}
\]
The $q$ parameter also characterizes the spectral hardness for hydrogen recombination, similarly to $\varepsilon  $. (See \secref{sec:sub:photoionization_recombination_timescales} for a physical interpretation.)

We can evaluate the right-hand side (RHS) of the slow ionization condition (\eqnref{eq:typeI_condition}) by considering the ratio between $t_\mathrm{rec}$ and $R/\mach \cs $, 
\begin{equation}
    \frac{R}{\mach \cs  t_\mathrm{rec}}
    = 
    \braket{
    \frac{\varphi q}{\varepsilon  x }
    }^{1/2}
    \braket{\frac{\cs }{\ceq }}^{-(1+\beta)}
    \mach^{-1}.
    \label{eq:trec_tcross}
\end{equation}
Also, we can compare the ratio between $t_\mathrm{ioni}$ and $t_\mathrm{rec}$ as 
\begin{equation}
    \begin{split}
    \frac{t_\mathrm{ioni}}{t_\mathrm{rec}}
    =  \braket{\frac{\varepsilon  q  x  }{\varphi}}^{1/2} 
    \braket{\frac{\cs }{\ceq }}^{-\beta},
\end{split}
\label{eq:tioni_trec_ratio:dimensionless}
\end{equation}
For $t_\mathrm{ioni} \ll t_\mathrm{rec}$, this ratio gives an approximate \ion{H}{1} abundance in the photoinoization-recombination equilibrium
\begin{equation}
    y_\mathrm{HI} 
    \approx  \braket{\frac{\varepsilon  q  x  }{\varphi}}^{1/2} 
    \braket{\frac{\cs }{\ceq }}^{-\beta} 
    ,
\quad \braket{\mathrm{for} ~ t_\mathrm{ioni} < t_\mathrm{rec}}.
\label{eq:y_HI_equi_reduced}
\end{equation}
 

\subsection{Type~I Regime}
\label{sec:typeI:regime}
Now we will find the expressions for the parameter space that satisfies the Type~I wind condition (\eqnref{eq:typeI_condition}).
Recalling that 
typical $\cs $ and flow velocity follow the three pattern summarized in \tref{tab:regime_summary} (regardless of the wind types), 
the slow photoionization condition \eqnref{eq:typeI_condition} is equivalent with the sum of these three conditions:
\begin{description}
    \item [Condition~I-1] $c_\mathrm{ch}^\mathrm{I} > \ceq  ~ \cap ~ t_\mathrm{ioni} > \mathrm{min}\left[t_\mathrm{rec} (\cs  = \ceq ), R/\ceq \right]$. 
    \item [Condition~I-2] $c_\mathrm{ch}^\mathrm{I} < \vg  ~ \cap ~ t_\mathrm{ioni} > \mathrm{min}\left[t_\mathrm{rec} (\cs  = \vg ), R/\mach_\mathrm{g}^\mathrm{I} \vg  \right]$. 
    \item [Condition~I-3] $ v_\mathrm{g } < c_\mathrm{ch}^\mathrm{I} < \ceq  ~ \cap ~ t_\mathrm{ioni} > \mathrm{min}\left[t_\mathrm{rec} (\cs  = c_\mathrm{ch}^\mathrm{I}), R/c_\mathrm{ch}^\mathrm{I} \right]$. 
\end{description}

First, we consider Condition~I-1. 
If $\varphi > \varepsilon  q^{-1}  x $ (cf. \eqnref{eq:trec_tcross}), 
the recombination timescale is faster than the replenishment of atomic hydrogen through advection,
and Condition~I-1 reduces to 
\[
\left\{ 
( x , \varphi)~|~ 
\varphi >  x , 
~ 
\varepsilon  q x  > \varphi, 
~ 
\varphi > \varepsilon  q^{-1}  x 
\right\}
\]
using Eqs.\eqref{eq:tioni_trec_ratio:dimensionless} and \eqref{eq:cchI_def}. 
In the opposite case $\varphi < \varepsilon  q^{-1}  x $, 
the Condition~I-1 reduces to 
\[
\left\{
( x , \varphi)~|~ 
\varphi >  x , 
~
t_\mathrm{ioni} > R/\ceq ,
~
\varphi < \varepsilon  q^{-1}  x 
\right\}
\] 
where 
\begin{gather}
    \frac{R}{t_\mathrm{ioni} \ceq } 
    =  x  ^{-1} 
    \frac{\Phi_{\rm EUV}}{\Phi_\mathrm{c,i}}
    = \varphi
    \varepsilon  ^{-1}  x ^{-1}
    \label{eq:t_ioni_VS_c_eq}
\end{gather}  
Hence, Condition~I-1 is equivalent to the sum of 
\begin{equation}
\left\{
( x , \varphi)~|~
\varphi >  x ,
~
\varphi < \varepsilon  q  x ,
~
\varphi > \varepsilon  q^{-1}  x 
\right\}
\end{equation}
and
\begin{equation}
\left\{
( x , \varphi)~|~
x > 1, 
~
x < \varepsilon  ,
~ 
x < \varepsilon  q^{-1} 
\right\}.
\end{equation}
After some math, 
we learn that 
\begin{equation}
\varepsilon   > 1 \quad \mathrm{or} \quad \varepsilon  q > 1
\label{eq:hard_spectrum_def}
\end{equation}
is necessary for $(x,\varphi)$ to exist, and in this case, Condition~I-1 reduces to 
\[
      x  < \varphi < \mathrm{max}\braket{\varepsilon  , \varepsilon  q} \times  x . 
\]
For $\{(\varepsilon , q)|~\varepsilon   < 1 , ~ \varepsilon  q < 1\}$, the mathematical set of Condition~I-1 becomes null, meaning that the Type~I winds will never be a high temperature flow at $\ceq $. 
Qualitatively, 
the Type~I winds with the equilibrium temperature ($\ceq $) form only when the spectrum is hard enough so that one photoionization deposits larger energy than $c_p m \ceq ^2$ ($\varepsilon   > 1$), or the gas can be heated to $\cs  = \ceq $ at a lower $\phieuv $ than is needed to get $t_\mathrm{ioni}/t_\mathrm{rec} < 1$. 
When the spectrum is soft, 
high-temperature flows ($\ceq $) cannot form unless the wind is Type~II.
We will see this point later in \secref{sec:typeII}.

Next, we consider Condition~I-2. 
The condition $c_\mathrm{ch}^\mathrm{I} < \vg $ reduces to $\varphi <  x ^{-1/2}$ (see \eqnref{eq:cchI_vg}), 
and thus the emission rate relevant here is those having $\varphi < 1$, as we are interested in the region outside the critical radius, $ x  > 1$. 
In this case, it is physically trivial that when the spectrum satisfies the hard spectrum condition \eqnref{eq:hard_spectrum_def}, 
$t_\mathrm{ioni} > \mathrm{min}\left[t_\mathrm{rec}(\cs  = c_\mathrm{ch}^\mathrm{I}), R/\mach_\mathrm{g}^\mathrm{I} \vg \right]$ is always met. 
This is also understood as follows: 
$\ceq  > \vg $ always holds in the radius of interest, $ x  > 1$, by definition, and  
\eqnref{eq:tioni_trec_ratio:dimensionless} indicates $t_\mathrm{ioni} > t_\mathrm{rec}$ for $\varepsilon  q > 1$. 
Similarly, when $\varepsilon   > 1$, 
\begin{equation}
        \frac{R}{t_\mathrm{ioni} \mach_\mathrm{g}^\mathrm{I} v_g}
        =   \varepsilon  ^{-1}  x ^{-1} < 1
    \label{eq:t_ioni_VS_v_g}
\end{equation}
and thus $t_\mathrm{ioni} > R/\vg $, with $ x $ of interest. 
Hence, Condition~I-2 simply reduces to $\varphi <  x ^{-1/2}$ for $\varepsilon  >1$ or $\varepsilon  q > 1$. 

The non-trivial case of Condition~I-2 is when the spectrum satisfies $\varepsilon   < 1$ and $\varepsilon  q < 1$, i.e., when the spectrum is soft. 
Using Eqs.(\ref{eq:vg_ceq}), (\ref{eq:trec_tcross}), (\ref{eq:tioni_trec_ratio:dimensionless}), and (\ref{eq:t_ioni_VS_v_g}), Condition~I-2 reduces to the sum of 
\[
\left\{( x , \varphi)~|~
    \varphi  <   x ^{-1/2},
    ~
    \varphi  < \varepsilon   q  x ^{1+ \beta},
    ~
    \varphi < q \varepsilon ^{-1} x^{- (1 - \beta )}
    \right\}
\]
and 
\[
\left\{( x , \varphi)~|~
    \varphi  <   x ^{-1/2},
    ~ 
     x  > \varepsilon   ^{-1}, 
    ~
    \varphi > q \varepsilon ^{-1} x^{- (1 - \beta)}
    \right\}. 
\]
After tedious mathematics, 
this condition reduces to the following two cases depending on $\varepsilon  $ and $q$:  
\[
    \left\{ 
    \begin{array}{ll}
         \varphi <  \varepsilon   q  x ^{1 + \beta }& \quad \mathrm{for}~ 1\leq  x  < \varepsilon  ^{-1} \\
         \varphi  <    x ^{-1/2}& \quad \mathrm{for}~  x  > \varepsilon  ^{-1} \\
    \end{array}
    \right.
\]
for $\left\{(\varepsilon , q)|~\varepsilon   < 1,~\varepsilon  q < 1, ~ q < \varepsilon  ^{(2\beta + 1)/2}\right\}$, 
and
\[
    \left\{ 
    \begin{array}{ll}
         \varphi  <   \varepsilon   q  x ^{1+\beta}& \quad\mathrm{for}~ 1 <  x  < (\varepsilon  q)^{-2/(3+2\beta)} \\
         \varphi  <    x ^{-1/2}& \quad\mathrm{for}~  x  > (\varepsilon  q)^{-2/(3+2\beta)} 
    \end{array}
    \right.
\]
for $\left\{ (\varepsilon , q)|~\varepsilon   < 1,~ \varepsilon  q < 1,~ \varepsilon  ^{(2\beta + 1)/2} < q \right\}$.

Finally, we evaluate Condition~I-3. 
Again, it is trivial that when the spectrum satisfies \eqnref{eq:hard_spectrum_def}, 
the Type~I condition is always met for $c_\mathrm{ch}^\mathrm{I} < \ceq $. 
Therefore, here we discuss the non-trivial cases, $\left\{(\varepsilon , q)|~\varepsilon   < 1,~ \varepsilon  q < 1\right\}$. 
To that end, we calculate Eqs.(\ref{eq:tioni_trec_ratio:dimensionless}) and (\ref{eq:trec_tcross}) with $\cs  = c_\mathrm{ch}^\mathrm{I}$ in advance,
\begin{gather}
    \left.\frac{t_\mathrm{ioni}}{t_\mathrm{rec}}\right|_{\cs  = c_\mathrm{ch}^\mathrm{I}} 
    = \braket{
    \frac{\phieuv }{\Phi_\mathrm{c} x  (\varepsilon  q)^{3/(3+2\beta)}}
    }^{-(3+2\beta)/6}
    \\ 
    \frac{R}{t_\mathrm{rec}c_\mathrm{ch}^\mathrm{I}}
    = \braket{
    \frac{\phieuv (\varepsilon  q^{-1})^{3/(2\beta-1)}}{\Phi_\mathrm{c} x  }
    }^{-(2\beta-1)/6}
\end{gather}
Using these relations and Eqs.(\ref{eq:cchI_def}) and (\ref{eq:cchI_vg}),
Condition~I-3 reduces to the sum of 
\[
\left\{
    ( x , \varphi)~|~
     x ^{-1/2} < \varphi  <  x ,
    ~
    \varphi < \braket{{q}\varepsilon  }^{\frac{3}{3+2\beta}}  x ,
    ~
    \varphi  < \braket{\frac{q}{\varepsilon  }}^{\frac{3}{2\beta-1}}  x 
    \right\}
\]
and 
\[
\left\{
    ( x , \varphi)~|~
     x ^{-1/2} < \varphi  <  x ,
    ~
    \varphi  < \varepsilon  ^{3/2}  x ,
    ~
    \varphi  > \braket{\frac{q}{\varepsilon  }}^{\frac{3}{2\beta-1}}  x 
    \right\}.
\]
Consequently, Condition~I-3 reduces to the following two cases depending on $\varepsilon  $ and $q$:
\[
\left\{
    ( x , \varphi)~|~
    \varphi >   x ^{-1/2},
    ~
    \varphi <  \varepsilon  ^{3/2}  x 
    \right\}
\]
for $\left\{(\varepsilon , q)|~\varepsilon   < 1,~ \varepsilon  q < 1,~ q < \varepsilon  ^{(2\beta+1)/2}\right\}$,
and 
\[
\left\{
    ( x , \varphi)~|~
    \varphi >   x ^{-1/2},
    ~
    \varphi <  (\varepsilon   q)^{3/(2\beta+3)}  x 
    \right\}
\]
for $\left\{(\varepsilon , q)|~\varepsilon   < 1,~ \varepsilon  q < 1,~ q > \varepsilon  ^{(2\beta+1)/2}\right\}$.

To summarize this section, winds become Type~I when satisfying the Type~I condition (\eqnref{eq:typeI_condition}), 
\[
    t_\mathrm{ioni} > \mathrm{min} \braket{t_\mathrm{rec}, \frac{R}{\mach \cs }}. 
\]
This condition covers a certain area on the $ x $--$\varphi$ plane, depending on the spectral hardness parameters, $\varepsilon  $ and $q$, which are uniquely set once an EUV spectrum is given.
Based on $\varepsilon  $ and $q$, we can classify the $ x $--$\varphi$ map into the following three different classes:
\paragraph{\bf H-class}
    Hard spectrum class. 
    This class is defined by $\{(\varepsilon  , q) | \varepsilon   > 1 \cup \varepsilon  q > 1\}$. 
    Hot ($\cs  = \ceq $), atomic flow is possible only in this case.    
    The Type~I regime covers 
    \[
        \varphi < \mathrm{max}\braket{1, q} \varepsilon     x  . 
    \]
    on the $ x $--$\varphi$ plane. 
\paragraph{\bf Sa-class}
    Soft spectrum with advection replenishment. 
    This class is defined by $\{(\varepsilon  , q) | \varepsilon   < 1, \varepsilon  q < 1, q < \varepsilon  ^{(2\beta+ 1)/2)}\}$. 
    The Type~I regime corresponds to 
    \begin{equation}
        \left\{ 
    \begin{array}{ll}
         \varphi <  \varepsilon   q  x ^{1 + \beta }
         & \quad \mathrm{for}~ 1\leq  x  < \varepsilon  ^{-1} \\
         \varphi <  \varepsilon  ^{3/2}  x & \quad \mathrm{for}~  x  > \varepsilon  ^{-1} \\
    \end{array}
    \right.
    \label{eq:typeI_regime_Saclass}
    \end{equation}
    Type~I flow is possible only with temperatures below the equilibrium value, i.e., $\cs  < \ceq $, and the atomic flow is maintained by advection in the steadily heated, free winds ($\cs \sim c_\mathrm{ch}^\mathrm{I}$). 
\paragraph{\bf Sr-class}
    Soft spectrum with recombination reproduction.
    This class is defined by $\{(\varepsilon  , q)| \varepsilon   < 1, \varepsilon  q < 1, q > 1 \}$.
    The Type~I regime is 
    \begin{equation}
    \left\{ 
    \begin{array}{ll}
         \varphi <  \varepsilon   q  x ^{1+\beta}& \quad\mathrm{for}~ 1 \leq   x  < (\varepsilon  q)^{-2/(3+2\beta)} \\
         \varphi <  \braket{\varepsilon  q}^{3/(3+2\beta)}  x  & \quad\mathrm{for}~  x  > (\varepsilon  q)^{-2/(3+2\beta)} 
    \end{array}
    \right.
    \label{eq:typeI_regime_Srclass}
    \end{equation}
    Atomic flow is maintained by recombination rather than replenishment through advection. 

\begin{figure*}
    \centering
    \includegraphics[clip, width = \linewidth]{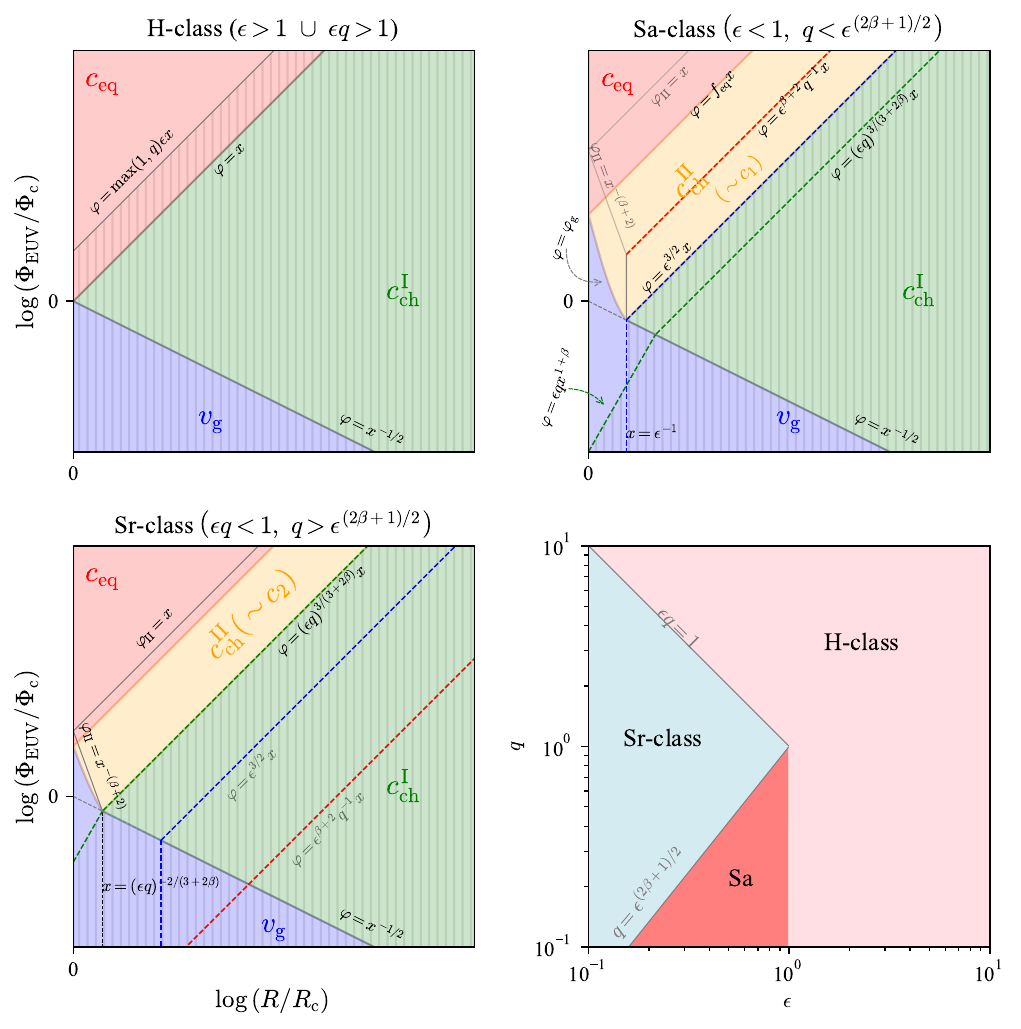}
    \caption{Same as Figures~\ref{fig:sqplane}, \ref{fig:H-class}, \ref{fig:Sa-class}, and \ref{fig:Sr-class} but with $\beta$ left unspecified. The expressions for the boundaries are also annotated.
    The top left, top right, and bottom left panels are those for the H-class, Sa-class, and Sr-class, respectively. 
    The bottom right is the spectral-hardness diagram.
    }
    \label{fig:classmaps}
\end{figure*}

The top left, top right, and bottom left panels in \fref{fig:classmaps} show the parameter spaces $(x, \varphi)$ for the H-class, Sa-class, and Sr-class, respectively. 
In each panel, the Type~I regime (\eqnref{eq:typeI_condition}) is represented by gray vertical lines. 
Generally, the Type~I regimes can be subdivided into three regimes: the isothermal wind regime where $\cs  \sim \ceq $ (red-shaded area); the steadily heated, free wind regime where $\cs  \sim c_\mathrm{ch}^\mathrm{I}$ (green-shaded area); and the gravity-inhibited wind regime where $\cs  \sim \vg $ (blue-shaded area). 
These three regimes correspond to Regions~A, B, and C in \begelman{}, respectively (cf. \tref{tab:regime_summary}). 
We follow these notations for the regime labels, but we designate the Type~I steadily heated, free wind regime as Regime~B-I to highlight its classification as a Type~I regime (cf. \tref{tab:sound_spees}).

Again, the class to which a given EUV spectrum is assigned is determined by the location of the spectrum's $\varepsilon  $ and $q$ on the $\varepsilon  $--$q$ plane (the bottom right panel in \fref{fig:classmaps}). 




\subsection{Characteristic Sound Speed: Type~II}
\label{sec:typeII}
The remaining parameter space uncovered by the Type~I regime is where photoionization is faster than recombination and the replenishment of the atoms through advection (\eqnref{eq:typeII_condition}), 
\[
    t_\mathrm{ioni} < \mathrm{min} \braket{t_\mathrm{rec}, \frac{R}{\mach \cs }}
\]
so that the flow includes the ionized region within $s < R$ (Type~II; right panel in \fref{fig:streamline_types}).
In this case, the characteristic sound speed for the Type~I winds, $c_\mathrm{ch}^\mathrm{I}$ (\eqnref{eq:cchI_def}), is unavailable to characterize the heating timescale of Type~II flows since it has been derived with the assumption of fully atomic winds within $s < R$. 
Thus, we need to redefine an alternative characteristic sound speed for Type~II flows, which we address in this section.

For Type~II winds, the total energy gain during the gas flowing to the height $\sim R$ is decomposed into two parts. 
The first component is the initially injected energy until a lunched atomic hydrogen undergoes photoionization once. 
The launched atomic hydrogen can survive, on average, for a timescale of $t_\mathrm{ioni}$ and thus can travel a distance of $\sim \mach \cs  t_\mathrm{ioni}$.
The surviving distance corresponds to $s_1$, the length of the neutral part depicted in the right panel of \fref{fig:streamline_types}.
It is worth noting that, by definition, the timescale of recombination and advection replenishment is guaranteed longer than the photoionization timescale in the Type~II wind, and thus $t_\mathrm{ioni}$ sets the average timescale for which a newly launched atomic hydrogen can persist in the wind. 

The second component is the additional supplied energy by reionizations of recombined hydrogen in the ionized region. 
Recombination can take place in the \ion{H}{2} region if the timescale is sufficiently shorter than the wind crossing timescale. 
The recombined hydrogen will undergo reionization due to the short ionization timescale, depositing energy to the gas. 
The \ion{H}{1 } abundance remains small since $t_\mathrm{ioni} < t_\mathrm{rec}$.

Hence, to define the characteristic sound speed for Type~II winds, $c_\mathrm{ch}^\mathrm{II}$, we separate the integration range of \eqnref{eq:c_ch_def} for the initially injected energy and secondary deposited energy as 
\begin{equation}
    c_p {c_\mathrm{ch}^\mathrm{II}}^2 
    =  \int_0^{s_\mathrm{II}} 
    \frac{\avesigma  \aveenergy   \phieuv }{4\pi m r^2} 
    \frac{\dd s}{c_\mathrm{ch}^\mathrm{II}} 
    + 
    \int_{s_\mathrm{II}}  ^R 
    \frac{t_\mathrm{ioni}}{t_\mathrm{rec}}
    \frac{\avesigma  \aveenergy   \phieuv }{4\pi m r^2} 
    \frac{\dd s}{c_\mathrm{ch}^\mathrm{II}} , 
    \label{eq:cch:typeII:integral_form}
\end{equation}
where $s_\mathrm{II}$ is the surviving length of launched hydrogen, $s_\mathrm{II} = c_\mathrm{ch}^\mathrm{II} t_\mathrm{ioni}$. 
The first and second terms on the RHS represent the initially injected energy and additional energy supply through reionizations, respectively. 
Here, we have approximated the \ion{H}{1} abundance to unity while the gas undergoes the first photoionization (the first term on the RHS) and to the value set by ionization-recombination equilibrium, $y_\mathrm{HI}\sim t_\mathrm{ioni}/t_\mathrm{rec}$ (\eqnref{eq:y_HI_equi_reduced}), in the ionized region (the second term on the RHS). 

The first term on the RHS of \eqnref{eq:cch:typeII:integral_form} is explicitly computed as
\begin{equation}
    c_p c_1^2 \approx \frac{\avesigma  \aveenergy   \phieuv }{4\pi m r^2} 
    t_\mathrm{ioni} = \frac{\aveenergy}{m}
    \label{eq:c1_def}
\end{equation}
where $c_1$ is the sound speed corresponding to the initially injected energy.
The above equation is rewritten to
\[
    \frac{c_1}{\ceq } = \varepsilon  ^{1/2}. 
\]
This indicates that for $\varepsilon  > 1$, heating due to the initially injected energy is sufficiently rapid that the launched gas can reach the equilibrium temperature within a short distance $s < \mach \cs  t_\mathrm{ioni}$.
On the other hand, for $c_1 < \vg $, or equivalently $\varepsilon  x < 1$, gravity is so strong that the initially injected energy is not enough to form an escaping wind. 
This suggests that in this range, the additional energy supplied through reionizations is indispensable to form escaping winds. 

The second integration on the RHS of \eqnref{eq:cch:typeII:integral_form} can also simplify to
\[
\int_{s_\mathrm{II}}  ^R 
    \frac{t_\mathrm{ioni}}{t_\mathrm{rec}}
    \frac{\avesigma  \aveenergy   \phieuv }{4\pi m r^2} 
    \frac{\dd s}{c_\mathrm{ch}^\mathrm{II}} 
    \approx
    \frac{t_\mathrm{ioni}}{t_\mathrm{rec}}
    \frac{\avesigma  \aveenergy   \phieuv }{4\pi m r^2} 
    \frac{R - c_\mathrm{ch}^\mathrm{II} t_\mathrm{ioni}}{c_\mathrm{ch}^\mathrm{II}} 
\]
and then the definition of the characteristic sound speed for Type~II winds reduces to
\begin{equation}
    c_p {c_\mathrm{ch}^\mathrm{II}}^2   
    = c_p c_1 ^2 
    + \frac{\avesigma  \aveenergy   \phieuv }{4\pi m r^2} \frac{t_\mathrm{ioni}}{t_\mathrm{rec}}
    \frac{R - c_\mathrm{ch}^\mathrm{II} t_\mathrm{ioni}}{c_\mathrm{ch}^\mathrm{II}} . 
    \label{eq:chII_original_def}
\end{equation}
This equation can be rewritten to 
\[
\begin{split}
    c_p {c_\mathrm{ch}^\mathrm{II}}^2 
    & =  \frac{\avesigma  \aveenergy   \phieuv }{4\pi m r^2} 
    \left[
    t_\mathrm{ioni} + 
    \frac{t_\mathrm{ioni}}{t_\mathrm{rec}}
    \braket{
    \frac{R}{c_\mathrm{ch}^\mathrm{II}} 
    - t_\mathrm{ioni}
    }
    \right]
    \\
    & = 
     \frac{\avesigma  \aveenergy   \phieuv }{4\pi m r^2} 
    \left[
    t_\mathrm{ioni} \braket{1 - \frac{t_\mathrm{ioni}}{t_\mathrm{rec}}} 
    + 
    \frac{t_\mathrm{ioni}}{t_\mathrm{rec}}
    \frac{R}{c_\mathrm{ch}^\mathrm{II}} 
    \right]. 
\end{split}
\]
One can observe from this form that the Type~II characteristic sound speed $c_\mathrm{ch}^\mathrm{II}$ smoothly transitions to the Type~I characteristic sound speed $c_\mathrm{ch}^\mathrm{I}$ since the boundary between Type~I and Type~II regimes is expressed by either $t_\mathrm{ioni} = t_\mathrm{rec}$ or $t_\mathrm{ioni} = R/c_\mathrm{ch}$ for all classes.
This continuity arises because $c_\mathrm{ch}^\mathrm{I}$ can be interpreted as a special case of $c_\mathrm{ch}^\mathrm{II}$ when the surviving distance $s_1$ is larger than $R$. 
It also emphasizes that $c_\mathrm{ch}^\mathrm{II}$ is defined only in the parameter space where 
\begin{equation}
    t_\mathrm{ioni} < t_\mathrm{rec} 
    \quad
    \cap
    \quad
    t_\mathrm{ioni} < \frac{R}{c_\mathrm{ch}^\mathrm{I}}
\label{eq:chII_available_condition}
\end{equation}
is met. 
The Type~II regime is always included in this parameter space,
and thus $c_\mathrm{ch}^\mathrm{II}$ is always defined in the Type~II regime (\eqnref{eq:typeII_condition}). 

\eqnref{eq:chII_original_def} can also be rewritten in dimensionless form as
\begin{equation}
    \braket{\frac{c_\mathrm{ch}^\mathrm{II}}{\ceq }}^2 
    = \varepsilon   
    + \braket{\frac{\varphi \varepsilon   q}{ x }}^{1/2}
    \braket{\frac{c_\mathrm{ch}^\mathrm{II}}{\ceq }}^{-(\beta+1)}
    \braket{1 - \varphi^{-1}\varepsilon   x \frac{c_\mathrm{ch}^\mathrm{II}}{\ceq }}
    .
    \label{eq:chII_original_def:dimensionless}
\end{equation}
The last factor of the second RHS term represents the length of the ionized region.
\eqnref{eq:chII_original_def:dimensionless} is also equivalently rewritten to
\begin{equation}
    \braket{\frac{c_\mathrm{ch}^\mathrm{II}}{\ceq }}^2 
    = 
    \varepsilon   \left[1 - \braket{\frac{x \varepsilon   q}{ \varphi }}^{1/2}
    \braket{\frac{c_\mathrm{ch}^\mathrm{II}}{\ceq }}^{-\beta}\right]
    + \braket{\frac{\varphi \varepsilon   q}{ x }}^{1/2}
    \braket{\frac{c_\mathrm{ch}^\mathrm{II}}{\ceq }}^{-(\beta+1)}. 
    \label{eq:chII_original_def:dimensionless:another_form}
\end{equation}
The relative strength of the second term to the first term in Eqs.\eqref{eq:chII_original_def:dimensionless} and \eqref{eq:chII_original_def:dimensionless:another_form} is determined by the ratio between the crossing timescale and recombination timescale, 
\begin{equation}
    \frac{R}{\cs  t_\mathrm{rec}}
    = 
    \braket{
    \frac{\varphi q}{\varepsilon  x }
    }^{1/2}
    \braket{\frac{\cs }{\ceq }}^{-(1+\beta)}
    \label{eq:trec_tcross_wo_mach}
\end{equation}
(cf. \eqnref{eq:trec_tcross}). 
Again, the RHSs of Eqs.\eqref{eq:chII_original_def:dimensionless} and \eqref{eq:chII_original_def:dimensionless:another_form} are guaranteed positive in the parameter space represented by \eqnref{eq:chII_available_condition}, which is rewritten to a dimensionless form as 
\begin{equation}
\left\{
    (x, \varphi) ~\left| ~
    \braket{\frac{\varepsilon  q  x  }{\varphi}}^{1/2} 
    \braket{\frac{\cs }{\ceq }}^{-\beta} < 1 
    ,
    ~
    \varphi > \varepsilon ^{3/2} x
    \right.
    \right\}
    \label{eq:chII_available_condition:dimensionless}
\end{equation}
This is consistently satisfied in the Type~II regime. 

The characteristic sound speed $c_\mathrm{ch}^\mathrm{II}$ resulting from \eqnref{eq:chII_original_def:dimensionless} is a monotonically increasing function of $\varphi$ at any distance $x$ and is a monotonically decreasing function of $x$ at any $\varphi$. 
This is understood by taking the partial differences of both sides of \eqnref{eq:chII_original_def:dimensionless} with respect to $\varphi$ and $x$, 
\[
    \begin{gathered}
    \begin{aligned}
        \braket{\frac{\partial u}{\partial \varphi}}_x 
        & = \frac{u}{\varphi} \left[\frac{1}{2} + \braket{\frac{\varphi}{\varepsilon x u} - 1}^{-1}\right] 
        \\ 
        & \quad \times 
        \left[
        \frac{2 u^2}{u^2 - \varepsilon}
        + \beta + 1
        + \braket{\frac{\varphi}{\varepsilon x u} - 1}^{-1} 
        \right]^{-1} 
        \quad \braket{> 0}
    \end{aligned}
        \\
    \begin{aligned}
        \braket{\frac{\partial u}{\partial x}}_\varphi
        &= - \frac{u}{x} \left[\frac{1}{2} + \braket{\frac{\varphi}{\varepsilon x u} - 1}^{-1}\right] 
        \\
        & \quad \times
        \left[
        \frac{2 u^2}{u^2 - \varepsilon}
        + \beta + 1
        + \braket{\frac{\varphi}{\varepsilon x u} - 1}^{-1} 
        \right]^{-1} 
        \quad \braket{< 0}
        ,
    \end{aligned}
    \end{gathered}
\]
where $u \equiv c_\mathrm{ch}^\mathrm{II}/\ceq$. 
This simply reflects that the heating timescale of Type~II winds gets shorter as $\phieuv $ increases 
and longer at large distances.

The Type~II characteristic sound speed is a metric for the heating timescale to be compared with $\ceq $ and $\vg $. 
Therefore, it is convenient to derive the equations, $\varphi = \varphi(x)$, at which $c_\mathrm{ch}^\mathrm{II}$ equals to $\ceq $ and $\vg $. 
Since $c_\mathrm{ch}^\mathrm{II}$ derived by \eqnref{eq:chII_original_def:dimensionless} is a monotonically increasing function of $\varphi$ at any distance, there is a unique $\varphi$ at which $c_\mathrm{ch}^\mathrm{II} = \ceq $ at any $x$. 
Substituting $c_\mathrm{ch}^\mathrm{II} = \ceq $ into \eqnref{eq:chII_original_def:dimensionless}, such $\varphi$ is derived as 
\[
    \varphi  = f_\mathrm{eq} (\varepsilon , q) x,
\]
where 
\[
    f_\mathrm{eq} (\varepsilon , q) 
    \equiv \frac{4 \varepsilon  q }
    {\braket{\sqrt{(\varepsilon ^{-1}-1)^2 + 4q} - (\varepsilon ^{-1}-1)}^2}
\]
Hence, the rapid heating condition, $c_\mathrm{ch}^\mathrm{II} > \ceq $, reduces to $\varphi > f_\mathrm{eq} x$. 
The factor $f_\mathrm{eq}$ is equal to unity when $\varepsilon  = 1$ or $\varepsilon  q = 1$, and the range of $f_\mathrm{eq}$ is tabulated in \tref{tab:f_eq} for other $(\varepsilon , q)$. 
\begin{table}[htbp]
    \centering
    \begin{tabular}{LcC}
    \text{Parameter space} & Class & \text{Range of }~f_\mathrm{eq} \\
    \hline
        \left\{(\varepsilon , q)|~\varepsilon  q > 1, ~\varepsilon  > 1\right\} 
        &  
        H
        &
        1< f_\mathrm{eq} < \mathrm{min}(\varepsilon , \varepsilon  q)
        \\
        \left\{(\varepsilon , q)|~\varepsilon  q > 1, ~\varepsilon  < 1\right\} 
        &  
        H
        &
        \varepsilon  < f_\mathrm{eq} < 1
        \\
        \left\{(\varepsilon , q)|~\varepsilon  q < 1, ~\varepsilon  > 1\right\} 
        &  
        H
        &
        \varepsilon  q < f_\mathrm{eq} < 1
        \\
        \left\{(\varepsilon , q)|~\varepsilon  q < 1, ~\varepsilon  < 1\right\} 
        &  
        Sa, Sr
        &
        f_\mathrm{eq} > 1
    \end{tabular}
    \caption{Possible values of $f_\mathrm{eq}$ taken in the parameter spaces, and the corresponding spectrum classes.}
    \label{tab:f_eq}
\end{table}
For the H-class, the Type~II regime $\varphi > \mathrm{max}(1, q)\varepsilon  x$ always satisfies $\varphi  > f_\mathrm{eq} x$, indicating that Type~II winds are always isothermal in the H-class. 
For the Sa- and Sr-classes, Type~II winds are not necessarily isothermal and can even be gravity-inhibited. 
This is because the soft spectrum has a relatively small energy deposition per photoionization, and a large $\phieuv $ is needed to increase photoionization to heat the gas to $\cs  = \ceq $.
We will discuss these points in more detail in \secref{sec:typeII:regime}.

Similarly, there is a unique $\varphi$ at which $c_\mathrm{ch}^\mathrm{II} = \vg $. 
Substituting $c_\mathrm{ch}^\mathrm{II} = \vg $ into \eqnref{eq:chII_original_def:dimensionless} reduces to
\begin{equation}
    \braket{\frac{\varphi q x^\beta}{\varepsilon }}^{1/2} 
    =
    x^{-1} \varepsilon ^{-1} - 1 
    + \braket{\frac{\varepsilon  q x^{1+\beta}}{\varphi}}^{1/2}.
    \label{eq:varphi_g_def}
\end{equation}
Note that the LHS is $R/\vg  t_\mathrm{rec}$ with $\cs  = \vg $. 
Solving this equation with respect to $\varphi$ gives the functional form of 
\begin{equation}
\begin{split}
    \varphi&  = \varphi_\mathrm{g} (x; \varepsilon , q, \beta)
    \\ 
    & \equiv 
    \braket{
    \frac{2\sqrt{q\varepsilon }x^{(\beta+1)/2}}
    {\sqrt{(x^{-1}\varepsilon ^{-1}-1)^2 + 4 q x^{(2\beta + 1)/2}} - (x^{-1}\varepsilon ^{-1}-1)}
    }^2
    , 
\end{split}
\label{eq:varphi_g}
\end{equation}
at which $c_\mathrm{ch}^\mathrm{II} $ equals to $\vg $. 
This function has the intersections with $\varphi = x^{-1/2}$ at $x = \varepsilon ^{-1}$ and $x = (\varepsilon  q)^{-2/(3+2\beta)}$ and asymptotically approaches $\varphi = \varepsilon  x^{1/2}$ for $x \gg \varepsilon ^{-1}$. 
It also connects to $f_\mathrm{eq} x$ at $x = 1$, which motivates us to define the secondary critical EUV emission rate as
\begin{equation}
\begin{split}
    \Phi_\mathrm{c}^\mathrm{II}
    & \equiv f_\mathrm{eq} \Phi_\mathrm{c} \\
    & = \frac{4 \varepsilon  q }
    {\braket{\sqrt{(\varepsilon ^{-1}-1)^2 + 4q} - (\varepsilon ^{-1}-1)}^2} \Phi_\mathrm{c},
\end{split}
\label{eq:secondary_critical_luminosity}
\end{equation}
Above this critical EUV emission rate, ionized winds can be isothermal at $x = 1$. 
However, since Type II winds in the H-class are guaranteed isothermal, the secondary critical EUV emission rate is physically meaningful only for the Sa and Sr classes.

The root of the implicit equation, Eq.\eqref{eq:chII_original_def:dimensionless}, can be explicitly expressed by simple forms in some special cases.
When the additional energy injection by reionization is negligible due to slow recombination,
i.e., 
\begin{equation}
    \braket{\frac{c_\mathrm{ch}^\mathrm{II}}{\ceq }}^{\beta+1}
    \gg 
    \braket{\frac{\varphi q}{ x  \varepsilon   }}^{1/2}, 
    \label{eq:slow_recombination_condition}
\end{equation}
(cf. \eqnref{eq:trec_tcross}), 
the total energy gain during the gas traveling to the height of $\sim R$ is primarily covered by the initially injected energy.
In this case, we can drop the second term on the RHS of \eqnref{eq:chII_original_def:dimensionless} to obtain a zeroth-order root as 
\begin{equation}
    c_\mathrm{ch}^\mathrm{II} \approx c_1 = \varepsilon  ^{1/2} \ceq .
    \label{eq:c2_in_slow_recombination}
\end{equation}
Substituting this root back into \eqnref{eq:slow_recombination_condition}, we get 
\begin{equation}
    \varphi \ll \varepsilon ^{\beta + 2} q^{-1} x, 
    \label{eq:c2_in_slow_recombination_valid_paramspace}
\end{equation}
a sufficient condition where \eqnref{eq:c2_in_slow_recombination} is available.

On the other hand, when recombination is fast enough so that the additionally deposited energy dominates over the initially deposited energy, i.e.,
\begin{equation}
    \braket{\frac{c_\mathrm{ch}^\mathrm{II}}{\ceq }}^{\beta+1}
     \ll 
    \braket{\frac{\varphi q}{ x  \varepsilon   }}^{1/2}, 
    \label{eq:rapid_recombination_condition}
\end{equation}
we can neglect the first term of the RHS in \eqnref{eq:chII_original_def:dimensionless:another_form}. 
Then, the zeroth order $c_\mathrm{ch}^\mathrm{II}$ is derived as
\[
    \braket{\frac{c_\mathrm{ch}^\mathrm{II}}{\ceq }}
    \approx 
    \braket{\frac{\varphi_\mathrm{II}}{ x   }}^{1/2(\beta + 3)},
\] 
where
\[
\begin{gathered}
    \varphi_\mathrm{II} \equiv \frac{\phieuv }{\Phi_\mathrm{c,rec}^\mathrm{II}} \\ 
    \Phi_\mathrm{c,rec}^\mathrm{II} \equiv 
    \frac{\Phi_\mathrm{c}}{\varepsilon  q} . 
\end{gathered}
\]
The normalization factor $\Phi_\mathrm{c,rec}^\mathrm{II}$ is the secondary critical EUV emission rate for recombination-dominated Type~II winds.
Here we have retrieved the Type~II characteristic sound speed (\eqnref{eq:cchII_def})
\[
    c_2  \equiv \ceq  \braket{\frac{\varphi_\mathrm{II}}{ x   }}^{1/2(\beta + 3)}. 
\]
Substituting $c_2 $ into Eq.\eqref{eq:rapid_recombination_condition}, 
we obtain a sufficient condition under which the above approximate root is available:
\begin{equation}
    \varphi \gg \varepsilon  ^{\beta + 2} q^{-1}  x , 
    \label{eq:rapid_recombination_condition_paramspace}
\end{equation}
which is opposite to \eqnref{eq:c2_in_slow_recombination_valid_paramspace}.

In the following subsections, we will evaluate the sound speeds and corresponding parameter spaces in the Type~II regime for each class using $c_\mathrm{ch}^\mathrm{II}$.

\subsection{Type~II Winds}
\label{sec:typeII:regime}
As mentioned in \secref{sec:typeII}, the remaining parameter space uncovered by the Type~I regime (the area covered by gray vertical lines in \fref{fig:classmaps}) is the Type~II regime. 
The condition of the Type~II regime is expressed by the fast ionization condition (\eqnref{eq:typeII_condition}), 
\[
    t_\mathrm{ioni} < \mathrm{min} \braket{t_\mathrm{rec}, \frac{R}{\mach \cs }}.
\]
Similarly to the Type~I regime, it can equivalently be broken down into the following three conditions: 
\begin{description}
    \item [Condition~II-1] $c_\mathrm{ch}^\mathrm{II} > \ceq  ~ \cap ~ t_\mathrm{ioni} < \mathrm{min}\left[t_\mathrm{rec} (\cs  = \ceq ), R/\ceq \right]$
    \item [Condition~II-2] $c_\mathrm{ch}^\mathrm{II} < \vg  ~ \cap ~ t_\mathrm{ioni} < \mathrm{min}\left[t_\mathrm{rec} (\cs  = \vg ), R/\mach_\mathrm{g}^\mathrm{II} \vg  \right]$
    \item [Condition~II-3] 
    $ v_\mathrm{g } < c_\mathrm{ch}^\mathrm{II} < \ceq $ $\cap$ $t_\mathrm{ioni} < \mathrm{min}\left[t_\mathrm{rec} (\cs  = c_\mathrm{ch}^\mathrm{II}), R/c_\mathrm{ch}^\mathrm{II} \right]$
\end{description}
Here, $\mach_\mathrm{g}^\mathrm{II}$ is the typical Mach number for gravity-inhibited Type~II winds, which we will derive the specific form later in this section. 

In this section, we analyze the fast ionization condition for each class to derive the sound speed and typical flow speed for Type~II winds and corresponding parameter spaces on the $x$--$\varphi$ plane. 

\subsubsection{Type~II winds in H-class}
\label{sec:typeII:regime:H}
For the H class, it is trivial that $c_\mathrm{ch}^\mathrm{II} $ always exceeds $ \ceq $ from a physical point of view as follows. 
The Type~I condition $t_\mathrm{ioni} > \mathrm{min}(t_\mathrm{rec}, R/\mach \cs )$ indicates that on average, the launched gas reaches $s = R$ with a hydrogen atom being photoionized less than once.
Such gas can still have a characteristic sound speed $c_\mathrm{ch}^\mathrm{I}$ exceeding $\ceq $ in this class. 
In the case where a launched hydrogen experiences photoionization at least once (Type~II), the gas acquires more energy than the Type~I case through recombination followed by reionization.
Thus, the corresponding Type~II characteristic sound speed must exceed $\ceq $.
Hence, the sound speed and the typical flow speed in the ionized region of Type~II flows are given by $\cs  = \ceq $. 
One can also directly prove $c_\mathrm{ch}^\mathrm{II} > \ceq $ mathematically from \eqnref{eq:chII_original_def:dimensionless} for $\varepsilon  > 1$ or $\varepsilon  q > 1$ (see Appendix~\ref{sec:math_proof}).

%

Since $c_\mathrm{ch}^\mathrm{II} > \ceq $, the parameter spaces satisfied by Conditions~II-2 and II-3 are null, and thus the fast ionization condition is equivalent with $\varphi  > \mathrm{max}(1, q)\varepsilon  x$ for the H-class. 
It indicates that the Type~II regime is always in Regime~A (isothermal winds) for the H-class. 

The fact of $c_\mathrm{ch}^\mathrm{II} > \ceq $ indicates that heating is always rapid enough to form isothermal winds in the Type~II regime. 
As discussed in Sections~\ref{sec:typical_flow_velocities} and \ref{sec:c_ch}, this parameter range has a shorter heating timescale than the gravitational timescale, and thus winds can escape without being influenced by gravity.
The typical flow speed is thus given by $\sim \ceq $.

\subsubsection{Type~II winds in Sa-class}
The Type~II regime in the Sa-class is expressed by 
\begin{equation}
    \left\{ 
    \begin{array}{ll}
         \varphi >  \varepsilon   q  x ^{1 + \beta }
         & \quad \mathrm{for}~ 1\leq  x  < \varepsilon  ^{-1} \\
         \varphi >  \varepsilon  ^{3/2}  x & \quad \mathrm{for}~  x  > \varepsilon  ^{-1} \\
    \end{array}
    \right.
    \label{eq:typeII_regime_Saclass}
\end{equation}
(cf. Eq.\eqref{eq:typeI_regime_Saclass}). 

First, we consider Condition~II-1. 
The fast ionization condition $t_\mathrm{ioni} < \mathrm{min}\left[t_\mathrm{rec} (\cs  = \ceq ), R/\ceq \right]$ is rewritten to $\varphi  > \mathrm{max}(1, q)\varepsilon  x = \varepsilon  x$ in this class. 
One can confirm that the parameter space of $c_\mathrm{ch}^\mathrm{II}>\ceq $, or equivalently $\varphi > f_\mathrm{eq} x$, is covered by the Type~II regime, $\varphi > \varepsilon  x$, in this class ($\varepsilon  < 1$). 
Consequently, Condition~II-1 reduces to $\varphi > f_\mathrm{eq} x$.

Next, we consider Condition~II-2. 
We first evaluate $\mach_\mathrm{g}^\mathrm{II}$. 
To form escape winds in this gravity-inhibited regime, the gas must obtain energy to reach $\cs  \sim \vg  $ until traveling to a height of $\sim R$. 
Thus, the self-consistent condition is 
\begin{equation}
\begin{split}
    c_p \vg ^2 
    & \approx c_p c_1^2
    + 
    \frac{\avesigma  \aveenergy   \phieuv }{4\pi m r^2} \frac{t_\mathrm{ioni}}{t_\mathrm{rec}}
    \frac{R - \mach_\mathrm{g}^\mathrm{II}\vg  t_\mathrm{ioni}}{\mach_\mathrm{g}^\mathrm{II}\vg }
    \\
&       = c_p \varepsilon  \ceq ^2
    \braket{1 - \frac{t_\mathrm{ioni}}{t_\mathrm{rec}}}
    + c_p \varepsilon  \ceq ^2
    \frac{R}{\mach_\mathrm{g}^\mathrm{II}\vg t_\mathrm{rec}}
\end{split}
     ,
     \label{eq:typeII:mach_g:self_consistent_condition}
\end{equation}
which reduces to
\[ 
    \frac{R}{\mach_\mathrm{g}^\mathrm{II}\vg t_\mathrm{rec}}
    \approx 
    x^{-1} \varepsilon ^{-1} - \braket{1 - \frac{t_\mathrm{ioni}}{t_\mathrm{rec}}}
    =
    x^{-1} \varepsilon ^{-1} - 1 + 
    \braket{\frac{\varepsilon  q x^{1+\beta}}{\varphi}}^{1/2}
\]

This ratio is larger than unity in the parameter space, 
\[
    \left\{
    (x, \varphi)~|~
    x < \frac{\varepsilon ^{-1}}{2}
    ~\cup~
    \left(
    \varphi < \frac{\varepsilon  q x^{1+\beta}}{(2-x^{-1}\varepsilon ^{-1})^2}
    ~\cap~
    x > \frac{\varepsilon ^{-1}}{2}
    \right)
    \right\}
\]
and in this case, the rapid ionization condition in the gravity-inhibited regime, $t_\mathrm{ioni} < \mathrm{min}\left[t_\mathrm{rec} (\cs  = \vg ), R/\mach_\mathrm{g}^\mathrm{II} \vg  \right]$, is
$t_\mathrm{ioni} < t_\mathrm{rec}$, i.e.,
$\varphi > \varepsilon  q x^{\beta + 1}$.
This is always satisfied in the Type~II gravity-inhibited regime. 
On the other hand, the above ratio is less than unity in
\[
    \left\{
    (x, \varphi)~|~
    x > \frac{\varepsilon ^{-1}}{2}, 
    ~
    \varphi > \frac{\varepsilon  q x^{1+\beta}}{(2-x^{-1}\varepsilon ^{-1})^2},
    \right\}
\]
and in this regime, the rapid ionization condition becomes 
\[
    \frac{t_\mathrm{ioni}}{t_\mathrm{rec}} < x^{-1}\varepsilon ^{-1} - \braket{1-\frac{t_\mathrm{ioni}}{t_\mathrm{rec}}}, 
\]
or equivalently $x < \varepsilon ^{-1}$. 
After all, the rapid ionization condition in the gravity-inhibited regime, $t_\mathrm{ioni} < \mathrm{min}\left[t_\mathrm{rec} (\cs  = \vg ), R/\mach_\mathrm{g}^\mathrm{II} \vg  \right]$, reduces to $x < \varepsilon ^{-1}$.


The other condition in Condition~II-2, $c_\mathrm{ch}^\mathrm{II} < \vg $, is equivalent with $\varphi < \varphi_\mathrm{g}$ (see \eqnref{eq:varphi_g}). 
To summarize Condition~II-2, 
the gravity-inhibited regime for Type~II winds is defined by
\[
    \left\{
    (x, \varphi)~|~
    x < \varepsilon ^{-1},
    ~
    \varphi > \varepsilon  q x^{\beta + 1},
    ~
    \varphi < \varphi_\mathrm{g}
    \right\}.
\]
In this regime, winds have a sound speed of $\sim \vg $ and a Mach number of 
\begin{equation}
    \mach_\mathrm{g}^\mathrm{II}
    \approx 
        \braket{\frac{\varphi q x^\beta}{\varepsilon  }}^{1/2}
        \left[
        x^{-1} \varepsilon ^{-1} - 1 
        + \braket{\frac{\varepsilon  q  x ^{\beta + 1} }{\varphi}}^{1/2} 
        \right]^{-1}
        \label{eq:mach_g:typeII}
\end{equation}
Mach number increases with $\varphi$ at individual distance owing to increasing heating with $\varphi$ and approaches unity at the $c_\mathrm{ch}^\mathrm{II} = \vg $ boundary (see \eqnref{eq:varphi_g_def}).

Finally, we evaluate Condition~II-3. 
This condition clearly corresponds to the remaining parameter space that is not covered by Conditions~II-1 and II-2, that is, 
\[
\left\{
    (x, \varphi)~|~
    \varphi < f_\mathrm{eq} x,
    ~
    \varphi > \varphi_\mathrm{g},
    ~
    \varphi > \varepsilon ^{3/2} x
    \right\}.
\]
In this space, the heating timescale is shorter than the gravitational timescale for Type~II winds but is not short enough to heat the gas to the equilibrium temperature until traveling to the height of $\sim R$. 
Therefore, $\cs  \sim c_\mathrm{ch}^\mathrm{II}$ and $\mach \sim 1$ in this regime.

The sound speed in this regime is obtained by solving the implicit equation Eq.\eqref{eq:chII_original_def:dimensionless} with respect to $c_\mathrm{ch}^\mathrm{II}/\ceq $.  
The expression of the root would have a complex functional form with arguments of $\varepsilon $, $\varphi \varepsilon  q/x $, and $\varphi/\varepsilon  x$, which might be unpractical to be presented. 
Instead, it would be convenient to show approximate forms of the root here. 
In the Sa-class, the parameter space where $\cs  = c_\mathrm{ch}^\mathrm{II}$ always contains the slow recombination regime \eqnref{eq:c2_in_slow_recombination_valid_paramspace}. 
This is because in this class, the second term on the RHS of \eqnref{eq:chII_original_def:dimensionless} is exactly zero at the Type~I and Type~II boundary ($\varphi = \varepsilon ^{3/2}x$), and \eqnref{eq:c2_in_slow_recombination_valid_paramspace} always exists within the Type~II regime. 
The approximated $c_\mathrm{ch}^\mathrm{II}$ is 
\[
    c_\mathrm{ch}^\mathrm{II} \approx c_1 = \varepsilon ^{1/2} \ceq .
\]
This approximation holds in the parameter space of 
\[
    \left\{
    (x, \varphi)~|~
    x > \varepsilon ^{-1},
    ~
    \varphi \ll \varepsilon ^{\beta + 2} q^{-1} x,
    ~
    \varphi > \varepsilon ^{3/2} x
    \right\}.  
\]
Here, the condition $x > \varepsilon ^{-1}$ originates from a requirement for winds to be escaping, $c_\mathrm{ch}^\mathrm{II} > \vg $. 
The corresponding area is illustrated by the orange-shading bounded by the red and blue dashed lines in the top right panel in \fref{fig:classmaps}, as an example.
Having this area where the additionally deposited energy through reionizations is negligible is a unique characteristic of the Sa-class. 
Note that in $x < \varepsilon ^{-1}$, gravity is so strong that the additional energy deposition through reionizations is always needed to form steadily escaping winds.

In this class, the fast recombination regime (\eqnref{eq:rapid_recombination_condition_paramspace}) is not necessarily contained in the free wind regime where $\cs  = c_\mathrm{ch}^\mathrm{II}$. 
This is physically understood as follows. 
Suppose the case where $\varepsilon $ has a value slightly smaller than unity, 
$c_\mathrm{ch}^\mathrm{II} $ needs only a small contribution from the additionally supplied energy through reionizations (the second term on the RHS of \eqref{eq:chII_original_def:dimensionless}) to reach $c_\mathrm{ch}^\mathrm{II} > \ceq $. 
Hence, in this class, it is possible that the fast recombination regime is not present in Regime~B-II (the free Type~II winds), and the slow recombination regime (\eqnref{eq:c2_in_slow_recombination_valid_paramspace}) directly connects to the boundary between Regime~A (isothermal winds) and Regime~B-II (free winds), at which $c_\mathrm{ch}^\mathrm{II} = \ceq $.

The resulting parameter space for the Sa-class is summarized by the top right panel of \fref{fig:classmaps}.

\subsubsection{Type~II winds in Sr-class}

The Type~II regime of the Sr-class is expressed by 
\begin{equation}
    \left\{ 
    \begin{array}{ll}
         \varphi >  \varepsilon   q  x ^{1+\beta}& \quad\mathrm{for}~ 1 \leq   x  < (\varepsilon  q)^{-2/(3+2\beta)} \\
         \varphi >  \braket{\varepsilon  q}^{3/(3+2\beta)}  x  & \quad\mathrm{for}~  x  > (\varepsilon  q)^{-2/(3+2\beta)} 
    \end{array}
    \right.  
    \label{eq:typeII_regime_Srclass}
\end{equation}
(cf. Eq.\eqref{eq:typeI_regime_Srclass}). 
The fast ionization condition for Condition~II-1 is $\varphi > \mathrm{max}(1, q)\varepsilon  x$ in this class. 
Since $f_\mathrm{eq} > 1$ for the soft-spectrum classes,
the Type~II regime meets the fast-heating condition, $\varphi > f_\mathrm{eq} x$. 
Therefore, Condition~II-1 is equivalent with $\varphi > f_\mathrm{eq} x$ for the same reason as in the Sa-class. 

Regarding Condition~II-2, we do a similar analysis to that we have done for Type~II winds in the Sa-class.
The condition $c_\mathrm{ch}^\mathrm{II} < \vg $ is again $\varphi < \varphi_\mathrm{g}$ (\eqnref{eq:varphi_g}),
and $\varphi_\mathrm{g}$ connects to the intersection between $\varphi = \varepsilon  q x^{1+\beta} $ and $\varphi  = (\varepsilon  q)^{3/(3+2\beta)} x$ at $x = (\varepsilon  q)^{-2/(3+2\beta)}$ (see \eqnref{eq:typeII_regime_Srclass}). 
Therefore, the gravity-inhibited regime for Type~II winds is defined by
\[
    \left\{
    (x, \varphi)~|~
    x < (\varepsilon  q)^{-2/(3+2\beta)},
    ~
    \varphi < \varphi_\mathrm{g}, 
    ~
    \varphi > \varepsilon   q x^{1+\beta}
    \right\}
    .
\]
In this regime, $\cs  \sim \vg $ and $\mach = \mach_\mathrm{g}$ (\eqnref{eq:mach_g:typeII}). 

The other parameter space corresponds to the regime of free Type~II winds, and Condition~II-3 reduces to 
\[
    \left\{
    (x, \varphi)~|~
    \varphi > (\varepsilon  q)^{3/(3+2\beta)} x,
    ~
    \varphi > \varphi_\mathrm{g}, 
    ~
    \varphi < f_\mathrm{eq} x
    \right\}
    .
\]
The typical sound speed and Mach number are $\cs  \sim c_\mathrm{ch}^\mathrm{II}$ and $\mach \sim 1$, respectively. 

We can derive an approximate form of $c_\mathrm{ch}^\mathrm{II}$ for this class instead of deriving the exact root from the implicit equation \eqref{eq:chII_original_def:dimensionless}. 
In contrast to the Sa-class, the Sr-class does not cover the slow recombination regime, as \eqnref{eq:c2_in_slow_recombination_valid_paramspace} is outside of the Type~II regime (\eqnref{eq:typeII_regime_Srclass}). 
This means additional deposited energy always contributes to the heating for Type~II winds in this class.
Hence, $c_\mathrm{ch}^\mathrm{II}$ can be approximated to  Eq.\eqref{eq:cchII_def}, 
\[
    c_\mathrm{ch}^\mathrm{II} \sim c_2  = \ceq \braket{\frac{\varphi_\mathrm{II}}{x}}  ^{1/2(\beta+3)}
\]
in the parameter space, 
\[
    \left\{( x , \varphi)~|~ 
        \varphi_\mathrm{II} <  x , 
        ~
        \varphi > (\varepsilon  q)^{3/(3+2\beta)}  x , 
        ~
        \varphi_\mathrm{II} >  x ^{-(\beta+2)} 
        \right\} 
\]
This parameter space is represented by the area bounded by the gray solid and green dashed lines in \fref{fig:classmaps}. 
One can observe that the gray line gives approximate boundaries with the gravity-inhibited and isothermal wind regimes. 
For reference, we also plot this boundary in the Sa-class map of \fref{fig:classmaps}. 
The gray line does not necessarily give approximate boundaries with the gravity-inhibited and isothermal wind regimes, since neglecting the contribution of the initially injected energy to heating is not always justified in the Sa-class.

\section{Comparison with Hydrodynamics Simulations}
\label{sec:comparison_HD_simulations}

To assess the predictability of our phenomenological model and the validity of the adopted assumptions, we briefly compare the analytic model with hydrodynamics simulations, where ray-tracing EUV radiative transfer and nonequilibrium thermochemistry are self-consistently treated. 
We stress that self-consistent coupling between hydrodynamics and thermochemistry is required to make meaningful comparisons. 
We provide the computational methods of the simulations in \secref{sec:sub:computational_methods} and compare the resulting disk structures with the analytical predictions of the model in \secref{sec:sub:comparison_with_simulations}. 

\subsection{Computational Methods}
\label{sec:sub:computational_methods}
For the methodology, we largely follow \citet{2018_Nakatani, 2018_Nakatanib}. 
Here, we provide a brief overview of the methods employed, referring interested readers to the original papers for more comprehensive details on the computational techniques. 

In our simulations, we start with an initially hydrostatic disk (in the poloidal directions) that is exposed to EUV radiation from a central source. 
The disk is assumed to be both axisymmetric and midplane symmetric, with an initially isothermal temperature profile in the vertical direction.  

The initial density distribution of the hydrostatic disk is 
\[
    \nh = n_{\rm m}(R) 
        \exp \left[ -\frac{z^2}{2h^2} \right] , \label{eq:inidenstr}
\]
where $n_{\rm m} (R)$ represents the number density at the midplane ($z= 0$),
and $h$ denotes the pressure scale height defined as the ratio of $\cs$ to the orbital frequency $\Omega \equiv \sqrt{GM_*/R^3}$. 

The midplane density is determined by specifying the initial surface density profile $\Sigma$, given by
\begin{equation}
    \Sigma = \Sigma_0 \braket{\frac{R}{R_0}}^{-1} 
    \label{eq:sigma}	
\end{equation}
with $R_0$ and $\Sigma_0$ being a reference radius and surface density at $R_0$, respectively.
The product of $\Sigma_0$ and $R_0$ is determined by the total disk mass
\begin{equation}
    M_{\rm disk} = \int_{R_{\rm min} }^{R_{\rm max}} 2\pi R \Sigma \, \dd R
		= 2\pi  (R_{\rm max} - R_{\rm min}) R_0\Sigma_0, 
		\label{eq:diskmass}
\end{equation}
where $R_{\rm min}$ and $R_{\rm max}$ are the inner and outer truncated radii, respectively, set to the same values as the inner and outer computational boundaries, which will be described below.

The initial temperature profile is given by 
\[
    T = 100 \Kelvin \braket{\frac{R}{1\au}}^{-1/2}. 
\]
This profile is selected based on our previous simulations. 
While the choice of this profile may alter the base's geometry, it does not notably influence the comparisons conducted in this section.
The initial value is immediately adjusted to account for the local thermochemical processes within the wind region and remains nearly constant in the stable disk region. 

The simulations are performed in 2D spherical polar coordinates $(r, \theta)$.
We solve the time evolution of gas density $\rho$,
velocities $\vec{v} = (v_r,~ v_\theta, ~ v_\phi)$,
total gas energy density,
and chemical abundances, using a modified version of PLUTO \citep{2007_Mignone}:  
\begin{gather}
    \frac{\partial \rho}{\partial t} + \nabla \cdot \rho \vec{v}                                             =  0 ,                  \\
    \frac{\partial \rho v_r}{\partial t} + \nabla \cdot \left( \rho v_r \vec{v} \right)              =  -\frac{\partial p}{\partial r}
                    -\rho \frac{GM_*}{r^2} + \rho \frac{v_\theta^2 + v_\phi^2}{r}                   ,                       \\
    \frac{\partial \rho v_\theta}{\partial t} + \nabla \cdot \left( \rho v_\theta \vec{v} \right)    = - \frac{1}{r}\frac{\partial p}{\partial \theta }
                    - \rho \frac{v_\theta v_r}{r} + \frac{\rho v_\phi^2}{r} \cot \theta                     ,                       \\
    \frac{\partial \rho v_\phi}{\partial t} + \nabla ^l \cdot \left( \rho v_\phi \vec{v} \right)     = 0     ,       \label{eq:euler_phi}            \\              
    \frac{\partial E}{\partial t} + \nabla \cdot \left(H \vec{v} \right)                                     = - \rho v_r \frac{ GM_* }{r^2} +\rho \left( \Gamma_\mathrm{EUV} -\Lambda    \right),                        \\
    \frac{\partial \nh y_\mathrm{HI} }{\partial t} + \nabla \cdot \left( \nh y_\mathrm{HI} \vec{v} \right)               = - \nh y_\mathrm{HI} k_\mathrm{ioni} + \alpha\nh^2 y_\mathrm{e} y_\mathrm{HII}       .                       
    \\
    \frac{\partial \nh y_\mathrm{HII} }{\partial t} + \nabla \cdot \left( \nh y_\mathrm{HII} \vec{v} \right)               =  \nh y_\mathrm{HI} k_\mathrm{ioni} - \alpha\nh^2 y_\mathrm{e} y_\mathrm{HII}   
    \\
    \text{and} \quad
    \frac{\partial \nh y_\mathrm{e} }{\partial t} + \nabla \cdot \left( \nh y_\mathrm{e} \vec{v} \right)               = \nh y_\mathrm{HI} k_\mathrm{ioni} - \alpha\nh^2 y_\mathrm{e} y_\mathrm{HII}          .                       
\end{gather}
\eqnref{eq:euler_phi} is written in the angular momentum conserving form.  
Here $p$ denotes the gas pressure; 
and $y_\mathrm{HII}$ and $y_\mathrm{e}$ are the chemical abundances of \ce{H+} and \ce{e-}, respectively;
and $\Lambda$ is the total cooling rate per unit mass (specific cooling rate). 
We incorporate radiative recombination cooling and Ly$\alpha$ cooling in the present study. 
The total gas energy density and enthalpy density are defined as 
\[
    \begin{gathered}
    E =  \frac{1}{2} \rho v^2 + \frac{p}{\gamma-1}\\
    H =  E + p, 
    \end{gathered}
\]
respectively, with $\gamma$ being the specific heat ratio ($\gamma = 5/3$) and the ideal equation of state. 
We use $M_\mathrm{disk} = 0.01 \Msun$ and $M_* = 1.0 \Msun$ throughout our simulations.

We run four models varying the EUV spectral hardness and $\phieuv$, covering high- and low-$\phieuv$ ranges for both soft and hard spectra.
We employ $\delta$-function spectra of the form $\delta(h\nu - h\nu_0)$, where $h\nu_0$ is the EUV photon energy. 
In the soft-spectrum runs, $h\nu_0 = 14.6\eV$, while in the hard-spectrum runs, $h\nu_0 = 72.9\eV$. 
For the soft-spectrum runs, $h\nu_0 = 14.6\eV$, which corresponds to the average energy of a blackbody spectrum with $T_\mathrm{eff} = 10^4\Kelvin$ above the Lyman limit. 
For the hard-spectrum runs, we chose $h\nu_0 = 72.9\eV$ arbitrarily for experimental purposes, corresponding to the average energy of a blackbody spectrum with $T_\mathrm{eff} = 3\times 10^5\Kelvin$.
We explore two orders of magnitude in the dimension of the EUV luminosity: $L_\mathrm{EUV} = 10^{28}\erg \sec^{-1}$ and $10^{30}\erg \sec^{-1}$ for the low- and high-luminosity cases, respectively.
The corresponding $\phieuv$ is determined by $L_\mathrm{EUV} = h\nu_0 \phieuv$. 
A summary of the input EUV parameters is provided in \tref{tab:numerical_models}.
The Model labels reflect both the magnitude of $L_\mathrm{EUV}$ (high: H, low:L) and spectral hardness (soft: S, hard: H).
\begin{table}[]
    \centering
    \begin{tabular}{l c c c}
    Model &  Emission Rate $\Phi_{40}$ & Photon Energy (eV) &  Luminosity $L_{30}$ \\ 
    \hline
    HS   &   4.3 & 14.6 & 1.0 \\
    HH   &   0.86 & 72.9 & 1.0 \\
    LS     &  $4.3\e{-2}$ & 14.6 & 0.01\\ 
    LH     &  $0.86\e{-2}$ & 72.9 & 0.01 \\
    \end{tabular}
    \caption{EUV parameters for each run. Here, $\Phi_{40} \equiv \phieuv/10^{40}\sec^{-1}$ and $L_{30} \equiv L_\mathrm{EUV}/10^{30}\erg\sec^{-1}$. In the model labels, the first letter indicates either high (H) or low (L) luminosity, while the second letter signifies either soft (S) or hard (H) spectrum. }
    \label{tab:numerical_models}
\end{table}
The EUV emission rate $\phieuv$ of Model~HS is the closest to the supposedly typical value for pre-main-sequence low-mass stars \citep[$\phieuv \approx 10^{40}$--$10^{41}\sec$; e.g.,][]{2014_Alexander}, a value commonly adopted in the literature.
We have also performed runs with luminosities ten times higher than Models~HS and HH. However, these models produced qualitatively similar structures to those in Models~HS and HH, as we will discuss in \secref{sec:sub:comparison_with_simulations}.

Our computational domain spans $r = [0.1, 20]\times R_{\rm g}( = [R_\mathrm{min}, R_\mathrm{max}])$ and $\theta = [0, ~\pi/2] {\rm \, rad}$ with a resolution of $N_r \times N_\theta = 256 \times 160$ cells. 
The grid is spaced logarithmically in the radial direction, whereas in the polar direction, two different resolutions are employed. The domain is divided into two regions at $\theta = 1\rad$, with each region having 80 cells uniformly spaced.
We have arrived at this grid resolution after conducting a convergence check for Models~HS and LH. Changing $N_r \times N_\theta$ to $128 \times 160$ and $512\times 320$, respectively, have yielded overall similar structures.

\subsection{Comparisons with the Phenomenological Model}
\label{sec:sub:comparison_with_simulations}
\begin{figure*}[htbp]
    \centering
    \includegraphics[clip, width = \linewidth]{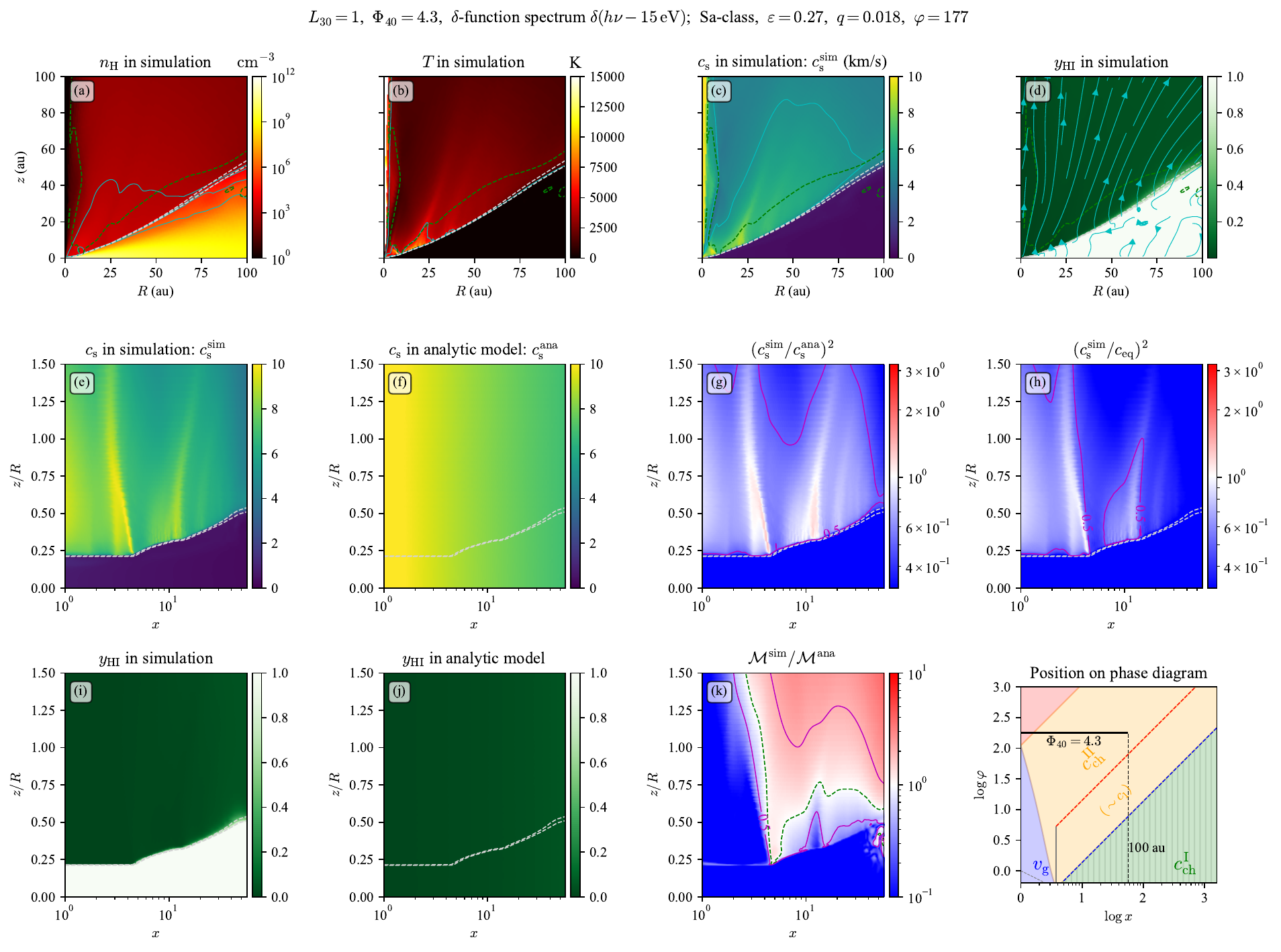}
    \caption{
    Snapshots of Model~HS, showcasing various physical quantities:
    (a--d) number density, gas temperature, isothermal sound speed, and \ion{H}{1} abundance from left to right, respectively. 
    The green dashed contour represents the isothermal sonic surface, 
    while the light-gray dashed contours indicate where EUV flux is absorption-attenuated by half and 90\% from the original value, i.e., $\chi_\mathrm{e} =\chi_\mathrm{i} = 0.5$ and 0.1, respectively.
    These contours roughly demarcate the wind region from the steady disk region. 
    In panel~(a), cyan contours represents $\nh = 10^3, 10^4, 10^5, 10^6 \cm^{-3}$; in panel~(b), $T = 1000, 5000\Kelvin$; and in panel~(c), $\cs = 5\kms$. 
    Streamlines are depicted in panel~(d). 
    Panels~(e) and (i) offer the same snapshots for $\cs$ and $y_\mathrm{HI}$, respectively, but in the $x$--$z/R$ coordinates for comparison with the analytically estimated values through the phenomenological model, presented in panels~(f) and (j). 
    Panel~(g) illustrates the difference between $\cs$ in the simulation and that predicted by the model, with the magenta contours representing where the analytic estimate falls within a factor of two from $\cs$ in the simulation. 
    The whitish region indicates where the analytic model accurately predicts $\cs$ (and thereby $T$) of the simulation. 
    For comparison with the canonical picture, where $\cs = \ceq$ is assumed, panel~(h) similarly displays the difference between $\cs$ in the simulation and $\ceq$. 
    A comparison between panels~(g) and (h) reveals that our phenomenological model improves the analytical predictions of temperature structures. 
    The light-gray dashed contours in panels~(e)--(j) also show where $\chi_\mathrm{e} = 0.1$ and 0.5, consistent with panels~(a)--(d).
    Panel~(k) presents the difference between the Mach number in the simulation to that predicted through the phenomenological model, with the green dashed contour again representing the isothermal sonic surface. 
    Lastly, the phase diagram for the spectrum of Model~HS is shown by the lower-right corner panel, with 
    a black horizontal bar indicates the location of this model on the phase diagram within the range displayed by panels~(a)--(k), i.e., $100\au$. 
    }
    \label{fig:model_H-S}
\end{figure*}
\begin{figure*}[htbp]
    \centering
    \includegraphics[clip, width = \linewidth]{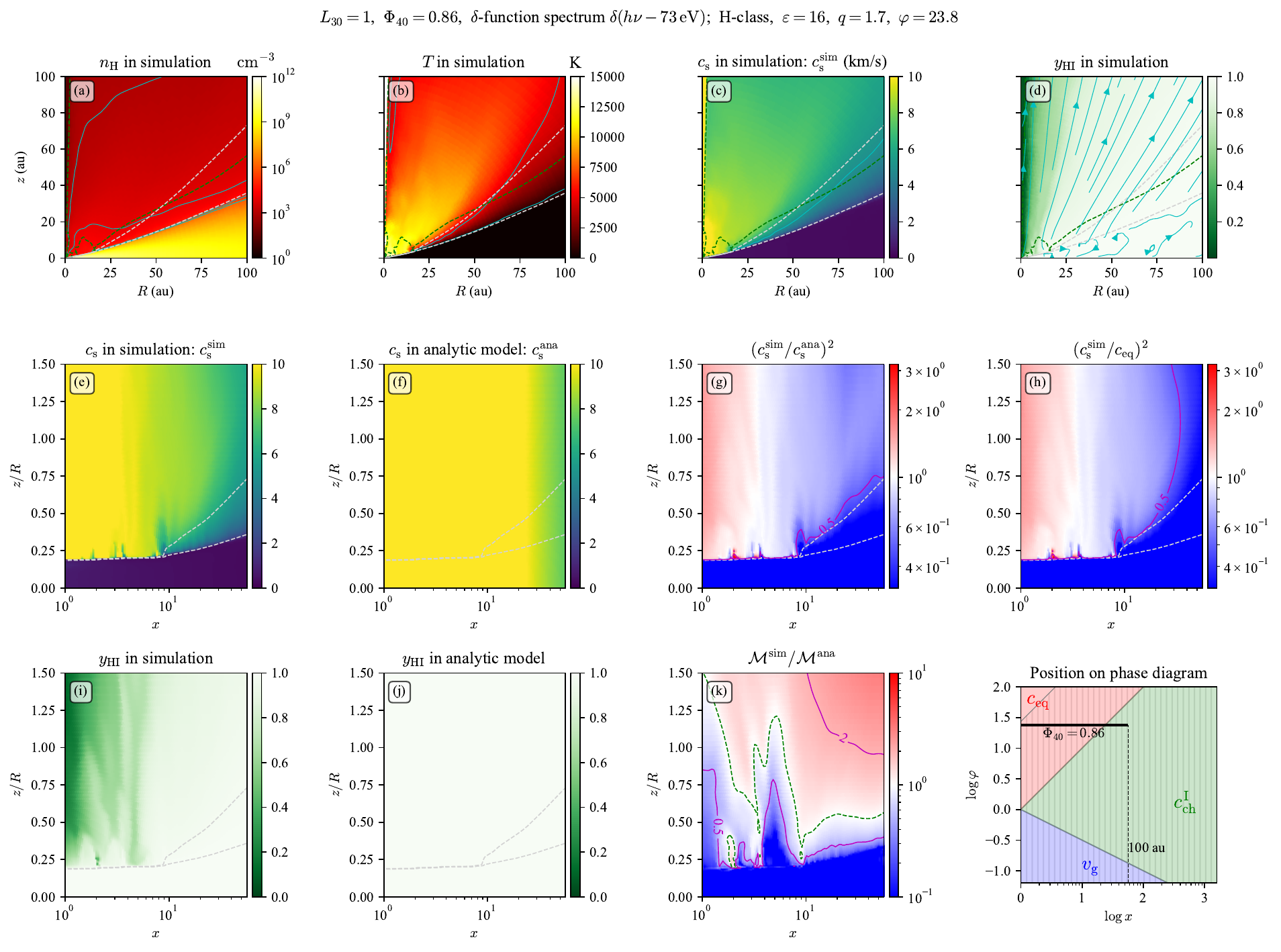}
    \caption{Same as \fref{fig:model_H-S} but for Model~HH. }
    \label{fig:model_H-H}
\end{figure*}
\begin{figure*}[htbp]
    \centering
    \includegraphics[clip, width = \linewidth]{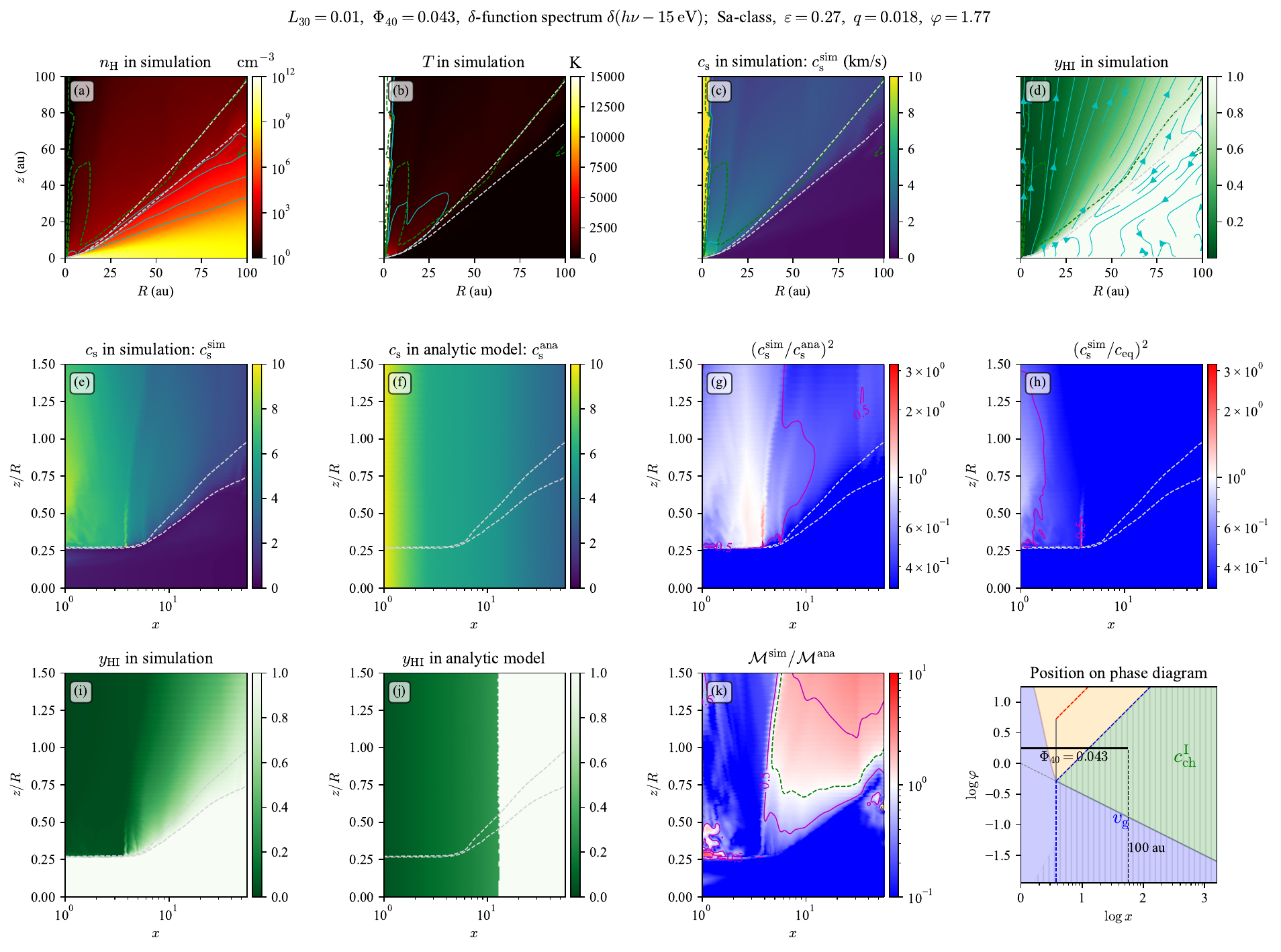}
    \caption{Same as \fref{fig:model_L-S} but for Model~LS.}
    \label{fig:model_L-S}
\end{figure*}
\begin{figure*}[htbp]
    \centering
    \includegraphics[clip, width = \linewidth]{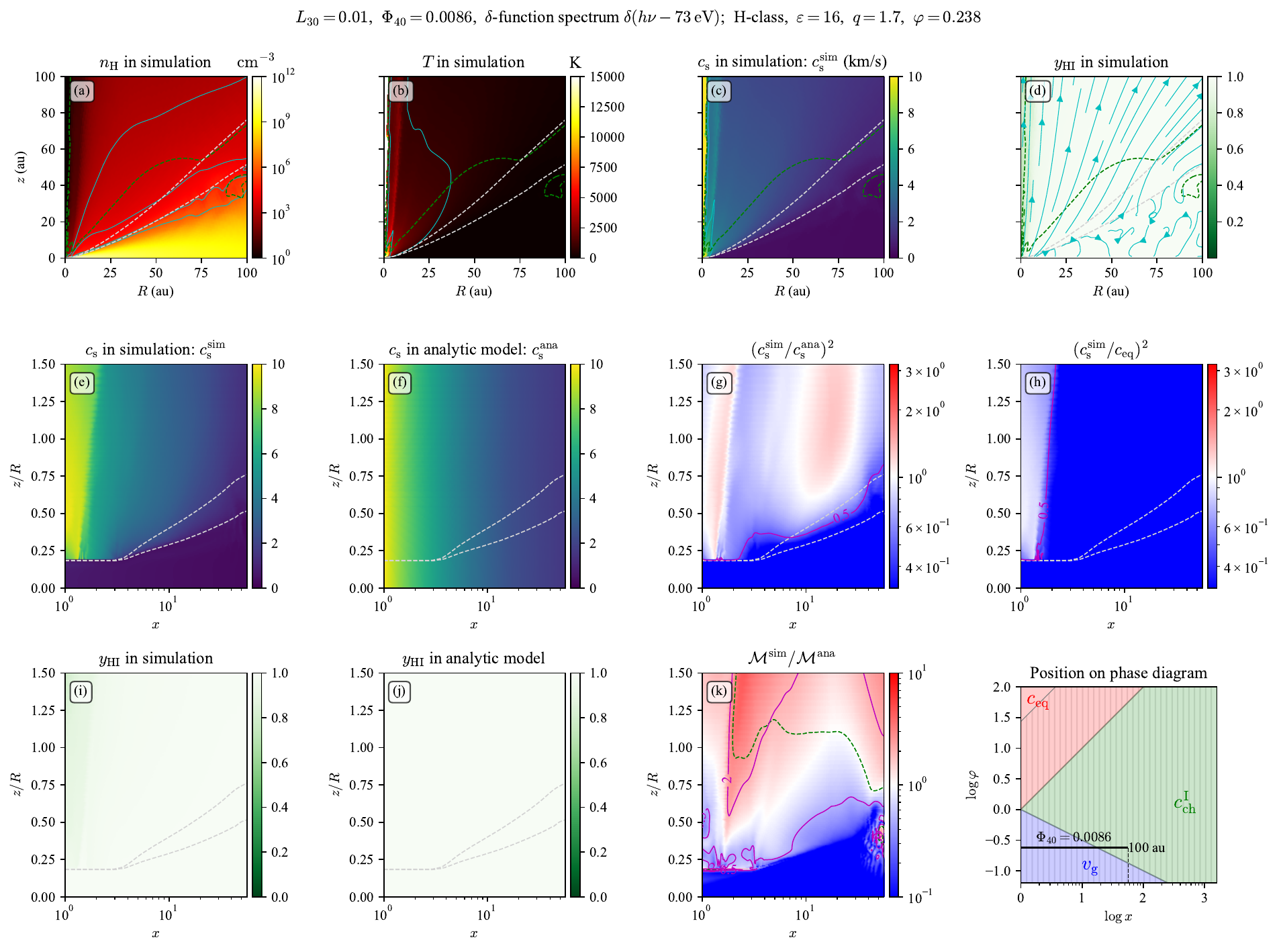}
    \caption{Same as \fref{fig:model_L-H} but for Model~LH.
    {}
    }
    \label{fig:model_L-H}
\end{figure*}
Figures~\ref{fig:model_H-S}, \ref{fig:model_H-H}, \ref{fig:model_L-S}, and \ref{fig:model_L-H} show snapshots of the simulations (panels~a--d) along with comparisons with analytical predictions by our phenomenological model (panels~e--k) for Models~HS, HH, LS, and LH, respectively. 
In these figures, we use the data from the gas structure when it has reached a nearly quasi-steady state.
However, in Model~HS, we employ time-averaged data to smooth out small fluctuations observed in the vicinity of the ionization front. 
In the phase diagram located at the lower-right corner of each figure, a black horizontal bar represents the spatial range displayed by panels~(a)--(k) for each model.

These figures directly illustrate the wide variety in the structures of $\nh$, $T$, and $y_\mathrm{HI}$, depending on $\phieuv$ and the spectral hardness, even when they share a common total input energy $L_\mathrm{EUV}$. 
Temperatures and isothermal sound speeds (panels~b and c) reach the equilibrium values ($T \approx10^4\Kelvin$ and $\cs\approx 10\kms$) only at relatively small $R$, indicating where the canonical picture of EUV photoevaporation applies. 
Temperatures overall decrease at larger distances due to relatively long heating timescales. 

Comparing Models~HS and HH, one can observe that H-class spectra tend to result in a larger volume of the high-temperature region, even though $\varphi$ decreases with spectral hardness. 
This reflects the fact that a larger fraction of the photon energy is consumed for photoionization with soft spectra. To heat the gas to high temperatures, soft spectra require multiple energy depositions through reionizations. 
However, this process is inefficient compared to energy deposition to an atomic gas. 

Panels~(e) and (f) demonstrate that our analytic model effectively captures the radial-dependent trend of $\cs$ (and consequently $T$) over a wide range of distances. 
In fact, the whitish region in panel~(g), where $\cs^2$ in the simulation is within the factor of two difference from $\cs^2$ predicted by the analytic model, covers a much broader region than the whitish region of panel~(h), which shows the ratio of $\cs^2$ in the simulation to the equilibrium value $\ceq^2$. 
This indicates that our model significantly improves the analytical prediction of $\cs$ compared to simply assuming $\cs = \ceq$ as in the canonical picture. 
This improvement holds true for a broad range of $\phieuv$ and the spectral hardness, as observed when comparing panels~(e) and (f) of Figures~\ref{fig:model_H-S}, \ref{fig:model_H-H}, \ref{fig:model_L-S}, and \ref{fig:model_L-H}. 
The enhancement is more noticeable as the wind temperature decreases with lower $L_\mathrm{EUV}$.

Our phenomenological model also adeptly captures the ionization structures. 
In panels~(d), (i), and (j), the greenish region represents the \ion{H}{2} region for each run, with its size varying significantly depending on $\phieuv$ and the spectral hardness (as evident by comparing the four models), as well as the radial distance to the source (as seen in panels~(i) and (j) in \fref{fig:model_L-S}).
As indicated by the analytical model, a harder spectrum tends to result in atomic winds even when $\phieuv$ is at the same levels. 
This is implied by comparing, for instance, Figures~\ref{fig:model_H-H} and \ref{fig:model_L-S}; 
the soft-spectrum run (Model~LS) shows a much more extended \ion{H}{2} region despite having $\sim 20$ times smaller $\phieuv$. 

Panels~(i) and (j) show that our analytic model predicts the radial extension of the \ion{H}{2} region and wind types across a broad range of $\phieuv$ and the spectral hardness with reasonable accuracy. 
Contrary to the canonical model, where the EUV-heated gas is typically assumed to be completely photoionized ($y_\mathrm{HI}\ll 1$) regardless of the distance to the radiation source and $\phieuv$, our numerical and analytic models highlight the necessity of incorporating the finite timescale of photoionization to discuss the ionization structure, especially for low-mass stars.

In panel~(k), we also compare the Mach number in the simulation, $\mach^\mathrm{sim}$, with the prediction from our phenomenological model, $\mach^\mathrm{ana}$. 
Despite some deviations, there is generally a close agreement between the simulation and the analytic estimate, even in the gravity-inhibited subsonic regions (Regime~C) of Model~LH (see \fref{fig:model_L-H}), capturing an overall radial-dependent trend in $\mach$. 
This suggests that our model provides a better estimate for $\mach$ across a wider parameter space, compared to the classical assumption of $\mach\sim 1$ everywhere beyond the critical radius. 

Nonetheless, there is still room for improvement in estimating $\mach$. 
Our model's treatment of a constant $\mach$ fails to fully predict the accelerating nature of photoevaporative winds. 
Additionally, discrepancies arise where the ionization front becomes parallel to the radial direction (see, e.g., the $x\lesssim 5$ region in panel~(i) of \fref{fig:model_H-S}). 
\footnote{The radiation flux needs to have a perpendicular component to the base to launch winds \citep{2022_NakataniTakasao}.}
This parallel feature may be artificial, possibly influenced by the computational boundaries; however, it is also observed across convergence-check runs with higher- ($N_r \times N_\theta = 512\times 320$) and lower-resolutions ($N_r \times N_\theta = 128 \times 160$) for Model~HS.
Should the parallel structure persist consistently across various numerical grid configurations, it would suggest that the structure has physically developed by radiation-hydrodynamical effects, at least within the adopted problem setup. 
If this is the case, the parallel feature implies that diffuse EUV might dominate over direct EUV in heating the \ion{H}{1} layer here, in contrast to the results of hydrostatic radiative transfer models \citep[e.g.,][]{2013_Tanaka}.
Further comparisons with carefully handled simulations are needed to examine numerical effects and explore the analytical predictability of $\mach$ in such regions, which we plan to address in future studies. 


We have also conducted simulations with luminosities ten times higher ($L_\mathrm{EUV} = 10^{31}\erg\sec^{-1}$) than Models~HS and HH (corresponding $\phieuv$ of $4.3\e{41}\sec^{-1}$ and $0.86\e{41}\sec^{-1}$, respectively). 
The resulting structures are qualitatively the same as those of Models~HS and HH other than forming larger isothermal regions (Regime~A winds). 
This outcome is expected based on the phase diagrams of Figures~\ref{fig:model_H-S} and \ref{fig:model_H-H} (see the lower-bottom panels). 
The disks of Models~HS and HH extend across Regimes~A and B, and increasing the luminosities does not alter this characteristic.

The mass-loss rates (within $R < 100\au$) obtained from the simulations exhibit a wide variation: $2\e{-10}\Msun\yr^{-1}$ for Model~HS, $2\e{-9}\Msun\yr^{-1}$ for Model~HH, $1\e{-11}\Msun\yr$ for Model~LS, and $2\e{-10}\Msun\yr^{-1}$ for Model~LH. 
These rates can vary significantly even when the same energy input rate $L_\mathrm{EUV}$ is applied. 
In general, models with soft spectra yield mass-loss rates approximately an order of magnitude smaller than those with hard spectra. 
This reflects the greater energy consumption by photoionization with soft spectra. 
While our current model is not specifically tailored for predicting mass-loss rates, it can be extended to cover this aspect. 
To that end, determining the appropriate base density for a wide range of $\phieuv$ and spectral hardness is essential, a task we aim to tackle in future studies.

In summary, our comparisons here highlight a significant enhancement in analytical predictions for the physical structures across a wide range of $\phieuv$ and spectral hardness.
This signifies a broadened understanding of EUV-driven photoevaporation for accretion disks through our phenomenological model.
However, for robustness, similar comparisons over an even broader range of $L_\mathrm{EUV}$, spectral hardness, and spectral shape are warranted, considering uncertainties in recombination timescales and assumptions regarding attenuation factors ($\chi_\mathrm{e}$ and $\chi_\mathrm{i}$) in our model (see discussions in \secref{sec:discussion:model_limitation}). 
These analyses would facilitate further model improvements and enhance its practical applications to general EUV spectra. 
While such more comprehensive comparisons are deferred to future studies, the brief comparisons provided here suffice to illustrate the remarkable generality of our model.


\section{On the Radial Extension of the H~II region }
\label{sec:limiting_radius}

Our model has shown that the canonical picture of EUV photoevaporation, where EUV photoheating drives isothermal ionized winds with temperature and velocity of $\sim 10^4\Kelvin$ and $\sim 10\kms$, respectively, holds only in Regime~A-II. 
The corresponding parameter space is formulated as $\varphi > \mathrm{max}(1,q)\varepsilon x$ for the hard-spectrum class and $\varphi > f_\mathrm{eq} x$ for the soft-spectrum classes. 
For soft-spectrum classes, ionized escaping winds are also feasible with lower temperatures (Regime~B-II). 
These findings suggest the presence of a maximum radial distance up to which launched winds can be photoionized before reaching heights of $\sim R$. 
Beyond this distance, the photoionization timescale is longer than the recombination timescale or advection timescale.
Consequently, hydrogen atoms remain only partially photoionized until they ascend to heights of $\sim R$ (cf. panels (d), (i), and (j) in \fref{fig:model_L-S}). 
On the scale beyond the maximum distance, the \ion{H}{2} region appears vertically extended rather than radially.
In this section, we quantify this distance limit in physical units and briefly discuss under which conditions the canonical picture of isothermal ionized winds remains valid.

We first consider the H-class. 
The boundary between Type~I and Type~II regimes is exclusively present within Regime~A, meaning that the canonical picture is always applicable to ionized winds in this class. 
The distance limit to which the \ion{H}{2} region extends is calculated from the Type~II condition (\eqnref{eq:typeII_condition_H_class}) as 
\[
    \begin{split}
    R_\mathrm{lim}    & =   R_\mathrm{c} \frac{\varphi}{\mathrm{max}(1,q)\varepsilon}
    =   \frac{\phieuv}{\mathrm{max}(1,q)}   \frac{\avesigma }{4\pi \ceq }
    \\ 
    &   \approx 53\au 
    \braket{\frac{\phieuv}{10^{40}\sec^{-1}}}
    \braket{\frac{\avesigma}{10^{-18}\cm^2}}
    \\
    &   \quad   \braket{\frac{\ceq}{10\kms}}^{-1} \times  \mathrm{min}(1,q^{-1}) .
    \end{split}
\]
The nominal values for the parameters are chosen to be typical for T~Tauri stars. 
This suggests that the canonical picture does not necessarily apply to disks around low-mass stars, but there are also atomic winds at the equilibrium temperature (Regime~A-I) or lower (Regime~B-I) for $R > R_\mathrm{lim}$. 
For massive stars, whose $\phieuv$ is several orders of magnitude higher than $10^{40}\sec^{-1}$, assuming the canonical picture of an EUV-irradiated disk is justified.
Since $R_\mathrm{lim}$ is essentially set by the ionization timescale, it is independent of the stellar gravity and is essentially set by $\phieuv$ and the spectral hardness.

The distance limit for the Sa-class is derived similarly as 
\[
    \begin{split}
    R_\mathrm{lim}    & =   R_\mathrm{c} \frac{\varphi}{f_\mathrm{eq}}
    =   \frac{\braket{\sqrt{(\varepsilon ^{-1}-1)^2 + 4q} - (\varepsilon ^{-1}-1)}^2}{4 q}   \frac{\phieuv  \avesigma }{4\pi \ceq }
    \end{split}
\]
When $(\varepsilon^{-1} -1 )^2 \gg 4q$, which is often the case for this class, $R_\mathrm{lim}$ is approximated to 
\[
    R_\mathrm{lim} \approx \frac{q}{(\varepsilon^{-1}-1)^2}\frac{\phieuv  \avesigma }{4\pi \ceq }
    = \frac{1}{(\varepsilon^{-1}-1)^2}\frac{3C^2 \phieuv  \alpha_\mathrm{eq} }{4\pi \ceq^2 } 
\]
In contrast to the H-class, this class has a regime where freely escaping ionized winds have temperatures lower than the equilibrium value (Regime~B-II). 
This motivates us to define the distance limit for the extension of lower-temperature ionized escaping winds (cf. Eqs.\eqref{eq:Sa-class:TypeIIcondition}, \eqref{eq:typeI_regime_Saclass}, or \eqref{eq:typeII_regime_Saclass}),
\[
    R_\mathrm{lim}^\mathrm{B}     =   R_\mathrm{c} \frac{\varphi}{\varepsilon^{3/2}}
    =   \frac{1}{\varepsilon^{1/2}}   \frac{\phieuv \avesigma }{4\pi \ceq }. 
\]
Note that these distance limits are available when $\phieuv$ exceeds the secondary critical EUV emission rate, i.e., $\varphi > f_\mathrm{eq}$. 
Hence, for Sa-class spectra, the canonical picture of EUV photoevaporation applies up to $R = R_\mathrm{lim}$, and beyond that, ionized nonisothermal winds extend up to $R\leq R_\mathrm{lim}^\mathrm{B}$ (cf. panels~(d), (i), and (j) in \fref{fig:model_L-S}). 


The radius limit of isothermal ionized winds for the Sr-class is the same as that for the Sa-class.
The secondary radius limit is derived as 
\[
    R_\mathrm{lim}^\mathrm{B}     =   R_\mathrm{c} \frac{\varphi}{(\varepsilon q)^{3/(3+2\beta)}}
    =   \frac{\varepsilon}{(\varepsilon q)^{3/(3+2\beta)}}   \frac{\phieuv\avesigma }{4\pi \ceq }.
\]
However, Sr-class might not be achieved with realistic EUV spectra, as mentioned in \secref{sec:sub:Sr-class}. 

The presence of the two limiting distances for the soft-spectrum classes suggests the potential to constrain $\avesigma$ and $\aveenergy$ by spatially resolving the radial extensions of the isothermal (Regime~A-II) and nonisothermal (Regime~B-II) ionized winds:
\[
    \frac{\varepsilon^{1/2} q}{(\varepsilon^{-1}-1)^2} \approx \frac{R_\mathrm{lim, obs}}{R_\mathrm{lim, obs}^\mathrm{B}}. 
\]
This comparison allows for the potential constraint of $\phieuv$ using the constrained spectrum parameters and observed radial extensions. 
Analyzing ionized wind tracers with different excitation temperatures could facilitate such analysis. 
We further discuss this point in \secref{sec:discussions:wind_observations}.

\section{Work Efficiency}
\label{sec:work_efficiency}

In photoevaporating systems, photoheating is responsible for supplying energy to increase the mechanical and thermal energies of the gas while partly being lost through cooling. 
The efficiency of this energy conversion process largely depends on the irradiating energy flux. 
By employing our model, we can make an order-of-magnitude estimation for this work efficiency and observe its variation with $\phieuv$ by comparing the magnitudes of the photoheating with the work required to liberate the gas from gravitational binding. 
This efficiency provides a physical reference for understanding the basic energy flow operating within photoevaporative winds in different regimes.

We define work efficiency $\eta$ as the ratio between the integration of the gravitational power per gas mass, 
\[
    \mathscr{P}_\mathrm{g} \equiv \frac{GM_*}{2r^2} \mach \cs 
\]
and the integrated specific heating rate $\Gamma_\mathrm{EUV}$ (Eq.\eqref{eq:gamma_EUV:2}): 
\begin{equation}
    \begin{split}
    \eta & \equiv  \braket{\int \frac{GM_*}{2r^2}\mach \cs \frac{\dd s}{\mach \cs} } \braket{\int \Gamma_\mathrm{EUV} \frac{\dd s}{\mach \cs}}^{-1} . 
\end{split}
\label{eq:eta:original}
\end{equation}
Again, the integration is taken along a streamline to the height of $\sim R$. 
The factor of $1/2$ in $\mathscr{P}_\mathrm{g}$ effectively accounts for the contribution of centrifugal force to the net gravity.

By the definition of the characteristic sound speed for Type~I winds (\eqnref{eq:c_ch_def}), $\eta$ is approximated to 
\begin{equation}
    \eta 
    \approx \eta^\mathrm{I} 
    \equiv \braket{\frac{\vg}{c_\mathrm{ch}^\mathrm{I}}}^2 \frac{\mach\cs }{c_\mathrm{ch}^\mathrm{I}}
    = \varphi^{-1} \mach \frac{\cs}{\ceq} 
\label{eq:eta:TypeI}
\end{equation}
The specific value of $\eta$ differs by the wind regimes (see \tref{tab:sound_spees}) and is summarized in \tref{tab:eta}. 
\begin{table}[htpb]
    \centering
    \begin{tabular}{l l R}
    Regime  &  State of winds & \eta \\
    \hline\hline
    A-I  &   Isothermal, atomic &    \varphi^{-1}\\
    B-I  &   Steadily-heated, free; atomic & \varphi^{-2/3} x^{-1/3}\\
    C-I  &   Gravity-inhibited, atomic&  1\\
    \hline
    A-II &   Isothermal, ionized& \eta_\mathrm{A}^\mathrm{II}\\
    B-II &   Free, ionized& x^{-1}(c_\mathrm{ch}^\mathrm{II}/\ceq)^{-2}\\
    C-II &   Gravity-inhibited, ionized &  1 \\
    \end{tabular}
    \caption{Order-of-magnitude estimation for work efficiency in each wind regime. }
    \label{tab:eta}
\end{table}
The efficiency generally decreases as the EUV emission rate increases. 
Note that for Type~I winds, $\eta$ is set by the local variables, namely $\mathscr{P}_\mathrm{g}$ and $\Gamma_\mathrm{EUV}$, as we assume all the relevant quantities remain constant along the streamline, thus giving $\eta^\mathrm{I} = \mathscr{P}_\mathrm{g}/\Gamma_\mathrm{EUV}$.

For Type~II winds, the work efficiency is approximated to 
\[
    \begin{split}
    \eta
    \approx\eta^\mathrm{II}
    \equiv  x^{-1}
    & 
    \left(
    \varepsilon   \left[1 - \braket{\frac{x \varepsilon   q}{ \varphi }}^{1/2}
    \braket{\frac{\cs}{\ceq }}^{-\beta}\right]
    \right.
    \\  &   \quad
    \left.
    + \braket{\frac{\varphi \varepsilon   q}{ x }}^{1/2}
    \braket{\frac{\cs}{\ceq }}^{-(\beta+1)}
    \mach^{-1}
    \right)^{-1} 
    \end{split}
\]
(cf. \eqnref{eq:chII_original_def:dimensionless:another_form}). 
In Regime~C-II (gravity-inhibited ionized winds), where $\cs\approx\vg$ and $\mach\approx\mach_\mathrm{g}^\mathrm{II}$, the work efficiency is again calculated to be unity, as in Regime~C-I. 
In Regime~B-II, using \eqnref{eq:chII_original_def:dimensionless}, $\eta^\mathrm{II}$ is expressed as
\[
    \eta^\mathrm{II} = x^{-1}\braket{\frac{c_\mathrm{ch}^\mathrm{II}}{\ceq}}^{-2}. 
\]
Given that $c_\mathrm{ch}^\mathrm{II}$ increases with the EUV emission rate, $\eta^\mathrm{II}$ monotonically decreases with increasing $\varphi$. 
For Regime~A-II, substituting $\cs\approx\ceq$ and $\mach\approx 1$ into $\eta^\mathrm{II}$, the work efficiency reduces to 
\[
    \eta^\mathrm{II} = 
    \eta_\mathrm{A}^\mathrm{II}
    \equiv
    x^{-1}
    \left(
    \varepsilon   
    \left[1 - \braket{\frac{x \varepsilon   q}{ \varphi }}^{1/2}
    \right]
    + \braket{\frac{\varphi \varepsilon   q}{ x }}^{1/2}
    \right)^{-1} .
\]
In the high flux limit of $\varphi/x \gg 1$, the work efficiency approximately scales as $\eta^\mathrm{II}\approx x^{-1/2}\varphi_\mathrm{II}^{-1/2}$. 
In comparison  to Type~I winds, the decrease in $\eta$ with respect to $\varphi$ is slower. 
This reflects that in this limit, energy is supplied in terms of reionizations, which is limited by recombination timescale proportional to $\varphi^{-1/2}$ (cf. \eqnref{eq:trec_base_density}).

In summary, the work efficiency generally declines as the EUV emission rate increases, indicating that more photo-energy is directed toward the gas's kinetic energy and is also drained through cooling as the temperature rises sufficiently to activate coolants. 
In the high flux limit of $\varphi/x \gg 1$, where winds attain an isothermal and ionized state (Regime~A-II), the efficiency approximately scales as $\propto \varphi^{-1/2}$ approaching zero. 
This indicates that a significant portion of the energy is lost by cooling, with only a small fraction contributing to the gas's mechanical and thermal energy. 
In the opposite limit, where the gas is gravity-inhibited, the efficiency becomes $\eta \approx 1$, meaning efficient conversion of supplied energy into the gas's mechanical and thermal energy. 
The highest efficiency is achieved when the gas is heated quasi-statically, keeping a practically hydrostatic structure.

We stress that the derived $\eta$ here is an order-of-magnitude estimation of energy conversion efficiency. 
The actual work efficiency defined in the form of \eqnref{eq:eta:original} should not reach unity even in the gravity-inhibited regime, as the deposited energy must partly be consumed to increase the enthalpy of the gas at least. 
Nevertheless, the order of $\eta$ and its $\varphi$-dependent trend would remain consistent in more strict estimations. 
The overall qualitative behaviors of the work efficiency would be applicable to other astrophysical photoevaporating objects in various scales, such as planetary atmosphere \citep{2009_MurrayClay, 2016_Owen}, molecular clouds, blackhole's accretion disk on the Bondi scale, and galactic minihalos.

\section{Discussions}       \label{sec:discussion}

We have constructed a model primarily tailored for EUV photoevaporation.
However, given that the underlying physics are essentially the same, with suitable modifications, the model can in principle be extended to account for other phenomena such as FUV- and X-ray-driven photoevaporation, as well as the photoevaporation of gas composed purely of metal species. 
As an initial demonstration of this versatility, in \secref{sec:discussion:critical_luminosities}, we approximately estimate the critical FUV and X-ray luminosities,
and based on these estimates, we discuss the parameter spaces corresponding to Regimes~A, B, and C of FUV- and X-ray-driven winds.
In \secref{sec:Xray_divergent_conclusion}, We explore how this discussion sheds light on a long-standing issue in the field of protoplanetary disk photoevaporation: the contradictory conclusions on the effectiveness of X-ray photoevaporation. 
We also examine the potential time-dependent variations of EUV photoevaporation classes arising from stellar and disk evolution in \secref{sec:discussion:time_dependent_evolution}. 
Finally, we address the model limitations and caveats of our model in \secref{sec:discussion:model_limitation}.

\subsection{FUV and X-ray Critical Luminosities}
\label{sec:discussion:critical_luminosities}

In protoplanetary disk dispersal, FUV and X-ray can also play important roles in driving photoevaporative winds as well as EUV.
These other channels of photoevaporation also have wind regimes corresponding to Regimes~A, B, and C. 
The parameter spaces could be analytically derived by constructing a similar phenomenological model to the ones in \begelman{} and the present study. 
Suitable modifications are, however, needed in such models, as the heating rates and equilibrium temperatures in FUV and X-ray photoevaporation differ from those in EUV photoevaporation, and the equilibrium temperatures should depend on the distance and, likely, the stellar luminosities. 
Nevertheless, we can roughly discuss the parameter spaces of FUV- and X-ray-driven photoevaporative winds by deriving the critical FUV and X-ray luminosities. 
In this section, we present such discussions, in particular for low-mass stars ($M_*\approx1\Msun$). 
It will show that hydrodynamical effects can significantly influence temperature determination and, thereby, mass-loss rates of FUV and X-ray photoevaporation in low-mass stars' systems.

FUV heats the gas through photoelectric effects on very small dust grains and polycyclic aromatic hydrocarbons (PAHs). 
For the heating rate, we adopt the specific photoelectric heating rate of \citet{1994_BakesTielens}: 
\[
    \Gamma_\mathrm{FUV}
           =   10^{-24} \erg \sec^{-1} \frac{G_{\rm FUV}  }{m} \epsilon_\mathrm{pe} 
\]
where $G_\mathrm{FUV}$ is the stellar FUV flux normalized by the Habing flux \citep[$G_0 \approx 1.6\e{-3}\erg\cm^{-2}\sec^{-1}$][]{1968_Habing}, 
and $\epsilon_\mathrm{pe}$ is photoelectric heating efficiency. 
To make our discussion simple, we assume that the PAH contribution dominates the heating and approximate $\epsilon_\mathrm{eq}$ to a constant with $\epsilon_\mathrm{eq}\approx 0.05$. 
This approximation would suffice for the purpose of this section, although strictly, the photoelectric heating efficiency depends on the ionization and recombination rates \citep[e.g.,][]{1994_BakesTielens}. 

Using $\Gamma_\mathrm{FUV}$, the characteristic sound speed for grain photoelectric heating is derived as 
\[
\begin{split}
    c_{\rm ch} ^\mathrm{FUV}
    & = \braket{\frac{\Gamma_\mathrm{FUV}R}{c_p}}^{1/3} 
    \approx \braket{\frac{10^{-24}\erg \sec^{-1} L_\mathrm{FUV}\epsilon_\mathrm{pe}}{4\pi c_p m G_0 R}}^{1/3} 
    \\
     &   \sim 
            2.7\kms 
            \braket{\frac{\epsilon}{0.05}}^{1/3}
            \braket{\frac{R}{1\au}}^{-1/3}
            \braket{\frac{c_p}{7/2}}^{-1/3}            
    \\  &   \quad\times
            \braket{\frac{m}{1.4m_{\rm H}}}^{-1/3}  
            \braket{\frac{L_{\rm FUV}}{10^{30} \erg \sec^{-1}}}^{1/3}
            .
\end{split}    
\]
where $L_\mathrm{FUV}$ is the FUV luminosity of the source. 
We obtain the critical FUV luminosity from comparisons of $c_\mathrm{ch}^\mathrm{FUV}$ with $v_\mathrm{g}$ as 
\begin{equation}
\begin{split}
    L_\mathrm{F,c}    
    & \equiv  
    \frac{4\pi c_p G_0 m R_\mathrm{c, F} v_\mathrm{c,F}^3}{10^{-24}\erg \sec^{-1} \epsilon_\mathrm{pe}}
    =   
    \frac{2 \pi  G_0 m GM_* c_\mathrm{eq}^\mathrm{F}}{10^{-24}\erg \sec^{-1} \epsilon_\mathrm{pe}}
    \\
    & \approx 
    1.9\e{31} \erg \sec^{-1} 
    \braket{\frac{m}{1.4m_\mathrm{H}}}
    \braket{\frac{M_*}{1\Msun}}
    \\  & \quad   \times 
    \braket{\frac{c_\mathrm{eq}^\mathrm{F}}{3\kms}}
    \braket{\frac{\epsilon_\mathrm{pe}}{0.05}}
\end{split}
\label{eq:critical_FUV_luminosity}
\end{equation}
where $R_\mathrm{c, F}$ is the critical radius for FUV-heated gas, $v_\mathrm{c, F}$ is the gravitational velocity at $R_\mathrm{c, F}$ (\eqnref{eq:vg_ceq}), and 
$c_\mathrm{eq}^\mathrm{F}$ is the isothermal sound speed of the FUV-heated gas at the equilibrium temperature at $R_\mathrm{c, F}$. 
FUV heating mainly works in the atomic and molecular layers of the disk, where strong coolants are abundantly present. 
It leads to a lower equilibrium temperature, $\sim3000$--$8000\Kelvin$ in atomic gas and $\sim 2500\Kelvin$ in fully molecular gas, and correspondingly increasing the critical radius, compared to those of the EUV-heated gas; for low-mass stars, $R_\mathrm{c, F} \sim 3$--$12\au$ \citep{2022_Pascucci}. 
The nominal value of $c_\mathrm{eq}^\mathrm{F}$ in \eqnref{eq:critical_FUV_luminosity} corresponds to molecular gas with an equilibrium temperature of $\sim 2500\Kelvin$. 

The nominal value of $L_\mathrm{F,c}$ is comparable to or slightly higher than typically adopted FUV luminosities for accreting low-mass stars, $L_\mathrm{FUV} \approx 10^{30}$--$10^{31}\erg\sec^{-1}$ \citep[e.g.,][]{2009_Gorti}. 
The fact that $L_\mathrm{FUV} > L_\mathrm{F,c}$ is a necessary condition for an FUV-heated gas to meet the fast-heating condition $c_\mathrm{ch}^\mathrm{FUV} > c_\mathrm{eq}^\mathrm{F}$ (Regime~A),
implies that FUV-irradiated disks of typical low-mass systems are mostly in Regime~B.
There, the deposited energy by FUV heating efficiently goes into the mechanical energy of the winds without being lost through cooling significantly.
The temperatures of winds in Regime~B are thus influenced by hydrodynamical effects and are approximately set by the characteristic sound speed, which is less than the equilibrium temperature.
The winds can typically form escaping winds without being inhibited by gravity, especially for outer radii, since $L_\mathrm{FUV} < L_\mathrm{F,c}$ is required for winds to be in Regime~C. 
We note, however, that $L_\mathrm{F,c} $ is inversely proportional to the ratio of the PAH abundance in the disk to the interstellar value. 
FUV-irradiated disks with lower PAH abundances, as often inferred for protoplanetary disks \citep[e.g.,][]{2007_Geers, 2013_Vicente}, are likely composed of the free-wind regime (Regime~B) and gravity-inhibited regime (Regime~C).

As for X-ray photoevaporation, the heating is caused by the excess energy remaining after the photoionization of various elemental species. 
In contrast to the EUV heating, not all primary nonthermal electron's energy goes to gas heating; the heating efficiency $f_\mathrm{h}$ is $\sim 10\,\%$ in atomic gas and $\sim 40\%$ in a molecular gas \citep{1996_Maloney}.  
Hence, we estimate the specific heating rate due to (unattenuated) X-rays as 
\[
\begin{split}
    \Gamma_\mathrm{X} & =   \frac{f_{\rm h}}{m}  \int  \sigma_\mathrm{X}(E) \frac{L(E)}{4\pi r^2} 
    \dd E
\end{split}
,
\]
where $\sigma_\mathrm{X}$ is the X-ray absorption cross-section, $L(E)$ is the specific X-ray luminosity, and $N_\mathrm{H}$ is the column density of hydrogen nuclei.
For $\sigma_\mathrm{X}$, we adopt the cross-section of \citet{2000_Wilms},
\[
    \sigma_\mathrm{X} = 2.27\e{-22} \braket{\frac{E}{1\keV}}^{-2.5},
\]
(the fit is adopted from \citet{2004_Gorti}).
We rewrite the heating rate to 
\[
\Gamma_\mathrm{X} 
       =   \frac{f_\mathrm{h}}{m} 
            \frac{L_\mathrm{X}}{4\pi r^2} \Bar{\sigma}_\mathrm{x0}
\]
using the total X-ray luminosity $L_\mathrm{X}$ and the energy-averaged, unattenuated cross-section
\[
    \Bar{\sigma}_\mathrm{x0}
    \equiv \dfrac{\int \sigma_\mathrm{X} L \dd E}{\int L \dd E}.
\]
Then, the characteristic sound speed for the X-ray heating is computed as 
\[
\begin{split}
    c_\mathrm{ch}^\mathrm{X} & 
    = \braket{\frac{\Gamma_\mathrm{X} R}{c_p}}^{1/3}
    \approx \braket{
    \frac{f_\mathrm{h}L_\mathrm{X} \Bar{\sigma}_\mathrm{x0} }{4 \pi c_p m R}
    }^{1/3}
    \\
    &   \sim    3.7 \kms 
    \braket{\frac{f_\mathrm{h}}{0.4}}^{1/3}
    \braket{\frac{L_\mathrm{X}}{10^{30}\erg\sec^{-1}}}^{1/3}
    \braket{\frac{c_p}{7/2}}^{-1/3}
    \\  &   \quad   \times
    \braket{\frac{m}{1.4m_\mathrm{H}}}^{-1/3}
    \braket{\frac{R}{1\au}}^{-1/3}
    \braket{\frac{\bar{\sigma}_\mathrm{x0}}{2\e{-22}\cm^2}}^{1/3}
    \end{split}
\]
Then, the critical X-ray luminosity is computed as 
\begin{equation}
    \begin{split}
    L_\mathrm{X,c} 
    &\equiv 
    \frac{2 \pi m GM_* c_\mathrm{eq}^\mathrm{X}}{f_\mathrm{h} \bar{\sigma}_\mathrm{x0}} \\
    & \approx 0.73\e{31}\erg \sec^{-1} 
    \braket{\frac{m}{1.4m_\mathrm{H}}}
    \braket{\frac{M_*}{1\Msun}}
    \\  &   \quad   \times
    \braket{\frac{c_\mathrm{eq}^\mathrm{X}}{3\kms }}
    \braket{\frac{f_\mathrm{h}}{0.4}}^{-1} 
    \braket{\frac{\bar{\sigma}_\mathrm{x0}}{2\e{-22}\cm^2}}^{-1}
    \end{split}
\label{eq:critical_Xray_luminosity:molecular}
\end{equation}
where $c_\mathrm{eq}^\mathrm{X}$ is the isothermal sound speed of X-ray-heated gas at the equilibrium temperature at the critical radius for the X-ray-heated gas where $c_\mathrm{eq}^\mathrm{X}$ equals the gravitational velocity 
(\eqnref{eq:vg_ceq}).

The nominal values of the variables in \eqnref{eq:critical_Xray_luminosity:molecular} have been chosen for molecular gas in a similar manner to \eqnref{eq:critical_FUV_luminosity}, assuming that relatively hard X-rays with $E \approx 1\keV$ would reach molecular layers. 
It means that \eqnref{eq:critical_Xray_luminosity:molecular} gives a criterion for the excitation of molecular winds by X-ray heating when the average energy of the given X-ray spectrum is $\approx 1\keV$. 
For atomic gas, softer X-rays contribute to the heating, but the nominal critical luminosity is likely almost the same at $L_\mathrm{X, c} \sim 1.4\e{31}\erg\sec^{-1}$, where we set $f_\mathrm{h} = 0.1$, $c_\mathrm{eq}^\mathrm{X} = 7\kms$, and $\bar{\sigma}_\mathrm{x0} = 1\e{-21}\cm^2$ assuming average X-ray energy of $\sim 0.5 \keV$ and an equilibrium temperature at the critical radius of $\sim 8000\Kelvin$. 

The estimated critical X-ray luminosity is comparable to a high-end X-ray luminosity observed towards T Tauri stars \citep[e.g.,][]{2007_Gudel}.  
If the intrinsic X-ray spectra of young low-mass stars had an average energy of observed X-rays of $\sim 1\keV$, most of the X-ray-irradiated disks would have a significant fraction of gravity-inhibited-wind regions beyond the critical radius ($\sim 2$--$12\au$), depending on the radial profile of the equilibrium temperature.  
This fact may give implications to a cause for the divergent conclusions on the effectivities of X-rays for driving photoevaporative winds in the literature (\secref{sec:Xray_divergent_conclusion}).

To summarize, low-mass stars need to have FUV and X-ray stellar luminosities comparable to or higher than the typical values in order to drive freely, escaping supersonic winds that are hardly inhibited by gravity. 
This is because the cross-sections for FUV and X-ray are generally small, and thus the heating rate, which is essentially set by the product of cross-section and local flux, also tends to get relatively small.
It means that cross-sections are a fundamentally important quantity to set the rapidness of heating.
While a lower cross-section lets photons reach a deeper interior of the disk and has the potential to drive dense winds, the photoheated gas is more susceptible to gravity due to the resulting long heating timescale, leading to weak subsonic winds. 
Hence, in contrast to often-adopted intuitive assumptions, lower cross-sections do not necessarily result in higher mass-loss rates, especially when the stellar luminosities are below the critical values.

From our discussions here, we can conclude hydrodynamical effects can overall influence temperature determination and, thereby, mass-loss rates of FUV- and X-ray-driven photoevaporative winds for low-mass stars. 
It highlights the importance of performing hydrodynamics simulations coupled with radiative transfer and chemistry for investigating FUV and X-ray photoevaporation around low-mass stars in general.  

\subsection{Implications to a Cause for Divergent Conclusions on X-ray Photoevaporation}
\label{sec:Xray_divergent_conclusion}
In the field of protoplanetary disk photoevaporation around low-mass stars, the debate over the effectiveness of X-ray heating in driving strong winds has been ongoing. 
Some suggest that X-rays can induce a substantial mass-loss rate of $\approx 10^{-8}\Msun \yr^{-1}$, whereas others have obtained significantly lower mass-loss rates, differing by more than an order of magnitude. 
The exact reasons for this divergence remain unclear, but it is presumed that the discrepancies stem from the different numerical methods and codes employed, 
specifically, whether the soft X-ray component is included and whether major coolants and adiabatic cooling are incorporated. 
Here, we will illustrate how our model, along with the critical X-ray luminosity (\eqnref{eq:critical_Xray_luminosity:molecular}), could provide a plausible and physically coherent explanation for the observed discrepancies across the previous models. 
Additionally, we will clarify why the diverse numerical methods contributed to the variations in mass-loss rates. 

The inquiry regarding the effectiveness of X-ray heating on disk dispersal began with \citet{2004_Alexander}, who used a 2D model to calculate the self-consistent hydrostatic density structure of an X-ray-irradiated disk with a simplified radiative transfer. 
Their findings suggested a negligible X-ray photoevaporation rate compared to EUV, indicating that X-ray-driven winds likely play a minor role in disk dispersal.

Subsequently, \citet{2009_Gorti} examined the X-ray heating's effectiveness on driving winds through a self-consistent hydrostatic model with radiative transfer and chemistry.
The authors obtained a similar conclusion to that of \citet{2004_Alexander}. 
Conversely, \citet{2009_Ercolano} reached a contradictory conclusion, positing that X-ray heating yields a larger mass-loss rate of $\gtrsim 10^{-9}\Msun\yr^{-1}$ than typical EUV photoevaporation rates ($\sim 10^{-10}\Msun\yr^{-1}$). 
They achieved this through a full 3D Monte Carlo radiative transfer \citep{2008_Ercolano}, 
attributing the discrepancies in the conclusions between the models to the interpretation of the calculation data and the hardness of the adopted X-ray spectra (see Section~5 of \citet{2009_Ercolano}). 
Notably, all models consistently indicated low mass-loss rates when adopting hard X-ray spectra, e.g., those that peak at $>1\keV$.

\citet{2010_Owen} conducted 2D hydrodynamics simulations of X-ray-driven photoevaporation with a simplified temperature treatment based on the radiative transfer calculations of \citet{2009_Ercolano}. 
The gas temperature is given by the equilibrium value, which is parametrized uniquely by the ionization parameter $\xi \equiv L_\mathrm{X}/\nh r^2$, and the temperature is given by referencing the $\xi$--$T$ table in the hydrodynamics simulations. 
They derived a large mass-loss rate of $\sim 10^{-8}\Msun\yr^{-1}$, which has been reproduced by follow-up studies using the same table-lookup technique \citep[e.g.,][]{2019_Picogna}.

Advancing these efforts, \citet{2017_Wang} and \citet{2018_Nakatanib} introduced updated hydrodynamical models, incorporating ray-tracing radiative transfer and mutually consistent thermochemistry and hydrodynamics. 
However, both studies also concluded that X-ray-driven winds make minor contributions to the mass-loss rate compared to UV-driven winds, though adopted UV and X-ray luminosities and spectra are diverse. 
It is noteworthy that both \citet{2017_Wang} and \citet{2018_Nakatanib} used a harder X-ray spectrum than that of \citet{2009_Ercolano} (see also Section~4.2 of \citet{2018_Nakatanib}). 

Further analysis by \citet{2022_Sellek} explored the impact of the X-ray spectrum hardness on the conclusions regarding the capability of driving winds through X-ray heating using a hydrostatic density grid while performing a full radiative transfer. 
Their work revealed that X-rays in the range of a few $100\eV$ were most effective, with more energetic ones ($\sim 1\keV$) providing insufficient heating to drive winds. 
This emphasizes the role of adopted irradiation X-ray spectra in the diverse conclusions on the effectiveness of X-ray heating among photoevaporation models.

While our phenomenological model for EUV photoevaporation does not directly address X-ray photoevaporation unless the equilibrium temperature is suitably modified as discussed in \secref{sec:discussion:critical_luminosities}, 
our model has fruitful insights into the factors contributing to the diverse conclusions on X-ray photoevaporation. 
Supporting the findings of \citet{2022_Sellek}, our model and discussions in \secref{sec:discussion:critical_luminosities} suggest that strong X-ray-driven winds tend to occur when the X-ray spectrum is relatively soft. 
This can be understood by the fact that the critical luminosity is generally smaller for larger average cross-sections (cf. \eqnref{eq:critical_Xray_luminosity:molecular}), i.e., softer X-ray spectra, 
which are advantageous in yielding a relatively short heating timescale. 

Another important criterion that our model introduces for interpreting the results of photoevaporation models is the estimated representative value of the critical X-ray luminosity (\eqnref{eq:critical_Xray_luminosity:molecular}). 
Typical X-ray luminosities of low-mass stars fall below this value, indicating that X-ray-driven winds are most likely in the regimes of steadily heated, free winds (Regime~B) and gravity-inhibited winds (Regime~C). 
This implies that the winds hardly reach the equilibrium temperature due to the excessively long heating timescale. 
Hence, hydrodynamical effects, namely internal energy loss by expansion and advection, are not negligible for X-ray-driven photoevaporation.
Determining the wind temperature based solely on the equilibrium value, as adopted in hydrostatic models and the $\xi$--$T$ table-lookup approach of the hydrodynamical models, effectively treats all regions as if they were in Regime~A. 
Consequently, this would lead to an overestimation of temperatures. 
This, in turn, can underestimate the gravity's effect on winds and subsequent pressure drop, ultimately resulting in an overestimation of mass-loss rates.

Considering the points raised above, it is not surprising that the coupled thermochemical-hydrodynamical models of \citet{2017_Wang} and \citet{2018_Nakatanib} uniformly found a minor contribution of X-ray-driven winds to photoevaporation. 
In \citet{2017_Wang}, it was assumed that all irradiating X-ray photons have the single energy of $1\keV$ with $L_\mathrm{X} \approx 2.5\e{30}\erg\sec^{-1}$, which is well below the estimated critical X-ray luminosity (\eqnref{eq:critical_Xray_luminosity:molecular}). 
On the other hand, \citet{2018_Nakatanib} (and a follow-up study by \citet{2021_Komaki}) used an observed X-ray spectrum towards TW~Hya \citep{2007_Nomura_II}, which is softer than that of \citet{2017_Wang} but harder than those of \citet{2009_Ercolano}. 
This X-ray spectrum yields $\sigma_\mathrm{x0} \approx 1\e{-21}\cm^2$, and X-ray heating primarily occurs mostly in a low column-density layer, where the gas is practically atomic. 
The corresponding critical X-ray luminosity is $L_\mathrm{X, c}\approx 0.56$--$1.3\e{31}\erg\sec^{-1}$ with $f_\mathrm{h}$ set to $0.1$, with uncertainties arising from the uncertain equilibrium sound speed $c_\mathrm{eq}^\mathrm{X}$. 
In any case, this critical luminosity is higher than the adopted X-ray luminosity $L_\mathrm{X} = 10^{30}\erg\sec^{-1}$
in their model, and this would explain why strong X-ray-driven winds are not observed. 

Our discussions here suggest a gravitationally influenced nature for (hard) X-ray-driven winds around low-mass stars. 
However, this does not mean that X-rays are incapable of driving winds; rather, they are easily inhibited by gravity with typical luminosities and tend to be subsonic at lower temperatures than the equilibrium values due to long heating timescales.
To obtain physically meaningful X-ray photoevaporation rates, it is imperative to conduct hydrodynamical simulations coupled with radiative transfer and chemistry, employing either a sufficiently large computational boundary or an appropriate outer boundary condition to prevent spurious reflection of subsonic flows at the outer computational boundary and avoid neglecting nonzero mass-loss rates. 
This aspect remains open to date, leaving X-ray photoevaporation rates still uncertain. 

\subsection{Possible Time-Dependent Class Variation by Stellar and Disk Evolution}
\label{sec:discussion:time_dependent_evolution}

The classification and dynamics of disk photoevaporation are subject to variations contingent upon the evolutionary stages of both the star and the disk. 
Various factors stemming from the evolution of both stellar and disk components wield influence over the classification and dynamics of disk photoevaporation. 
This section provides a qualitative exploration of representative factors that contribute to such variations.

The accretion of disk material onto the star can generate strong UV and X-ray emissions by liberating the gravitational energy of the accreting gas.
As for FUV, the accretion-generated component can dominate the stellar component at early stages ($\sim 1\Myr$) where the accretion rate is relatively high \citep{2009_Gorti}.  
The FUV luminosities likely exceed the critical value (\eqnref{eq:critical_FUV_luminosity}) at this stage if the disk still retains sufficiently abundant PAHs/very small grains. 
This leads to driving a vigorous FUV photoevaporation. 
However, as the system evolves and the accretion rate diminishes, stellar emissions become more dominant.
The luminosities are relatively small compared to the early stages, correspondingly resulting in weaker photoevaporation.
Additionally, if small grains are depleted in later stages, FUV photoevaporation can further attenuate. 

On the other hand, strong accretion implies strong inner MHD winds capable of shielding UV and X-ray photons that could otherwise reach the wind region $x > 1$ \citep[e.g.,][]{2010_Turner, 2018_Takasao, 2022_Takasao, 2018_Fang, 2023_Fang}.  
The column density of the inner wind is higher when the system is younger and has a stronger wind. 
Only hard photons manage to penetrate the inner region with an attenuated total flux, leading to a hard spectral class and weakened photoevaporation. 
However, as the system evolves, the surface density reduces relatively quickly at inner radii since MHD winds are more efficient in removing mass there \citep{2016_Bai, 2016_Suzuki}.
Then, inner winds and disks become more transparent to UV and X-ray radiation, the spectrum transitions to softer.
This transition facilitates photons reaching the outer region without significant attenuation in the inner region, potentially exciting vigorous winds.
This interaction between radiation and inner winds is supported by observational evidence \citep{2020_Pascucci}. 
Hence, as well as the time-variation of accretion-powered UV and X-ray, accounting for spectral class and wind regime variations is crucial for understanding disk dispersal.

Moreover, stellar evolution, particularly for stars exhibiting time-varying surface magnetic activity, significantly influences the classification and dynamics of disk photoevaporation \citep{2021_Kunitomo, 2023_Nakatani}. 
For instance, intermediate-mass stars may possess a surface convective zone at ages below $\sim 1\Myr$, resulting in strong EUV and X-ray emissions from magnetic activity. 
During this phase, the spectral class is likely H, and strong photoevaporation may ensue due to potentially high UV and X-ray luminosities.
However, as these stars lose their surface convective zones around ages $\sim 1$--$10\Myr$, EUV and X-ray emissions can significantly weaken. 
It results in softer EUV emissions from moderately hot photospheres, and photoevaporation is weakened due to a low EUV emission rate. 

In conclusion, the long-term variability of stellar luminosity, stellar activity, disk accretion rate, and grain abundance significantly shape the classification and dynamics of photoevaporation at different stages of disk dispersal.
This indicates that the wind velocity, temperature, and ionization state significantly vary according to the system's evolutionary state, 
underscoring the importance of considering temporal variations in photoevaporation for understanding overall disk evolution and interpreting observational data on winds.


\subsection{Model Limitation and Caveats}   \label{sec:discussion:model_limitation}
Our model is inherently subject to uncertainties originating from the assumptions we have adopted for simplicity and therefore has caveats to interpret the results. 
Here, we will address these points in this section. 
We emphasize that, regardless of these model limitations and caveats, our model offers valuable insights into the fundamental physics underlying EUV-driven photoevaporative winds and systematic variations of wind characteristics across a broad parameter space. 

\paragraph{Incorporation of Helium photoionization} 
Our model focuses solely on EUV heating associated with hydrogen photoionization, neglecting potential contributions from helium photoionization. 
While the ionization potentials of \ion{He}{1} and \ion{He}{2} are relatively high (approximately $24.6\eV$ and $54.4\eV$, respectively), their overall impact on the photoheating rate and ionization state of hydrogen in photoevaporative winds is expected to be modest.
Despite the about an order of magnitude higher absorption cross-section per \ion{He}{1}/\ion{He}{2} than the cross-section per \ion{H}{1} in the EUV range, 
the elemental helium abundance of $\approx 10\%$, limits the total photoheating rate increment to only a few-fold at most when compared to the heating rate used in our model ($\Gamma_\mathrm{EUV}$; \eqnref{eq:gamma_EUV}). 
Similarly, the electron consumption resulting from the recombination of ionic helium species is projected to have a perturbing impact on the ionization state of hydrogen within photoevaporative winds. 
Consequently, the order estimation in the present study is expected to remain largely unaltered even with the incorporation of helium contributions. 

\paragraph{Self-attenuation effects in photoevaporative winds} We have approximated the absorption attenuation factors, $\chi_\mathrm{e}$ and $\chi_\mathrm{i}$ (see Eqs.\eqref{eq:gamma_EUV} and \eqref{eq:kioni}), as near unity, thereby assuming optically thin photoevaporative winds. 
However, especially in cases involving atomic flows, the photoevaporative winds in the inner region can shield to prevent EUV from reaching the outer region.
This self-attenuation effect can reduce the attenuation factors significantly.
In this case, the actual flux reaching the disk is smaller than the value used in our model for a given $\phieuv$, and thus the actual heating timescale can correspondingly become longer if we incorporate corrections due to self-attenuation. 
This would lead to an increase in the critical EUV emission rate $\Phi_\mathrm{c}$ by several times and corresponding adjustments in the boundaries of wind regimes on the phase diagram. 
Hence, the derived critical EUV emission rate should be interpreted as a lower limit, suggesting that the derived radius limits presented in \secref{sec:limiting_radius} should be considered upper limits. 

Since the self-attenuation is expected to be increasingly pronounced at outer radii, the actual boundary between Regimes~A and B may not adhere to a simple linear relation as depicted in our model but rather may be described by a convex downward function.  
A similar is true for the boundary between Regimes~B and C; the gravity-inhibited regime can be somewhat radially extended than the predictions of our model. 

Additionally, in the 2D cross-sectional view of a photoevaporating disk, the wind region would exhibit a vertically stratified structure with Regimes~A, B, and C present in distinct layers. 
This stratification arises due to the irradiating spectrum becoming harder as it is attenuated. 
In the most attenuated layer, both the photon number flux and the average cross-section are small, resulting in the layer being in a gravity-inhibited regime. 
In a moderately attenuated layer, the photoheated layer can exhibit characteristics of Regimes~A and B if $\phieuv$ is sufficiently high. 
We will explore this aspect in more detail in \secref{sec:comparison_HD_simulations}.


\paragraph{Inclusion of neutral coolants} 
Our model primarily relies on coolants effective in \ion{H}{2} regions, such as forbidden lines of ionic metals, Ly$\alpha$, and radiative recombination, which results in an equilibrium temperature of $\approx10^4\Kelvin$ for both atomic and ionized gas. 
However, as also discussed in \secref{sec:discussion:critical_luminosities}, atomic and molecular cooling, such as infrared emission from \ion{O}{1} and \ce{H2}, can be significant in the neutral layer of a disk.
This can lead to a substantially lower equilibrium temperature on the order of $\sim 10^2$--$ 10^3\Kelvin$. 
Thus, the equilibrium temperature would, in principle, depend on whether the gas is in a neutral or ionized state, which is determined by the luminosity of the ionizing radiation. 

This realization suggests the existence of a new Type~I wind regime characterized by winds exhibiting equilibrium temperatures typical of neutral gas.
This regime would fall within the Type~I regime and bridge connections to the Type~II regime at the high-flux parameter space.
The prevalence of this regime is expected, particularly for H-class EUV spectra, which yield a low $\avesigma$, creating the base within the deep interior of the neutral layer. 
Therefore, an important next step for the phenomenological model is to incorporate the effects of neutral coolants to formulate the new Type~I regime and determine the corresponding parameter space within the phase diagram. 

\paragraph{Investigation of the inner region}
While our study primarily focuses on the outer radius $x > 1$, where vigorous winds are anticipated due to weak gravity, investigating the inner region $x < 1$ is also necessary. 
This disk corona, i.e., nearly hydrostatic gas in $x <1$, has the potential to produce weak winds whose escaping rate exponentially decays inward, potentially serving as a source for time variation. 
Investigating this region could give valuable insights into achievable velocities and temperatures therein, which could provide surface mass-loss rates available for population synthesis studies and be used to discuss the observability of infrared wind tracers. 

\paragraph{Uncertainty in recombination timescale}
There remains uncertainty in the recombination timescale used in this study (\eqnref{eq:trec_base_density}), particularly regarding the applicability of the base density formulation (\eqnref{eq:base_density}) across different EUV emission rates and spectra. 
If the base density has another formulation within certain parameter spaces, it could impact the boundary of Types~I and II regimes. 
Although our key findings --- e.g., the presence of two wind types with Type~II being achieved at a relatively high $\phieuv$ --- qualitatively remain unchanged by such modifications, it is necessary to examine the general applicability of \eqnref{eq:base_density} by comparing the base density with numerical calculations. 

Additionally, our model has not accounted for the density reduction due to the expansion of evaporating gas, leading to an overestimation of additional energy supply through reionizations within the ionized winds.
Addressing this is another crucial aspect that needs to be incorporated into an updated model.

\paragraph{Integration with MHD winds}
Magnetohydrodynamics (MHD) winds are another important disk-dispersal process, especially during the early stage of protoplanetary disk evolution when the disk is relatively massive \citep[e.g.,][see also \secref{sec:discussion:time_dependent_evolution}]{2014_Turner, 2022_Pascucci}. 
Our model could be applied to the determination of MHD winds' thermochemical states if the winds have sufficiently low densities, allowing photochemical processes to dominate other heating sources \citep[e.g., ambipolar diffusion ][]{2001_Garciab, 2001_Garcia}. 
The velocity achieved by MHD winds would set whether the gas can reach the equilibrium temperature. 
In high-velocity winds, the gas may not have enough time to be heated, resulting in relatively low temperatures.  
Thus, corresponding modifications are necessary to the definition of $c_\mathrm{ch}$ (Eq.\eqref{eq:c_ch_def}) for the temperature estimation with accounting for this effect. 

Similarly, since the base density of MHD winds is not related to the photoionization and recombination rates in general, the estimation of the recombination timescale also needs to be modified. 
With the updated recombination timescale, the same criteria (Eqs.\eqref{eq:typeI_condition} and \eqref{eq:typeII_condition}) would still be applicable to infer the ionization state of the winds. 
Overall, while our model is not directly applicable to MHD winds, it could be adapted by treating other heating processes, MHD winds' velocities, and base densities. 

\subsection{Prospects for Potential Applications to Photoevaporative Wind Observations}
\label{sec:discussions:wind_observations}

Since the main scope of this paper is to broaden our understanding of EUV-driven photoevaporation in a physically consistent manner, we will leave a detailed discussion on the observational applications and tests of our model for future studies. 
Nevertheless, we describe here the potential utility of our model in investigating photoevaporative winds and stellar parameters and outline the steps that need to be addressed to that end. 

We have demonstrated that EUV-driven winds have a temperature gradient in the radial direction, even within the \ion{H}{2} region. 
When the spectrum is soft, such as those softer than a blackbody spectrum with an effective temperature of $T_\mathrm{eff} \lesssim 5\e{4}\Kelvin$, the \ion{H}{2} region extends across the isothermal and non-isothermal wind regimes (Regimes~A-II and B-II).
On the other hand, with hard spectra, the thermal balance dictates the temperature of the \ion{H}{2} region, resulting in isothermal winds. 
In both cases, there is a maximum radius beyond which the \ion{H}{2} region cannot extend (\secref{sec:limiting_radius}), and the winds transition to being atomic with lower temperatures. 

These findings suggest that the observed size of the ionized region size can vary depending on the ionized gas tracers used, which have different excitation temperatures. 
By comparing the sizes of the ionized region using different tracers, we can estimate the average EUV energy of the source and then determine the source's EUV emission rate $\phieuv$, following a similar procedure as outlined in \secref{sec:limiting_radius}. 

The expected structure of emissions from various tracers can be outlined as follows.
When the EUV spectrum is relatively hard, high-temperature ($\approx 10^4 \Kelvin$) ionized gas tracers (e.g., [\ion{Ne}{2}]\,12.81$\mum$, [\ion{O}{2}] and [\ion{O}{3}] optical lines) exhibit a compact emission (Regime~A-II). 
This region is surrounded by a more extended emission of neutral gas tracers (e.g., [\ion{O}{1}]\,6300\AA, [\ion{O}{1}]\,63$\mum$, and [\ion{C}{2}]\,158$\mum$) with a decrease in temperature towards the radial direction (Regimes~A-I and B-I). 
In contrast, when the spectrum is soft, the high-temperature ionized region is encircled by more extended emissions from intermediate-temperature ($\approx 5000$--$10^4\Kelvin$) ionized gas tracers (e.g., [\ion{S}{2}] and [\ion{N}{2}] optical lines; Regime~B-II),
followed by lower-temperature neutral gas tracers extending further outward (Regime~B-I). 
Note that $\phieuv$ must exceed the critical EUV emission rates (cf. Eqs.\eqref{eq:critical_luminosity} and \eqref{eq:secondary_critical_luminosity}) to yield the high-temperature ionized region.

Recent JWST observations have revealed relatively compact [\ion{Ar}{2}] emissions compared to [\ion{Ne}{2}] and [\ion{Ne}{3}] emissions for T~Cha \citep{2024_Bajaj, 2024_Sellek}. 
These differences in emission sizes, however, are concluded to be likely due to distinct photoionization processes for each tracer, specifically EUV ionization versus X-ray ionization. 
Considering that low-mass stars commonly exhibit X-rays, generalizing our model to the X-ray wavelengths appears to be a crucial step to making it available for inferring spectra and luminosities of low-mass stars based on the different emission extensions of wind tracers. 

Nevertheless, our model can further be expanded to develop an analytic model for X-ray photoevaporation, as discussed in \secref{sec:discussion:critical_luminosities}. 
By integrating our EUV model with such an X-ray photoevaporation model, an analytical approach can be formulated to estimate the line intensities of ionized gas tracers, similar to the approach by \citet{2009_HollenbachGorti}. 
Such a model provides analytical predictions generalized by incorporating hydrodynamical effects on temperature and ionization state determination, 
thereby updating the currently existing models that assume the equilibrium temperatures determined by thermochemical balance.  

Additionally, those models are of great use for investigating free-free emissions observed towards young sources and estimating $\phieuv$ based on the observed free-free flux density \citep{2012_Pascucci, 2021_Ricci}. 
The resulting spatial distribution of free-free emissions is valuable for interpreting the spatially resolved data of future ngVLA observations and exploring the origins of these emissions. 


\section{Summary and Prospects}   \label{sec:summary}

In the canonical picture of EUV photoevaporation, the photoevaporative winds consist of ionized gas at equilibrium temperatures $\approx 10^4\Kelvin$, corresponding to wind speeds $\approx 10\kms$ beyond the gravitational radius \citep{1994_Hollenbach} or critical radius \citep{2003_Liffman}. 
We have delved into the conditions under which this canonical picture holds true, constructing a phenomenological model based on the analytic approach of \citet{1983_Begelman}. 
The resulting generalized picture applies across the whole range of conditions where photoevaporation is possible, and yields estimates for the temperature, speed, and ionization state that EUV-driven disk winds exhibit as functions of four parameters: EUV luminosity, spectral hardness, distance from the radiation source, and the source's mass. 
These are the main findings: 
\begin{enumerate}
    \item   As in the model of \citet{1983_Begelman}, photoevaporative winds occur in three hydrodynamical regimes: isothermal (Regime~A); free-flowing (Regime~B), and gravity-inhibited (Regime~C). 
    Isothermal winds occur when the heating timescale is shorter than both the wind crossing and gravitational timescales, whereas gravity-inhibited winds result when the heating timescale is longer than the gravitational timescale. 
    In all other cases, the winds transition into free winds.

    \item The canonical picture applies only in a region of the parameter space characterized by high EUV fluxes. 
    This parameter region is denoted Regime~A-II and is represented by red shading without vertical gray stripes in the phase diagrams of Figures~\ref{fig:H-class}, \ref{fig:Sa-class}, \ref{fig:Sr-class}, and \ref{fig:classmaps}, as summarized in \tref{tab:sound_spees}. 

    \item   To form isothermal winds, the EUV emission rate must at least exceed the critical value (\eqnref{eq:critical_luminosity}), which is defined by the stellar mass, equilibrium temperature, and spectral hardness. 
    
    \item For lower EUV fluxes, the winds are cooler than the equilibrium temperature due to a longer heating timescale. 
    This results in two distinct regimes: 
        \begin{itemize}
            \item At larger disk radii where gravity is weak, the low-temperature winds escape freely without significant gravitational influence (Regime~B; green and orange regions in the figures). 
            \item Close-in, gravity is strong and inhibits the winds, leading to a substantial pressure drop (Regime~C; blue regions in the figures). 
        \end{itemize}
        
    \item The phase diagrams divide into three classes depending on the hardness of the EUV spectrum, as parameterized by $\varepsilon$ and $q$ (Eqs.~\eqref{eq:tildes} and \eqref{eq:q}).    
        \begin{itemize}
            \item   {\it H-class}: Characterized by hard spectra, depositing a relatively large energy per photoionization. 
            Ionized winds in this class are always characterized by isothermal flows at the equilibrium temperature (Regime~A-II), with the possibility of isothermal atomic winds in a limited parameter space (Regime~A-I). 
            Spectra from young low-mass stars with strong surface magnetic activity are expected to fall into this class. 

            \item   {\it Sa-class}: 
            One of the two soft-spectrum classes, where advection is more effective in maintaining atomic hydrogen in the winds than recombination. 
            With little energy deposited per photoionization, multiple reionizations are necessary before the gas heats to the equilibrium temperature. 
            This results in cooler ionized winds that either escape freely (Regime~B-II) or are gravity-inhibited (Regime~C-II). 
            Spectra in this class may be rarer than the H-class spectra, with main-sequence intermediate-mass stars potentially falling in the Sa class.

            \item   {\it Sr-class}: Same as Sa-class, but with recombination more effective than advection in maintaining atomic hydrogen in the winds. 
            The key distinction is that Sa-class has a parameter region where energy deposition through reionizations is negligible in ionized winds (see \secref{sec:sub:Sa-class}).
            Sr-class appears to occur rarely if ever in the context of protoplanetary disk dispersal. 
        \end{itemize}
\end{enumerate}

We have evaluated the model's predictions by comparing them against hydrodynamics simulations treating the radiative transfer and disequilibrium thermochemistry (\secref{sec:comparison_HD_simulations}). 
There is broad concurrence between the analytic model and the numerical results across orders of magnitude in the EUV emission rate and a wide range in spectral hardness.  Refining the model and ensuring its applicability across various EUV environments will nevertheless require more exhaustive assessments with diverse EUV spectra and flow geometries. 

To demonstrate the practical utility of the derived typical temperature and speed of the winds, 
we have estimated the limiting radius of the \ion{H}{2} region to which ionized winds can extend (\secref{sec:limiting_radius}). 
We have shown that this limiting radius is determined by the EUV emission rate, average absorption cross-section, isothermal sound speed at the equilibrium temperature, and spectral hardness. 
Our analysis suggests that the EUV emission rate can potentially be constrained by comparing the radial sizes of the \ion{H}{2} regions observed in ionized gas tracers with different excitation temperatures --- an avenue we intend to explore in future investigations. 
Furthermore, our analysis has revealed that while applying the canonical picture of EUV photoevaporation to disks around massive stars is well justified, the same does not hold for disks around young low-mass stars. 
Therefore, when interpreting observations of ionized wind tracers or conducting numerical simulations of photoevaporation for low-mass stars, it is crucial to take into account the finite timescales of photoheating and photoionization. 

We have also estimated the work efficiency, or ratio of the work necessary to move the gas against gravity to the energy deposited in that gas through photoheating (\secref{sec:work_efficiency}). 
We find that the efficiency is of order unity for gravity-inhibited winds and is highest when the gas is quasi-statically heated. 
Conversely, the efficiency drops as the EUV emission rate increases and more energy is directed towards gas kinetic and thermal energy or is lost to radiative cooling. 
At the highest fluxes, the efficiency scales as $\phieuv^{-1/2}$, where $\phieuv$ represents the rate at which the source emits EUV photons. 
The scaling in the efficiency aligns with that found by numerical simulations in the context of planetary atmospheric escape \citep{2009_MurrayClay, 2016_Owen}, 
suggesting that our results could be applied to understanding EUV photoevaporation across various classes of astrophysical objects (Mitani et al. in prep). 

The analytic model has the potential to be extended to FUV and X-ray photoevaporation with suitable modifications, providing a tool to explore the effectiveness of these energy bands in disk dispersal over an even wider parameter space (\secref{sec:discussion:critical_luminosities}). 
Additionally, the model has potential to explain divergent conclusions in the literature regarding X-ray photoevaporation (\secref{sec:Xray_divergent_conclusion}). 
We acknowledge that uncertainties persist in the analysis, leaving room for further refinements, including those discussed in \secref{sec:discussion:model_limitation}. 
Specifically, it is crucial to validate the predictions by comparing them against numerical simulations that treat the radiative transfer and time-dependent photochemistry.
Additionally, it will be essential to update the model to accommodate cases where the spectrum covers wider ranges in photon energy.
With these caveats in mind, we have discussed prospected applications of our model to observations \secref{sec:discussions:wind_observations}.

Since the model is analytic, it can be extended to enable exploring the detectability of photoevaporative winds and deriving mass-loss rates across a wide parameter space, which we hope will yield fresh insights into disk dispersal from both observational and theoretical perspectives. 
Moreover, the model provides a solid foundation for understanding photoevaporation in astrophysical objects including planets, molecular clouds, and galactic mini-halos.



\acknowledgments %
We thank the reviewer for the practical comments. 
R.N. is supported by the Japan Society for the Promotion of Science (JSPS), Overseas Research Fellowship. 
S.T. was supported by the JSPS KAKENHI grant Nos. JP21H04487, JP22K14074, and JP22KK0043.
This research was performed in part at the Jet Propulsion Laboratory, California Institute of Technology, under contract 80NM0018D0004 with the National Aeronautics and Space Administration and with the support of the NASA Exoplanets Research Program through grant 17-XRP17\_2-0081 to N.J.T.
Numerical computations were carried out on the Cray XC50 at the Center for Computational Astrophysics, National Astronomical Observatory of Japan.
\copyright 2024. All rights reserved.

\vspace*{1cm}
\software{Numpy \citep{numpy}, Matplotlib \citep{matplotlib}, Astropy \citep{astropy:2013,astropy:2018,astropy:2022}, SciPy \citep{scipy}
}
\vspace*{5cm}



\bibliographystyle{aasjournal}
\bibliography{bibsamples}

\begin{appendix}
\section{Basic notes for the analytical model}
\label{sec:list_of_variables}
We list symbols used in our phenomenological model in \tref{tab:variables}. 
\begin{table*}[]
    \centering
    \begin{tabular}{L l r}
    \text{Expression}
    &
    Description
    &
    Definition
    \\
    \hline
    \alpha  &   Case~B recombination rate coefficient   &   \eqnref{eq:recombination_coeff}  \\
    \alpha_\mathrm{eq}  &   $\alpha$ at the equilibrium temperature ($\approx 2\e{-13}\cm^3\sec^{-1}$)&   \eqnref{eq:recombination_coeff}  \\
    \beta   &   Power index for the temperature profile of $\alpha$ &   \eqnref{eq:recombination_coeff}  \\
    C       &   Empirical dimensionless factor for base density   &   \eqnref{eq:base_density}   \\
    c_\mathrm{s}    &   Typical isothermal sound speed of a wind    &   \\
    c_\mathrm{eq}   &   Isothermal sound speed at the equilibrium temperature ($\approx 10\kms$)    &   Beginning of \secref{sec:critical_radius}   \\
    c_\mathrm{ch}   &   Characteristic sound speed  &   \eqnref{eq:c_ch_def} \\
    c_\mathrm{ch}^\mathrm{I}    &   Characteristic sound speed for Type~I winds &   \eqnref{eq:cchI_def} \\
    c_\mathrm{ch}^\mathrm{II}    &   Characteristic sound speed for Type~II winds &   \eqnref{eq:chII_original_def} \\
    c_p     &   Specific heat at constant pressure  &   \eqnref{eq:critical_radius}\\
    c_1     &   Approximated $c_\mathrm{ch}^\mathrm{II}$ in the slow-recombination limit    &   \eqnref{eq:c2_in_slow_recombination} \\
    c_2     &   Approximated $c_\mathrm{ch}^\mathrm{II}$ in the rapid-recombination limit   &   \eqnref{eq:cchII_def}    \\
    \Gamma_\mathrm{EUV} &   Specific EUV photoheating rate  &   \eqnref{eq:gamma_EUV} and \eqnref{eq:gamma_EUV:2}\\
    \delta  &   Ratio of attenuated flux to the unattenuated flux   &   \eqnref{eq:gamma_EUV}    \\
    \langle\Delta E\rangle_i    &   Frequency-average deposited energy per photoionization  &   \eqnref{eq:gamma_EUV}    \\
    \aveenergy   &   $\langle\Delta E\rangle_i $ at $N_\mathrm{H} = 0$   &   \eqnref{eq:gamma_EUV:2}\\
    \varepsilon &   Normalized $\bar{E}_0$ by $mc_p \ceq^2$  &   \eqnref{eq:tildes}   \\
    F_\nu       &   Specific photon number flux     &   \secref{sec:euv_heating_rate}   \\
    F_0     &   Unattenuated total photon number flux   &   \secref{sec:euv_heating_rate}   \\
    f_\mathrm{eq}   &   Ratio of the secondary critical EUV emission rate to $\Phi_\mathrm{c}$      &   \eqnref{eq:secondary_critical_luminosity}\\  
    k_\mathrm{ioni} &   Photoionization rate coefficient    &   \eqnref{eq:tioni} \\
    M_*     &   Mass of the radiation source    &   \secref{sec:critical_radius}    \\
    m       &   Gas mass per hydrogen nucleus   &   \eqnref{eq:gamma_EUV}\\
    \mach   &   Typical Mach number of a wind   &   \eqnref{eq:slow_recombination_condition} \\
    \nh     &   Hydrogen nuclei number density  &   \secref{sec:euv_heating_rate}    \\
    N_\mathrm{HI}   &   \ion{H}{1} column density & \secref{sec:euv_heating_rate}   \\
    \Phi_\mathrm{EUV}   &   EUV emission rate of the source     &   \secref{sec:euv_heating_rate}   \\
    \Phi_\mathrm{c} &   Critical EUV emission rate &    \eqnref{eq:critical_luminosity}  \\
    \Phi_\mathrm{c}^\mathrm{II} &   Secondary critical EUV emission rate, $f_\mathrm{eq}\Phi_\mathrm{c}$ &  \eqnref{eq:secondary_critical_luminosity}    \\
    \Phi_\mathrm{c,rec}^\mathrm{II} &   Secondary critical EUV emission rate for recombination-dominated Type~II winds &  \eqnref{eq:cchII_def}    \\
    \varphi &   Normalized EUV emission rate, $\Phi_\mathrm{EUV}/\Phi_\mathrm{c}$    &   \eqnref{eq:cchI_def} \\
    \varphi_\mathrm{II} &   EUV emission rate normalized by $\Phi_\mathrm{c,rec}^\mathrm{II}$   &   \eqnref{eq:cchII_def}    \\
    \varphi_\mathrm{g}  &  Boundary of Regions~B-II and C at which $c_\mathrm{ch}^\mathrm{II} = v_\mathrm{g}$   &   \eqnref{eq:varphi_g}\\
    \psi    &   Photoevaporation parameter in \citet{1989_Bertoldi} &   \eqnref{eq:trec_tcross}  \\
    q   &   Spectrum hardness parameter for recombination   &   \eqnref{eq:q}  \\  
    r               &   Spherical radius from the source    &   \secref{sec:euv_heating_rate}   \\
    R               &   Cylindrical distance from the source  &   \secref{sec:critical_radius}  \\
    R_\mathrm{c}    &   Critical radius     &   \eqnref{eq:critical_radius}  \\
    \langle\sigma\rangle    &   Frequency-average cross-section &   \eqnref{eq:gamma_EUV}    \\
    \bar{\sigma}_0  &  $\langle\sigma\rangle$ at $N_\mathrm{H} = 0$   &   \eqnref{eq:gamma_EUV:2}\\
    t_\mathrm{g}    &   Gravitational timescale & \eqnref{eq:gravitational_timescale}\\
    t_\mathrm{ioni} &   Photoionization timescale   &   \eqnref{eq:tioni}\\
    t_\mathrm{rec}  &   Recombination timescale     &   \eqnref{eq:trec_base_density}    \\
    v_\mathrm{g}    &   Gravitational velocity  &   \eqnref{eq:vg_ceq}   \\
    x       &   Normalized distance, $R/R_\mathrm{c}$    &   \secref{sec:critical_radius}\\
    \chi_\mathrm{e} &   Attenuation factor due to absorption for photoheating & \eqnref{eq:gamma_EUV:2}  \\
    \chi_\mathrm{i} &   Attenuation factor due to absorption for photoionization    &   \eqnref{eq:tioni}    \\
    y_\mathrm{HI}   &   Atomic hydrogen abundance   &   \eqnref{eq:gamma_EUV}
    \end{tabular}
    \caption{Symbols for the fundamental variables used in our phenomenological model (Sections~\ref{sec:analytic_theory}--\ref{sec:cs}). We omit those that appear in later sections.}
    \label{tab:variables}
\end{table*}

Throughout this paper,
we use ``region'' to denote spatial extents, such as the \ion{H}{2} region, region of $x > 1$, or region where $s < R$. 
The word ``phase diagram'' refers to the $x$--$\varphi$ map. 
In contrast, the $\varepsilon$--$q$ plane is referred to as the ``spectral-hardness  diagram.''
We use ``parameter space'' to indicate a certain area of the phase diagrams and the spectral-hardness diagram. 
The term ``regime'' is employed to describe the parameter spaces where the same qualitative properties are shared, such as Regime~A, Type~I regime, among others. 
Therefore, ``parameter space'' and ``regime'' are utilized in a different context from ``region'' consistently.

\section{Mathematical proof for Type~II characteristic sound speed exceeding the equilibrium sound speed in H-class}
\label{sec:math_proof}

In \secref{sec:typeII:regime:H}, we mention that $c_\mathrm{ch}^\mathrm{II} > c_\mathrm{eq}$ generally holds for the H-class ($\left\{(\epsilon,q)|~\epsilon > 1 ~\cup~ \epsilon q > 1\right\}$), and it can be mathematically proven from \eqnref{eq:chII_original_def}. 
We demonstrate this proof in this Appendix.

When $\epsilon > 1$, it is straightforward to see that $c_\mathrm{ch}^\mathrm{II} > c_\mathrm{eq}$, as it necessitates
\[  
    \braket{\frac{c_\mathrm{ch}^\mathrm{II}}{c_\mathrm{eq}}}^2 > \epsilon
\]
according to \eqnref{eq:chII_original_def}. 
On the other hand, when $\epsilon < 1$ and $\epsilon q > 1$, the parameter space of interest is $\varphi > \mathrm{max}\braket{1, q} \epsilon x = \epsilon q x$. 
In this range, 
\[
\begin{split}
    \braket{\frac{c_\mathrm{ch}^\mathrm{II}}{c_\mathrm{eq}}}^2 
    & = \epsilon  
    + \braket{\frac{\varphi \epsilon  q}{ x }}^{1/2}
    \braket{\frac{c_\mathrm{ch}^\mathrm{II}}{c_\mathrm{eq}}}^{-(\beta+1)}
    \braket{1 - \varphi^{-1}\epsilon  x \frac{c_\mathrm{ch}^\mathrm{II}}{c_\mathrm{eq}}}
    \\ & 
    > \epsilon + \epsilon q \braket{\frac{c_\mathrm{ch}^\mathrm{II}}{c_\mathrm{eq}}}^{-(\beta+1)}
    \braket{1 - \frac{1}{q}\frac{c_\mathrm{ch}^\mathrm{II}}{c_\mathrm{eq}}}
\end{split}
\]
is required from \eqnref{eq:chII_original_def}. 
The RHS of this inequity is a monotonically decreasing function of $c_\mathrm{ch}^\mathrm{II}$ that takes a value of $\epsilon q$ at $c_\mathrm{ch}^\mathrm{II}/c_\mathrm{eq} = 1$, 
while the left-hand side (LHS) takes unity at the same $c_\mathrm{ch}^\mathrm{II}$. 
This indicates that the intersection between the functions of the LHS and RHS is present at $c_\mathrm{ch}^\mathrm{II}/c_\mathrm{eq} > 1$. 
Denoting $c^\prime $ as the coordinate of the intersection, 
the above inequity is rewritten as
\[
    \frac{c_\mathrm{ch}^\mathrm{II}}{c_\mathrm{eq}} > \frac{c^\prime}{c_\mathrm{eq}} > 1. 
\]
Thus, $c_\mathrm{ch}^\mathrm{II} > c_\mathrm{eq}$ is proven for Type~II winds in the H-class. 

\end{appendix}

\end{document}